# Reactive polar mesogenic self-assembly approach enables domain-programmable polymer ferroelectrics

*Fan Ye,*[1,2‡] *Minghui Deng,*[1‡] *Yuyang Zheng,*[1] *Xiujuan Liu,*[1] *Xiuhu Zhao,*[3,4] *Haowei Jiang,*[5] *Yanyun Hou,*[1] *Bingyu Zou,*[1] *Neng-Ang Peng,*[1] *Shuo Zhao,*[6] *Kutay Sağdıç,*[3,4] *Danqing Liu,*[3,4] *Yang Shen,*[6] *Yan-Qing Lu,*[7] *Satoshi Aya,*[1,2]* *Mingjun Huang*[1,2]*

[1] School of Emergent Soft Matter, State Key Laboratory of Advanced Papermaking and Paperbased Materials, South China University of Technology, Guangzhou 510640, China.

[2] Guangdong Provincial Key Laboratory of Functional and Intelligent Hybrid Materials and Devices, Guangdong Basic Research Center of Excellence for Energy & Information Polymer Materials, South China University of Technology, Guangzhou 510640, China.

[3] Department of Chemical Engineering and Chemistry, Eindhoven University of Technology, Eindhoven 5612 AE, The Netherlands.

[4] Institute for Complex Molecular Systems (ICMS), Eindhoven University of Technology, Eindhoven 5612 AE, The Netherlands.

[5] National Engineering Research Center of Novel Equipment for Polymer Processing, Department of Mechanical and Automotive Engineering, South China University of Technology, Guangzhou 510641, China.

[6] State Key Laboratory of New Ceramic Materials, School of Materials Science and Engineering, Tsinghua University, Beijing 100084, China.

[7] National Laboratory of Solid-State Microstructures, Key Laboratory of Intelligent Optical Sensing and Manipulation, College of Engineering and Applied Sciences, and Collaborative Innovation Center of Advanced Microstructures, Nanjing University, Nanjing 210023, China.

ABSTRACT: Ferroelectric polymers combine switchable polarization with the processability of soft materials, but their development has been dominated by poly(vinylidene fluoride) and related fluoropolymers, whose crystalline polar phases restrict mechanical compliance and domain design with spatial precision. Here we establish a generic design principle for creating intrinsically flexible ferroelectric liquid-crystal polymers through reactive polar mesogenic self-assembly. The approach creates polyfluoroalkyl-free polymer films in which robust ferroelectric order arises from liquid-crystalline molecular organization rather than crystalline phase formation. By transferring ferroelectric order from fluid mesogenic states into polymer networks, the resulting materials combine mechanical adaptability with programmable polar architectures. Especially, the photoalignment technology enables these polar states to be organized into pixelated domain architectures. This work establishes a design space towards soft ferroelectric polymers that integrate molecularly programmed polar order, mechanical tunability and environmentally conscious chemistry, expanding the design space of adaptive materials for flexible electronics, wearable systems and soft robotics.

## Introduction

Ferroelectricity, a symmetry-broken state characterized by spontaneous electric polarization ($\boldsymbol{P}_s$) and field-switchable polarity, underpins a broad range of technologies, ranging from non-volatile memory,[1] ferroelectronics,[2] computation,[3] and nonlinear optics[4] to thermal management[5]. Incorporating intrinsic ferroelectricity into polymers is both a fundamental and technological challenge.[6–8] It requires molecular fragments to organize into polar packing while retaining the processability and mechanical compliance of soft matter. Such materials are particularly important for flexible and wearable ferroelectric devices (**Figure 1a**).[8,9]

Polymer ferroelectricity has been dominated by poly(vinylidene fluoride) (PVDF) and related fluoropolymers, which remain among the few established classes of intrinsically ferroelectric polymers (**Figure 1b**).[8–11] In these materials, ferroelectricity generally originates from local symmetry-broken crystalline orders. Their crystalline positional orders impose rigidity and plasticity that constrains independent modulation of mechanical compliance and the spatial organization of polarization, particularly in the low-modulus regime.[12–14] Their reliance on fluoroalkyl chemistry also raises growing environmental concerns.[15] Beyond mechanical and chemical limitations, a central unresolved challenge in ferroelectrics is whether local polar domains can be precisely manipulated and assembled into extended, complex and programmable domain architectures.[16–18] Addressing these issues requires not simply a fluoroalkyl-free ferroelectric polymer system, but a material platform in which ferroelectric order, domain architecture, and mechanical property can be designed independently (**Figures 1a** and **1b**).[7,9,13]

Such a platform will enable a green and unprecedented adaptive ferroelectric optoelectronic devices whose properties are governed by hierarchical polar structures.

Liquid-crystalline polymers (LCPs) and elastomers (LCEs), where anisotropic liquid-crystalline (LC) mesogens align along the ensemble-averaged orientation, called the director $\mathbf{n}$, offer a useful conceptual blueprint.[19–22] Distinct from crystalline ferroelectric ceramics and PVDF-based polymers, the reactive mesogenic approach enables LC networks combine programmable alignment with fluid-processable precursors and elastic compliance (**Figure 1b**), which involves a three-step process.[19,23] First, material composition, crosslinking density and LC phase behavior could be encoded in a fluid monomer precursor. Then, director fields and topological patterns are written by surface alignment. Finally, this information is transferred into a solid-state network by *in-situ* photopolymerization.[20] However, conventional LCPs and LCEs are non-ferroelectric due to inversion symmetry of mesogenic dipoles (i.e., $\mathbf{n} \equiv -\mathbf{n}$), and therefore only responsive to high electric fields.[24,25] Conventional chiral smectic C (SmC*) liquid crystalline polymers (LCPs) lack *in-situ* processability and domain engineering capability due to their helical supramolecular smectic architecture.[26–28] The recent discovery of ferroelectric nematic ($N_F$) LCs has extended ferroelectricity to liquid-matter systems.[29–31] Breaking of inversion symmetry (i.e., $\mathbf{n} \neq -\mathbf{n}$) gives rise to macroscopic spontaneous polarization to align with the director ($\mathbf{P} = P_0 \cdot \mathbf{n}$). They retain the fluidity, surface-alignability and field responsiveness of liquid crystals, while exhibiting spontaneous polarization comparable to that of PVDF-based polymers.[31–34]

These characteristics suggest a direct route to polymer ferroelectrics that ferroelectric order could first be generated and spatially organized in a fluid $N_F$ precursor, and subsequently captured by polymer networks (**Figure 1c**).[35–37] Such a polar mesogenic self-assembly approach differs fundamentally from conventional polymer ferroelectrics. In principle, it would allow polarization to be designed, aligned and topologically patterned before polymer formation, enabling programmable domain structures from fluid ferroelectric states. The central bottleneck and challenge are therefore not simply polymerizing a liquid crystal, but creating a reactive ferroelectric mesogen that can simultaneously form a stable ferroelectric phase, retain switchable polar order, and survive ferroelectric polymer network without erasing the programmed polar field. Reactive monomers that meet these coupled requirements, and thereby enable ferroelectric liquid-crystalline polymers (ferro-LCPs), have not yet been established.[38,39] Here we realize this approach by developing a series of reactive ferroelectric nematic mesogens (RFMs). We identify an anomalous polar-polar transition between the ferroelectric nematic and ferroelectric smectic order as temperature increases, which is absent in its monomer state. We successfully employ photoalignment technique to organize these polar states into arrayed polar topological patterns. The resulting materials combine switchable polarization, mechanical adaptability and pixelated domain functionality, establishing reactive polar mesogenic self-assembly as a route to eco-friendly, domain-programmable ferroelectric polymers.

**Designing reactive mesogen as fluid ferroelectric precursor**

Although the molecular library of $N_F$ LCs grows rapidly, RFMs remain almost unexplored. Berrow *et al.* recently reported a reactive acrylate-functionalized $N_F$ monomer, but the ferroelectric order could not be preserved during its homopolymerization.[39] The design of RFMs poses several coupled challenges. The monomers must retain the stable polar packing needed for $N_F$ order after introducing polymerizable groups, exhibit a sufficiently accessible phase window for *in-situ* photopolymerization, and form polymer networks with enough segmental mobility to support polarization switching. The rigid polar mesogenic cores that favor ferroelectric ordering tend to produce polymers with high glass-transition temperatures ($T_g$), suppressing the mobility required for polarization switching. Conversely, introducing long aliphatic chains to reduce $T_g$ can destabilize the $N_F$ phase.[40–42] These competing requirements make RFMs difficult to design, but also define the molecular criteria for realizing switchable ferro-LCPs.

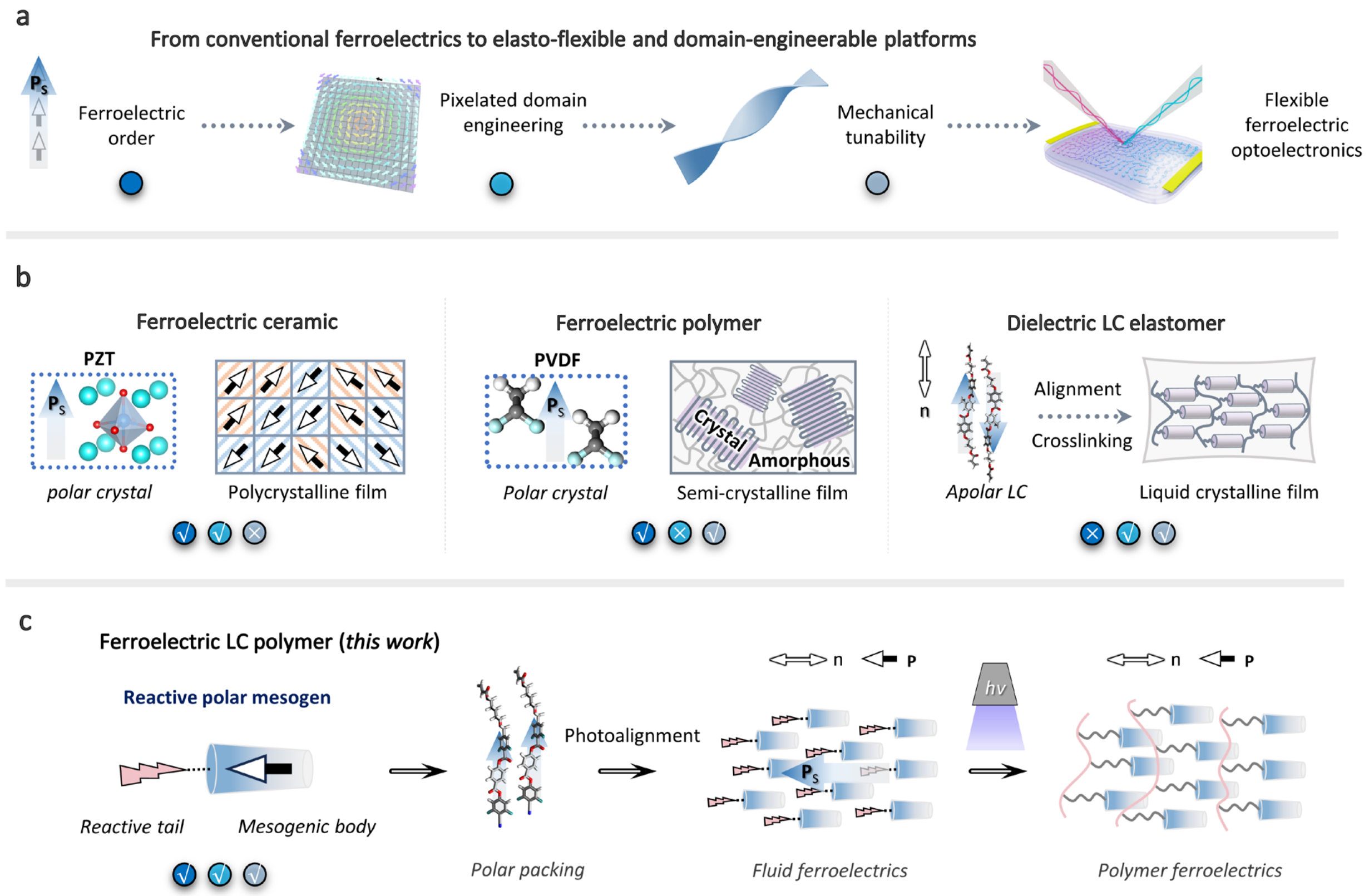

**Figure 1.** Designable polar domain structures and pixelated polar-order-coupled functionality for flexible ferroelectric optoelectronics. **a,** Conceptual design of flexible ferroelectrics as domain-engineerable platforms. **b,** Comparison of conventional ferroelectric materials and dielectric liquid-crystalline elastomers (LCEs). Ferroelectric ceramics, such as lead zirconate titanate (PZT), provide robust polarization but are intrinsically brittle. The representative semi-crystalline ferroelectric polymer poly(vinylidene fluoride) (PVDF) exhibits high flexibility but presents significant challenges for domain engineering. Conventional LCE enables precise spatial control of the LC director field but lack ferroelectric order. **c,** Schematic illustration of the direct transformation of reactive ferroelectric liquid crystals into ferroelectric liquid-crystalline polymers

(ferro-LCPs) via *in-situ* photopolymerization. The ferro-LCPs developed herein demonstrate that reactive polar-mesogen-directed self-assembly as an eco-friendly and domain-programmable strategy for creating ferroelectric polymers.

Leveraging recent insight into design principles of $N_F$ molecules, we design a series of RFMs that combine a highly polar aromatic core, a flexible alkyl spacer and a terminal acrylate group (**Figure 2a**). The RFMs possess large dipole moments of 11–14 D (**Figures 2b and S1a**), which should reinforce dipolar correlations and help preserve ferroelectric order. A benzyl ether linkage between the polar mesogenic core and the spacer introduces a bent molecular conformation (**Figure 2b**). Under free rotation around the principal molecular axis, this subtle chemical design, shifting in the position of one oxygen atom, yields a more pear-shaped molecular envelope, a geometry considered favorable for $N_F$ phase formation.

Polarized light microscopy (PLM; **Figures S2-S8**) and differential scanning calorimetry (DSC; **Figures 2d and S9**) reveal that this design produces a broad family of low-melting polar liquid-crystalline precursors. Except for **RFM-2**, the RFMs melt at approximately 60 °C and clear below 105 °C, allowing cell filling and photopolymerization at moderate temperatures (**Figure S9**). On cooling, most RFMs form the $N_F$ phase, except **RFM-7**. Remarkably, $N_F$ ordering is retained even in **RFM-8**, which bears the longest aliphatic spacer. The $N_F$ phase is enantiotropic in **RFM-3**, **RFM-4** and **RFM-6**, and monotropic in **RFM-2**, **RFM-5** and **RFM-8** (**Figures 2d and S9**). Before entering the $N_F$ phase, all RFMs generally pass through a similar phase behavior of isotropic (Iso) – nematic (N) – antiferroelectric splay nematic ($N_S$) or smectic Z ($SmZ_A$) phases. Notably, PLM observations of the antiferroelectric phases show zig-zag defects with chevron layering, confirming the lamellar structure of $SmZ_A$ phase analogous to the DIO case.[43] The observed phase sequences demonstrate that ferroelectric ordering can tolerate both polymerizable acrylates and extended flexible spacers, providing fluid precursors with phase windows compatible with subsequent solidification.

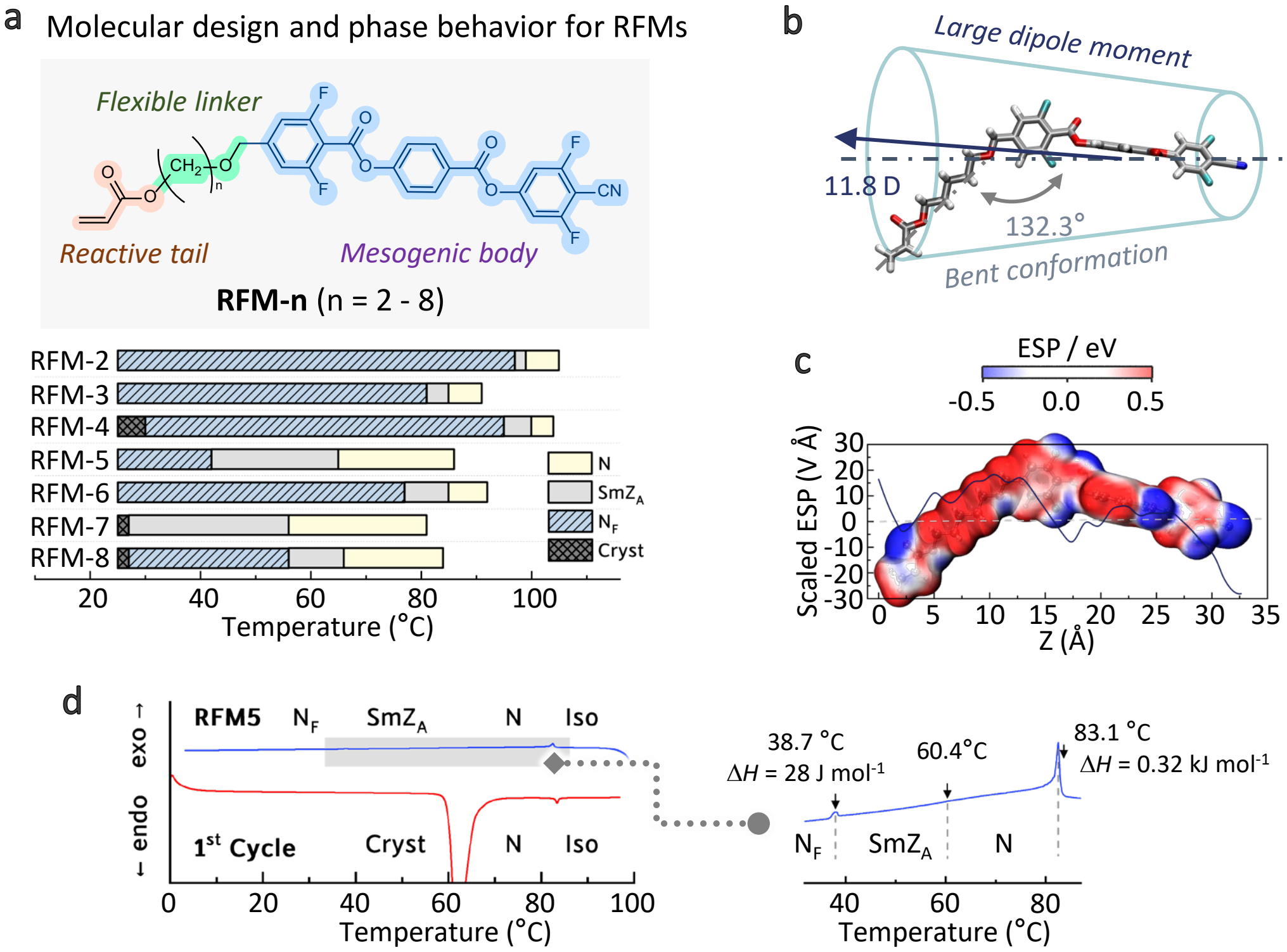

**Figure 2.** Molecular structure and phase behavior of RFMs. **a**, Chemical structure of the designed RFMs (top) and corresponding phase diagram (bottom). The phase-transition temperatures were determined by PLM upon cooling at rate of 10 K min$^{-1}$. **b**,**c**, DFT calculations of representative **RFM-5**. A large molecular dipole moment (11.8 D) and a bent-shaped conformation (**b**), combined with a modulated spatial charge distribution (**c**), are critical for stabilizing ferroelectric order in **RFM-5** bearing long aliphatic tails. **d**, DSC thermograms of **RFM-5,** together with an expanded view of the exothermic region recorded during the 1$^{st}$ cooling scan. Scan rate: 20 K min$^{-1}$.

Spacer length additionally produces an odd - even modulation of the transition temperatures and phase stability, correlating with an analogous variation in the DFT-calculated molecular dipole moment (**Figure S1a**). We speculate that this modulation arises from alternating orientations of the acrylate carbonyl group, causing the terminal –C(O)O– dipole to flip relative to the mesogenic core (**Figure S1b**). The results point to a potential coupling between the highly polar mesogenic core and the weak terminal dipole, despite their separation by a long aliphatic spacer. Electrostatic-potential (ESP) calculations further show an alternating charge distribution along the molecular long axis, e.g., the representative example of **RFM-5** (**Figure 2c**), in agreement with the Madhusudana model.[44,45] Negative regions associated with the fluorine atoms and ester groups favor lateral interactions between parallel molecules (**Figures 2c and S10**).[44,46] The acrylate group introduces an additional negative region and improves the periodic potential oscillation between the positive and negative. These combined features, i.e., large dipole magnitude, bent molecular conformation and favorable electrostatic patterning, stabilize the polar self-assembly of $N_F$ monomers. The resulting RFM library therefore establishes a set of fluid ferroelectric precursors suitable for testing whether pre-existing polar order can be captured by *in-situ* polymerization.

**Capturing fluid ferroelectric order by *in-situ* photopolymerization**

To determine whether ferroelectric order can survive network formation, the RFMs were formulated with 0.5 wt% Irgacure 651 and 0.5 wt% butylated hydroxytoluene (BHT) for *in-situ* photopolymerization (**Figure S11**). The additives produced neglectable change in the phase sequences of the RFMs, and the formulated mixtures retained their second-harmonic-generation (SHG) response in the polar phases (**Figure S2a**). The mixtures therefore provide fluid ferroelectric precursors in which the phase behavior and polar order remain compatible with radical photopolymerization. Following *in-situ* photopolymerization, the majority of PolyRFMs, i.e., odd-numbered PolyRFMs retain SHG activity, confirming that the polar phase is thermodynamically stable throughout repeated heating and cooling cycles (**Figures S12b-h and Table S1**). Such an odd-even numbered effect is consistent with RFMs, which maybe arise from the molecular conformations and dipole-dipole interactions. Afterwards, representative sample **RFM-5** was selected for detailed investigation because of its low clearing temperature and moderate spacer flexibility. The PLM textures of **RFM-5** mixed with 0.5 wt% Irgacure 651 and 0.5 wt% BHT demonstrate a uniform alignment of ferroelectric domains (**Figure 3a**): a homogeneous bright texture characteristic of the non-polar N phase at high temperatures; chevron-like layering texture in the intermediate temperature range corresponding to the $SmZ_A$ phase; and a banded texture with distinct domain walls corresponding to the $N_F$ phase at low temperatures.[31,43]

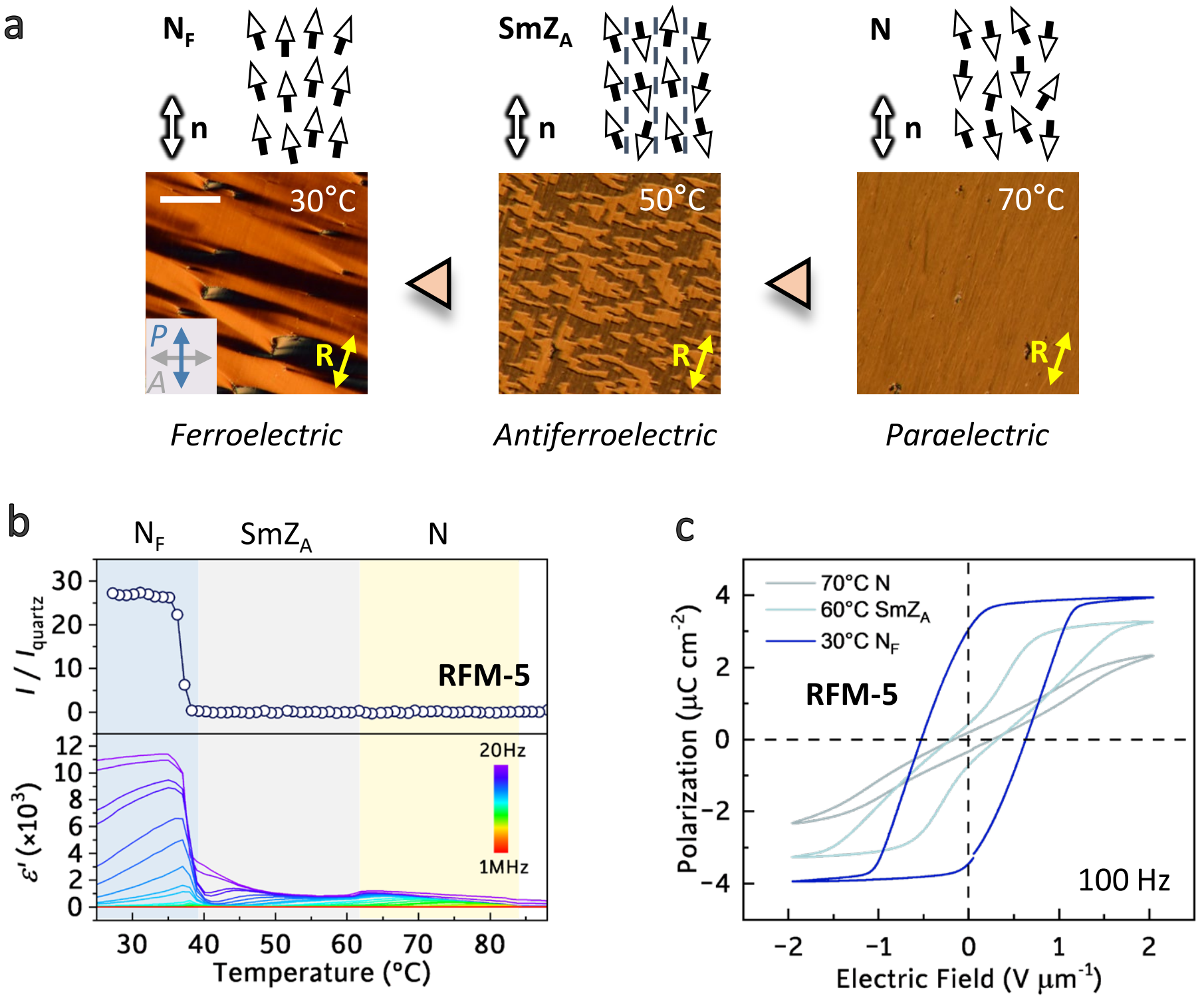

**Figure 3.** Ferroelectricity of fluid ferroelectric precursor. **a**, PLM textures of fluid ferroelectric precursor prepared by mixing **RFM-5** with Irgacure 651 and BHT, observed under crossed polarizers, and corresponding schematic illustration of dipole packings in **RFM-5**. Cell thickness: 2 mm with parallel alignment. Scale bar, 100 μm. **b**,**c**, Polarization characterizations of **RFM-5**.

The strong SHG intensity, giant dielectric permittivity versus temperatures (**b**), and typical ferroelectric hysteresis loops (**c**) confirm its ferroelectricity.

Ferroelectricity in **RFM-5** is verified by SHG, dielectric spectroscopy, and *P*-*E* hysteresis measurements (**Figures 3b and 3c**). For instance, strong SHG appears across the $SmZ_A$-$N_F$ transition at approximately 39 °C, together with a giant increase in apparent dielectric permittivity, $\varepsilon' > 10^4$, in the $N_F$ phase (**Figure 3b**). The *P*-*E* loop in the non-polar N phase (e.g., 70 °C) is a slim, nonlinear double loop with low hysteresis and large polarization ($P_m \sim 2$ μC $cm^{-2}$, **Figure 3c**). Into the $SmZ_A$ phase (e.g., 60 °C), a double hysteresis loop indicates the antiferroelectricity ($P_m \sim 3$ μC $cm^{-2}$, **Figure 3c**). Finally, a characteristic parallelogram hysteresis loop appears at low temperatures (e.g., 30 °C) confirms the ferroelectric switching in the $N_F$ phase ($P_S \sim 4$ μC $cm^{-2}$, **Figure 3c**). The **RFM-5** mixture therefore combines processing stability over repeated heating and cooling between 25 and 120 °C with preservation of ferroelectric order before polymerization.

Subsequently, time-dependent polymerization kinetics of the **RFM-5** mixture is monitored by *in-situ* FTIR spectroscopy under 365-nm UV irradiation at 30 mW $cm^{-2}$ and 36 °C. Time-dependent conversion of acrylate groups was quantified by monitoring the decrease of the absorption band at approximately at 810 $cm^{-1}$, assigned to the out-of-plane vibration of the vinyl C-H group (**Figure S11a**).[47,48] The reaction reached >80% conversion within 30 s, with little further change after 60 s (**Figure S11b**). Size exclusion chromatography (SEC) measurements show that the resulting polymers have degrees of polymerization ($D_p$) and polydispersity indices (*Đ*) comparable to conventional end-on apolar-LCPs (**Figure S11c and Table S1**).[49,50] In the following studies, the LCP films were prepared in liquid-crystal cells thinner than 10 mm and photopolymerized by bidirectional UV irradiation for 3 min to ensure high conversion.

**Polar order and emergent polar phase transitions in ferro-LCPs**

On the basis of two-dimensional wide-angle X-ray diffraction (2D WAXD, **Figures 4a, 4b, S13, S14**) and DSC measurement during heating and cooling cycles (**Figures 4c and S15**), we identify distinct phase transitional behaviors for the polymerized LCPs. The representative LCP obtained from **RFM-5**, called **PolyRFM-5** hereafter, exhibits three enantiotropic LC phases upon changing temperature. The 2D WAXD measurements on an unidirectionally aligned **PolyRFM-5** film/fiber are further carried out to resolve their structural order (**Figure 4a**), with the corresponding integrated one-dimensional WAXD profiles shown in **Figure 4b**. The first Iso – SmA phase transition occurs at c.a. 204 °C, as confirmed by an endothermic peak upon heating (212.6 °C, $\Delta H$ = 1.74 J $g^{-1}$) and an exothermic peak upon cooling (203.6 °C, $\Delta H$ = 1.71 J $g^{-1}$) in the DSC trace (**Figure 4c**). Upon cooling from the isotropic state, **PolyRFM-5** first forms an orthogonal lamellar phase, i.e., the smectic-A (SmA) phase (**Figures 4a and S13**). As shown in **Figure 4a** at 100 °C, this phase is characterized by a sharp small-angle reflection ($q_1$ = 1.81 $nm^{-1}$) along the meridional direction and two broad wide-angle diffraction spots along the equatorial direction. The director of LC mesogens is aligned parallel to the drawing direction in the LCP fiber. A second-order reflection ($q_2$ = 3.65 $nm^{-1}$), corresponding to approximately half of $q_1$, is also clearly observed in the one-dimensional WAXD profile (**Figure 4b**). The corresponding *d*-spacing of $q_1$ is comparable to the molecular length of **RFM-5** (calculated to be 3.17 nm at DFT: B3LYP-D3BJ/cc-pVTZ).[51] Because this phase is SHG-inactive (**Figure 4d**), we assign it to the traditional apolar SmA phase.

A second ordered phase appears below approximately 70 °C (69.6 °C, $\Delta H$ = 3.51 J $g^{-1}$) upon cooling, as indicated by an exothermic peak in the DSC trace (**Figure 4c**). The 2D and 1D WAXD patterns are broadly similar (**Figures 4a and 4b**), indicating that the SmA order is retained. The full width at half maximum of the smectic reflection decreases around 75 °C (**Figure 4b**). In contrast to the high-temperature SmA phase, this lower-temperature SmA phase produces a strong SHG signal (**Figure 4d**), allowing us to assign this phase to a ferroelectric SmA phase, i.e., $SmA_F$. The temperature dependence of dielectric permittivity ($\varepsilon'$) in **PolyRFM-5** exhibits a similar trend with SHG (**Figure 4d**). The $\varepsilon'$ increases dramatically upon cooling and shows the maximum at approximate 70 °C, indicating the strong dipolar fluctuations near SmA – $SmA_F$ ferroelectric phase transition. A large dielectric permittivity ($\varepsilon'$ = 35.4) is obtained at 70 °C, even higher than that in traditional PVDF polymers.[52] A more solid evidence of ferroelectricity in this $SmA_F$ phase is the appearance of distinct polarization reversal current peaks appear at 60 ˚C as shown in **Figure 4e**. The polarization reversal current gradually vanishes above 90 ˚C (**Figure S16**), indicating the apolar nature for high-temperature SmA phase. The spontaneous polarization ($P_s$) in $SmA_F$ phase of **PolyRFM-5** achieves ~5 μC $cm^{-2}$ (**Figure S16c**), similar as in **RFM-5** monomer (in the $N_F$ phase), implying that the majority of mesogenic dipoles were reorientated under strong electric field.

Further cooling induces a third transition near 55 °C, accompanied by substantial weakening and broadening of the $q_1$ reflection and disappearance of the higher-order $q_2$ peak (**Figures 4a and 4b**). These changes indicate a loss of quasi-long-range smectic order and a transition to a type of nematic order, although the transition is not clearly resolved by DSC (**Figures 4c and S15**). This nematic phase is SHG-active, being reasonably assigned to the $N_F$ phase. Its ferroelectricity is further validated by the detection of apparent polarization reversal current and similarly high $P_s$ values as in $SmA_F$ phase (**Figures 4e and S16c**). The coercive field ($E_c$) of **PolyRFM-5** in the ferroelectric LC phases ($SmA_F$ and $N_F$ phases) was near 45 V $mm^{-1}$ **(Figure S17)**, exhibiting a comparable level to traditional PVDF-based polymer ferroelectrics, although much higher than **RFM-5** monomer (~ 0.6 V $\mu m^{-1}$, **Figure 3c**).[53]

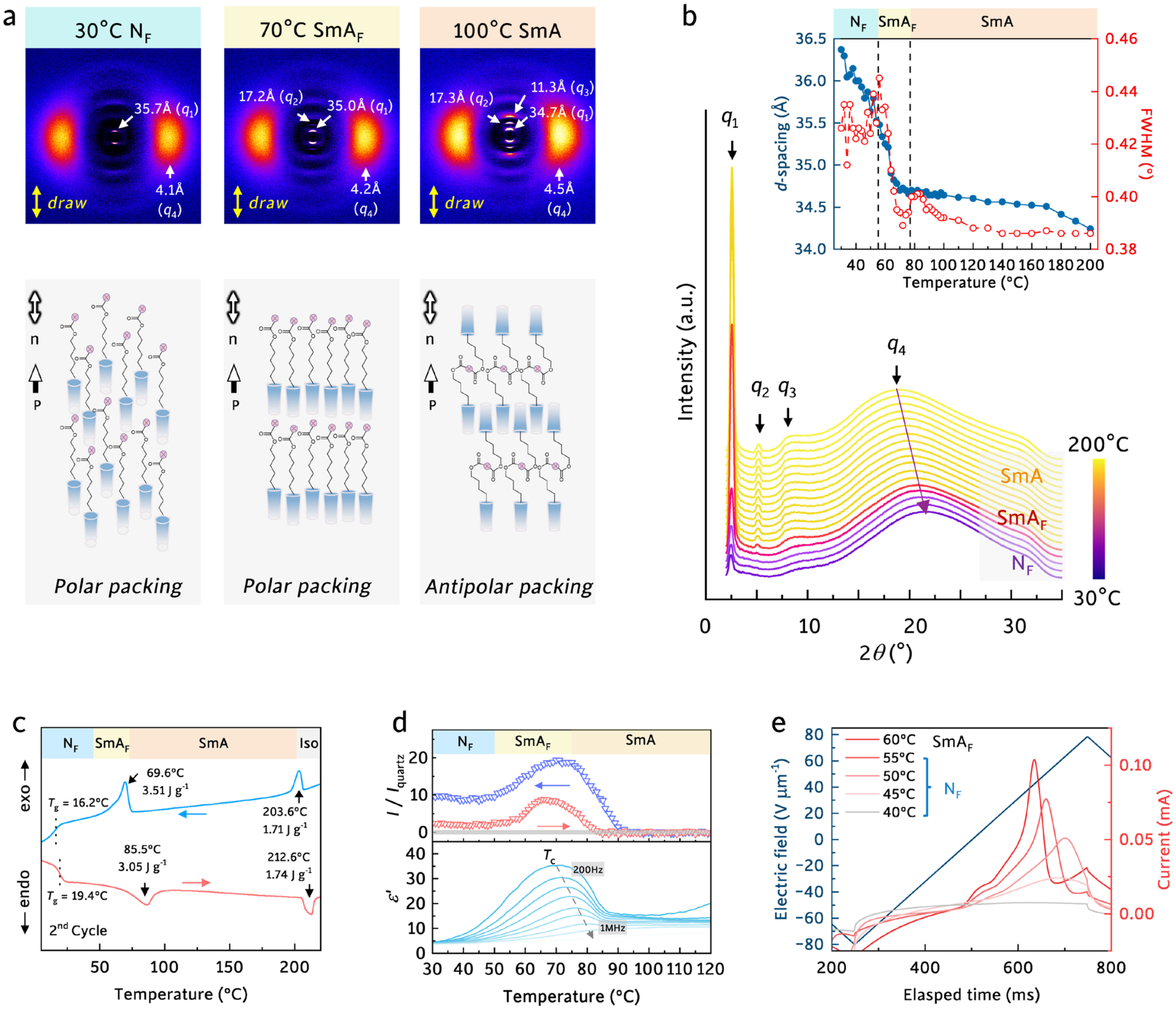

**Figure 4.** Phase behavior study and polarization characterization of *in-situ* photopolymerized PolyRFM-5. **a**, Representative 2D WAXD patterns of a drawn **PolyRFM-5** fiber in different LC phases collected during cooling from 120 °C. The schematic illustrations of molecular packing in corresponding antipolar SmA, polar SmA$_F$, and polar N$_F$ phase are shown at bottom. **b**, 1D WAXD profiles of **PolyRFM-5** film in different LC phases upon the cooling from 200 ˚C. The inset shows the temperature dependence of corresponding *d*-spacing and full width at half maximum (FWHM) of $q_1$ peak for **PolyRFM-5**. **c,** DSC thermograms of **PolyRFM-5** during the 2$^{nd}$ cycle. Scan rate: 10 K min$^{-1}$. **d,** SHG signal of **PolyRFM-5** recorded during the 1$^{st}$ heating (red) and cooling (blue)

cycle in a 2-μm-thick LC cell under parallel alignment (top), and temperature dependence of dielectric permittivity during the 1st heating processes in a 5-μm-thick LC cell without any alignment layers (bottom). **e**, Polarization reversal current profiles of **PolyRFM-5** measured under a 1-Hz triangle wave at various temperatures.

Such an anomalous ferroelectric $SmA_F$-$N_F$ transitional behavior, in which the higher-symmetry $N_F$ state emerges at lower temperatures, likely reflects frustration between polar interactions and positional packing constraints in the end-on ferro-LCP architecture. While dipolar correlations favor collective polar orientational order, the polymer backbone and long side chains promote smectic layering. The enhanced positional constraint is evident from the broad stability of SmA order in **PolyRFM-5**, whereas the monomer **RFM-5** exhibits only the nematic order over a comparable temperature range, even near the isotropic transition. Upon cooling, the inter-mesogen distance decreases from 4.5 to 4.2 Å (**Figure 4b**), strengthening dipolar interactions and promoting ferroelectric order in the ferro-LCPs.[31,45,54]. While longitudinal dipole–dipole interactions between the polar mesogens stabilize ferroelectric order along the director in both the smectic and nematic states, transverse electrostatic repulsion between laterally adjacent dipoles disfavors excessively dense syn-parallel packing within smectic layers. The $SmA_F$ phase can therefore be stabilized when smectic positional order and polar alignment are compatible, that is, when the transverse repulsive cost remains sufficiently low. Namely, the $SmA_F$ phase is experimentally observed to be stable only in a proper temperature range (**Figure 4a**). Upon further cooling, polar order strengthens and the intermolecular distance decreases, amplifying the electrostatic penalty of dense lateral packing (indicated by the arrow in **Figure 4b**). Once this repulsion exceeds the energetic gain associated with smectic layering, the layered structure becomes unstable and the system relaxes by disrupting positional order while retaining collective ferroelectric alignment (e.g., near 55 °C for **polyRFM-5**). Relative sliding between adjacent polar mesogens enables ferro-LCP to favor a more stable staggered nematic packing configuration in the form of the $N_F$ phase (**Figure 4a**). This provides a plausible mechanism for the observed $SmA_F$-$N_F$ phase transition in ferro-LCPs, and is consistent with some re-entrant behavior reported in small-molecule $N_F$ fluid systems.[40,45]

**Optically-processed domain engineering in ferro-LCPs**

The fine polar microstructures in fluidic ferroelectric precursors are intrinsically reconfigurable. To extend this control to deterministic spatial control of **P**, a central challenge in ferroelectric materials, we used photoalignment to define non-polar director templates at the substrate surface. LC cells spin-coated with SD1 (1.5 wt% in N,N-dimethylformamide, DMF) were exposed to predesigned patterns using a digital micromirror device (DMD)-based lithography system with a 405-nm LED light source, and *in-situ* photopolymerization under the same conditions used above (36 °C, 365 nm, 10 mW cm$^{-2}$, 1 min). This process allowed us to examine whether the surface-imposed orientational information could be preserved and translated into polar textures in the polymerized ferro-LCPs.[33] **Figure 5** demonstrates representative, complex photoaligned templates and their corresponding PLM textures for fluidic ferroelectric precursors and freshly polymerized ferro-LCPs. The continuous, defect-free textures indicate that the imposed director fields are transferred with high fidelity into the ferroelectric polymer film, allowing predefined orientational templates to serve as blueprints for programmed polar architectures. These results establish reactive polar mesogenic self-assembly as a route to programmable large-scale and complex domain structures in ferroelectric thin films: photoalignment prescribes the global topology, while annealing refines the local polar order without erasing the imposed pattern. This integration of pattern transfer, polar-domain reconfiguration, mechanical flexibility and modulus tunability enables complex ferroelectric architectures that are difficult to access in conventional solid-state inorganic and semicrystalline polymer ferroelectrics.

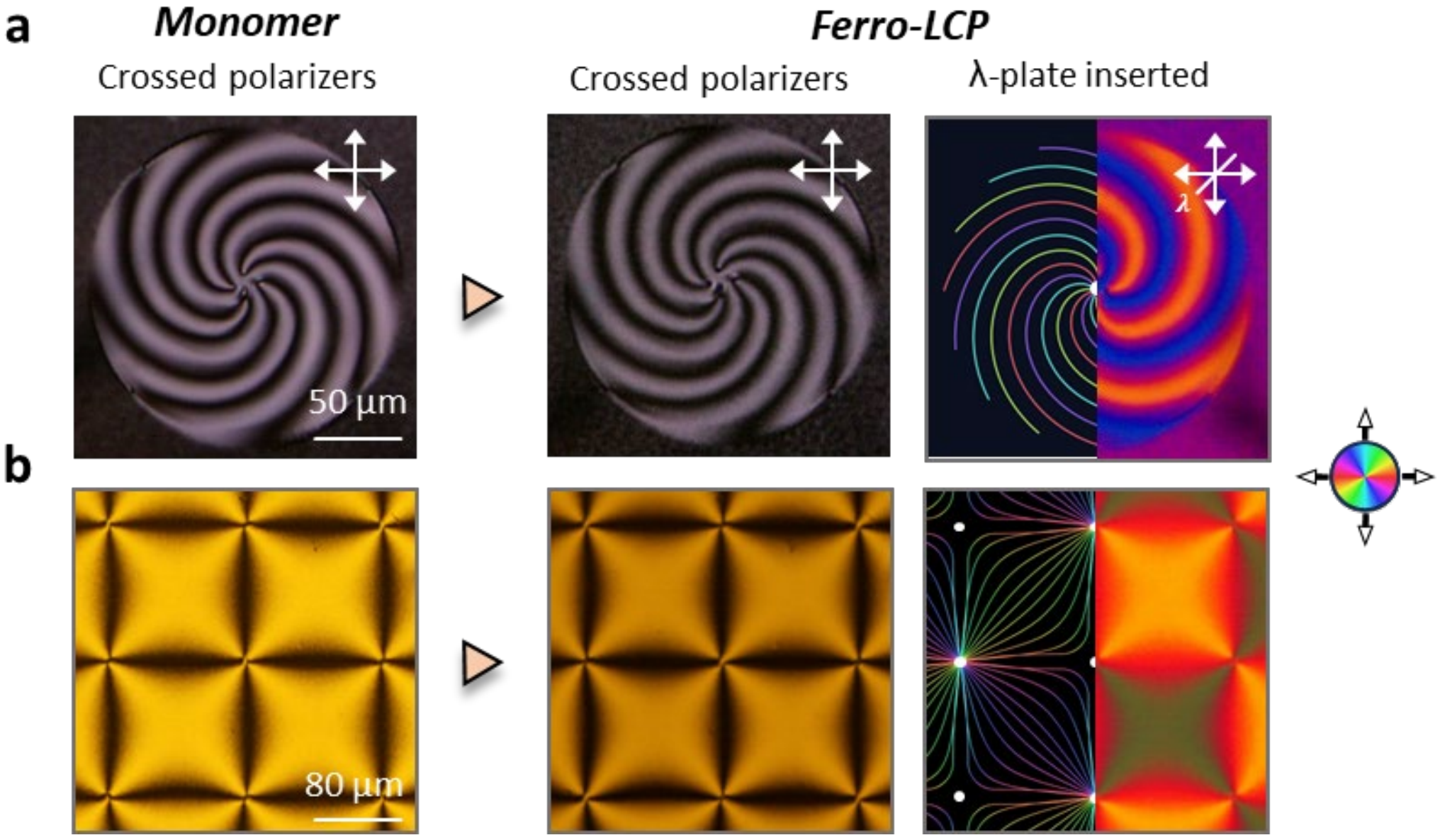

**Figure 5.** Photoalignment-programmed domain engineering in **PolyRFM-5**. **a**,**b,** Designed photoalignment templates and the corresponding PLM textures of $N_F$ monomers and polymerized ferro-LCP films, showing excellent programmability of polar domain structures. PLM images were taken under crossed polarizers, Left: PLM textures of $N_F$ monomers. Middle: PLM textures of **PolyRFM-5**. Right: PLM textures of **PolyRFM-5** with λ plate inserted.

**Approach to flexible ferroelectric liquid-crystalline film and elastomer**

The *in-situ* photopolymerization approach of reactive polar mesogens provides a straightforward route to ferro-LCPs. The mechanical modulus can be tuned, e.g., by systematically varying crosslinker concentration, spacer flexibility, and network architecture, while preserving ferroelectricity. These ferro-LCPs thus represent a chemically programmable precursor platform for fabricating ferro-LCEs. Experimentally, incorporating a low concentration of crosslinker (e.g., 3 wt% 1,6-hexanediol diacrylate) preserves the high pliability of **PolyRFM5** while maintaining ferroelectric switching capability comparable to that of the ferroelectric polymer PVDF (**Figure S18**).

Beyond tunable mechanics and polarization control across length scales, our ferro-LCPs/LCEs platform also offer opportunities to mitigate concerns associated with per- and polyfluoroalkyl substances. Notably, the fluorinated motifs in PolyRFMs are aromatic fluorides rather than polyfluoroalkyl segments, and the molecular design strategy does not intrinsically rely on fluorinated groups. The fully fluorine-free ferro-LCPs/LCEs are achievable through appropriate molecular design. One remaining limitation is the relatively high glass transition temperature of the current polyacrylate-based ferro-LCPs (e.g., $T_g \sim 20$ °C in **PolyRFM-5**), which is expected to hinder polarization reversal kinetics. Lowering the glass transition temperature therefore represents a promising strategy for substantially reducing the driving electric field required in future devices.

**Conclusion**

We have developed flexible, domain-engineerable polymer ferroelectric materials via a polar mesogenic approach for the first time. By integrating acrylate polymerizable groups and long flexible alkyl spacers into $N_F$ LC monomers, we achieved a direct transformation from fluid ferroelectric to flexible polymer ferroelectric through *in-situ* photopolymerization. The resulting ferroelectric LCPs and LCEs exhibit both thermodynamically stable ferroelectric smectic phase at higher temperatures and ferroelectric nematic phase at lower temperatures. Furthermore, predefined polarization structures can be written into the photo-alignment layer, enabling pixelized control of polar domain structures in the fluid ferroelectric state and further transferring to ferroelectric LCPs and LCEs. This work addresses key challenges in traditional polymer and solid-state ferroelectrics, including concerns over polyfluoroalkyl substances, uncontrolled polarization domains, and difficulties in elastic and softness design, while establishing new physical phenomena and material properties that open pathways toward flexible optoelectronic technologies.

# Supporting Information

# Reactive polar mesogenic self-assembly enables domain-programmable ferroelectric liquid-crystalline polymers

Fan Ye,[1,2‡] Minghui Deng,[1‡] Yuyang Zheng,[1] Xiujuan Liu,[1] Xiuhu Zhao,[3,4] Haowei Jiang,[5] Yanyun Hou,[1] Bingyu Zou,[1] Neng-Ang Peng,[1] Shuo Zhao,[6] Kutay Sağdıç,[3,4] Danqing Liu,[3,4] Yang Shen,[6] Yan-Qing Lu,[7] Satoshi Aya,[1,2*] Mingjun Huang[1,2*]

[1] School of Emergent Soft Matter, State Key Laboratory of Advanced Papermaking and Paperbased Materials, South China University of Technology, Guangzhou 510640, China.

[2] Guangdong Provincial Key Laboratory of Functional and Intelligent Hybrid Materials and Devices, Guangdong Basic Research Center of Excellence for Energy & Information Polymer Materials, South China University of Technology, Guangzhou 510640, China.

[3] Department of Chemical Engineering and Chemistry, Eindhoven University of Technology, Eindhoven 5612 AE, The Netherlands.

[4] Institute for Complex Molecular Systems (ICMS), Eindhoven University of Technology, Eindhoven 5612 AE, The Netherlands.

[5] National Engineering Research Center of Novel Equipment for Polymer Processing, Department of Mechanical and Automotive Engineering, South China University of Technology, Guangzhou 510641, China.

[6] State Key Laboratory of New Ceramic Materials, School of Materials Science and Engineering, Tsinghua University, Beijing 100084, China.

[7] National Laboratory of Solid-State Microstructures, Key Laboratory of Intelligent Optical Sensing and Manipulation, College of Engineering and Applied Sciences, and Collaborative Innovation Center of Advanced Microstructures, Nanjing University, Nanjing 210023, China.

* Corresponding author: satoshiaya@scut.edu.cn; huangmj25@scut.edu.cn

‡ These authors contribute equally to this work.

## 1. General materials and methods

**Materials.** All commercial reagents and solvents were used as received, unless stated otherwise. All the commercial solvents were obtained from Energy Chemical. Reactions were carried out under a nitrogen atmosphere using a magnetic stirring hotplate and monitored by thin layer chromatography (TLC) using an appropriate solvent system. The glass-backed silica gel TLC plates was purchased from Titan (60 F254, Titan, China) and visualized using UV light at wavelengths of both 254 nm and/or 365 nm. Column chromatography was performed using SepaBean™ flash chromatography system from Santai Technologies (40 - 63μm particle size).

**Nuclear Magnetic Resonance (NMR) Spectroscopy.** $^{1}H$, $^{19}F$, and $^{13}C$ NMR spectra were recorded on JNM-ECZ500 (JEOL) operating at 500 MHz, 126 MHz, and 471 MHz, respectively, using the TMS (trimethylsilane) as an internal standard for $^{1}H$ NMR and the deuterated solvent for $^{13}C$ NMR. The absolute values of the coupling constants are given in Hz, regardless of their signs. Signal multiplicities were abbreviated by s (singlet), d (doublet), t (triplet), q (quartet), quint (quintet), sext (sextet), and dd (double–doublet), respectively.

**Density Functional Theory (DFT) Calculations.** All DFT calculations were performed with the Gaussian 16W series of programs (Gaussian 16W, Revision A.03). The geometric optimizations and electrostatic potential (ESP) calculation were performed with a B3LYP-GD3BJ/cc-pVTZ basis set. The dipole moments were calculated at a B3LYP-GD3BJ/maug-cc-pVTZ level. The ESP surfaces were analyzed using Multiwfn (version 3.8 dev) program and visualized by VMD 1.9.3 program. The 3D data was reduced into 1D according to the previous works.[1,2]

**Size Exclusion Chromatography (SEC) Measurements.** SEC measurements were performed on a Shimadzu LC-10A system with Shodex KD 804 and KD-802.5 SEC columns in series at 50°C and a flow rate of 1 $mL \cdot min^{-1}$. Dimethyl formamide (DMF) with 0.2 M LiBr was used as the eluent. The in situ photopolymerized samples were dissolved in DMF at a concentration of 2.0 mg $mL^{-1}$ prior to injection. The system was calibrated with PS standards (Tosoh Bioscience), and molecular weights were analyzed using EcoSEC Elite-WS software.

**Polarizing Light Microscope (PLM) Observation.** The textures of RFMs and PolyRFMs were observed under crossed polarizers (Olympus BX-51). All samples were placed on a hot stage (Instec MK700) and photographed using a ToupCam E3ISPM09000KPB-E3 camera. Home-made 2-μm or 5-μm thick planar aligned LC cells were used for capturing the PLM textures.

**Time-dependent Fourier Transform Infrared (FTIR) Spectroscopy.** Time-dependent FTIR was performed using Varian 670-IR that was equipped with a microscopy setup over a range of 4000-600 $cm^{-1}$. The **RFM-5** (with 0.5 wt% photoiniator and 0.5 wt% inhibitor) was placed on a temperature-controlled stage and equilibrated at 36 ˚C. *In-situ* photopolymerization was initiated using a 365 nm UV light source with an intensity of 30 mW $cm^{-2}$ measured at the sample position. The sample was first irradiated continuously for 1 min, followed by a 1 min dark interval to evaluate post-illumination conversion. The sample was then heated to 100 ˚C and cooled back to the polymerization temperature, after which a second 1 min UV exposure was applied to consume any remaining acrylate groups. Time-dependent conversion of acrylate groups was quantified by monitoring the decrease of the absorption band at approximately at 810 $cm^{-1}$, assigned to the out-of-plane vibration of the vinyl C-H group.

**Second Harmonic Generation (SHG) Measurements.** A pulsed laser (MPL-III-1064-20 mJ, Changchun New Industries Optoelectronics Tech. Co., Ltd.; wavelength 1064 nm, energy 20 mJ, repetition rate 100 Hz) is used as the light source for SHG. The polarization state of the fundamental light is adjusted by either l/2 or l/4 plate before entering samples. The incident light is kept as a parallel beam with a diameter of 1 mm. The transmitted SH light is collected by a photomultiplier tube (DH-PMT-D100V, Daheng New Epoch Technology, Inc.). The temperature dependence of the SH signal is recorded with the help of a homemade Labview program, which enables the control of temperature (temperature fluctuation less than 0.1 °C) and the data logging of the SH signal at desired temperature intervals. Homemade 2-mm thick cells with planar alignment were used for measuring the SH signals as a function of temperature. The reference sample is either an x-cut lithium niobate (LN) wafer or a y-cut quartz wafer.

**Dielectric Spectroscopy Measurements.** Dielectric spectroscopy was measured by an LCR meter (E4980A, Keysight) at frequencies between 20 Hz and 1 MHz using LC cells with Indium Tin Oxide (ITO)-coated electrode area of 0.2 $cm^2$ (7 W $sq^{-1}$) and thickness of 5 mm, applying an out-of-plane electric field. The voltage applied to the ITO cell is 100 mV. The collection of the frequency and temperature sweeping data was automated by homemade software written in LabVIEW.

***P-E*** **Hysteresis Loops Measurements.** The polarization reversal measurements were conducted using a ferroelectric property analyzer (TOYO FCE10-S, Japan) with a Sawyer-Tower circuit. RFMs were injected into 5-mm ITO cells without polymer alignment layers, applying triangular waves. For **PolyRFM-5**, the *in-situ* photopolymerization was carried out and polarization reversal currents were acquired upon triangular waves. To obtain spontaneous polarization values, we extract the peaks in the current flow, and integrate as $P_{\mathrm{s}} = \int \frac{I_{\mathrm{P}}}{2A} \mathrm{d}t$, where $A$ is the active electrode area of the sample cell.

**Differential Scanning Calorimetry (DSC) Measurements.** The thermal properties of the RFMs and PolyRFMs were characterized by DSC using a DSC 2500 (TA instruments). To prepare the samples, approximately 5 mg of samples were loaded into hermetic aluminum pans. DSC data were acquired under a nitrogen atmosphere at a constant scanning rate of 10 $K \cdot min^{-1}$.

**Wide-angle X-ray Diffraction (WAXD) Measurements.** WAXD experiments were performed using Rigaku HomeLab diffractometer equipped with Cu Ka radiation (l = 0.15405 nm). For RFM monomers, samples were filled into homemade thin-walled quartz LC cells (thickness: 100 mm) with parallel alignment layers to record the 2D diffraction patterns. For ferro-LCPs, the unidirectionally aligned **PolyRFM-5** film was fabricated via *in-situ* polymerization of a $N_F$ phase that had been pre-aligned unidirectionally within a 100-mm-thick planar LC cell. The *in-situ* photopolymerized films were peeled off carefully from the 100-mm LC cells after being frozen at 0 °C for 1 hour. Then the samples were dried with nitrogen gas and mounted between polyimide tapes to further tests. The **PolyRFM-5** fiber was subsequently obtained by uniaxially and quickly hot-drawing LCP film at 120 °C.

**Photoalignment procedure.** Glass substrates were cleaned in an ultrasonic bath with diluted detergent for 15 min, rinsed with ultrapure water ten times, and subsequently treated by UV–ozone for 20 min followed by plasma cleaning for 5 min. A 1.5 wt% SD1 solution in dimethylformamide (DMF) was spin-coated onto the substrates (650 rpm for 7 s, 4000 rpm for 50 s, and 5500 rpm for 7 s) and baked at 100 °C for 10 min. Two coated substrates were assembled into liquid-crystal cells using UV-curable adhesive. Photoalignment patterns were written using a digital micromirror device (DMD)-based digital mask lithography system (OPTRON-80) equipped with a rotatable polarizer to control the polarization direction of the incident UV light. Under UV irradiation, SD1 molecules align perpendicular to the polarization direction of the light, guiding the alignment of the subsequently filled liquid crystals.

## 2. Supplementary Figures and Tables

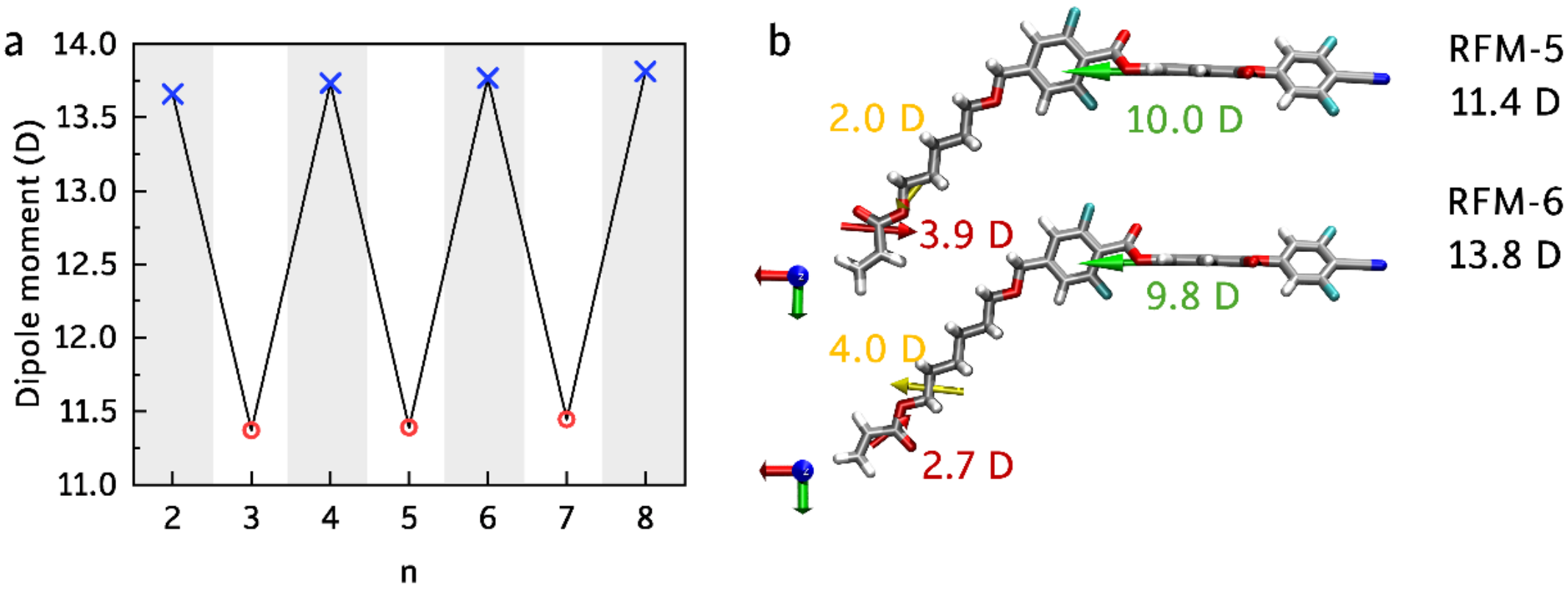

**Figure S1.** DFT-calculated molecular dipole moments at the DFT: B3LYP-D3BJ/maug-cc-pVTZ level. **a**, Odd–even dependence of the longitudinal molecular dipole moment magnitude on linker length. **b**, Contribution of the terminal acrylate group to the overall molecular dipole moment in odd- (**RFM-5**) and even-numbered (**RFM-6**) RFMs. The arrows indicate the directions of the segmental dipole moments of the mesogenic core (green), aliphatic chain (yellow), and terminal acrylate group (red).

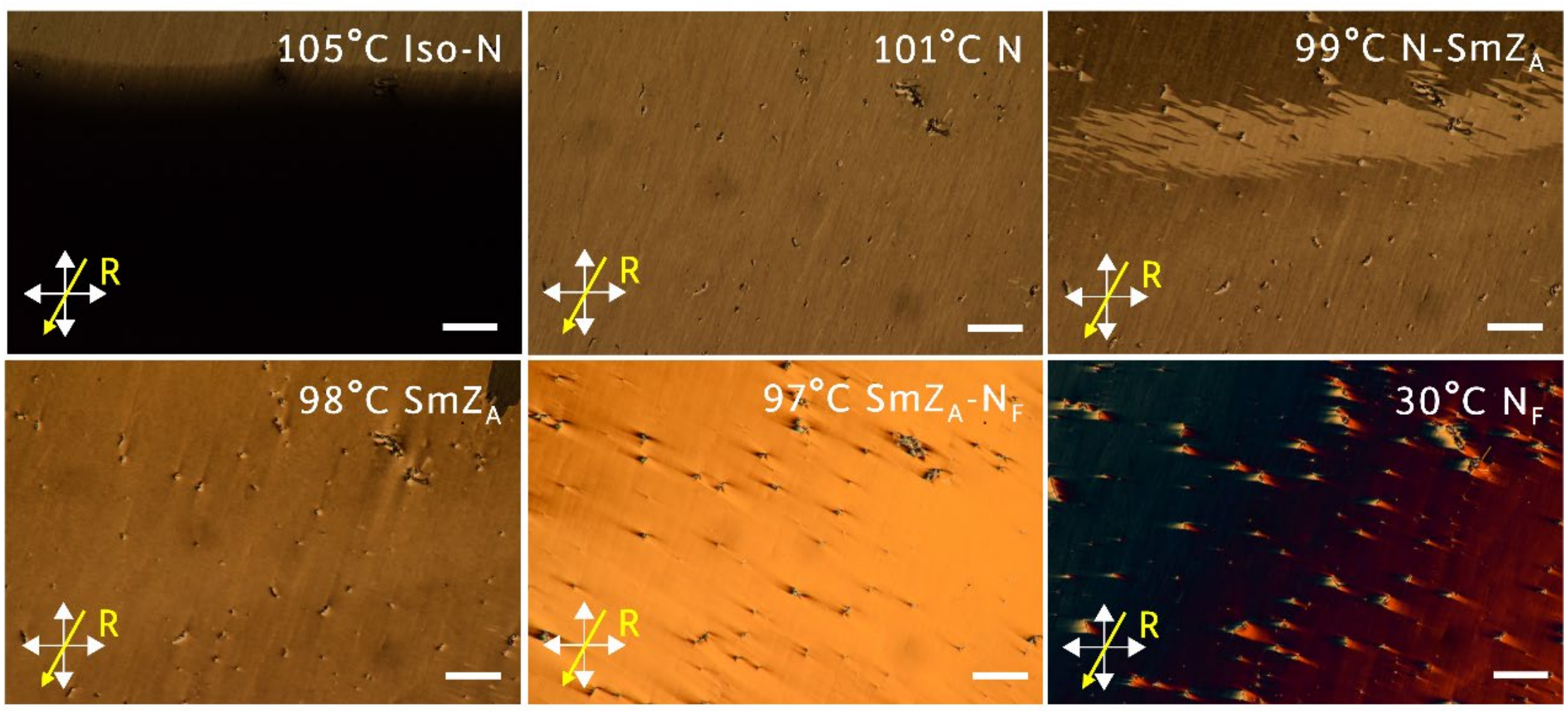

**Figure S2.** PLM textures of **RFM-2** observed under crossed polarizers in a 2-μm-thick planar LC cell. R denotes the rubbing direction. Scale bar, 100 μm.

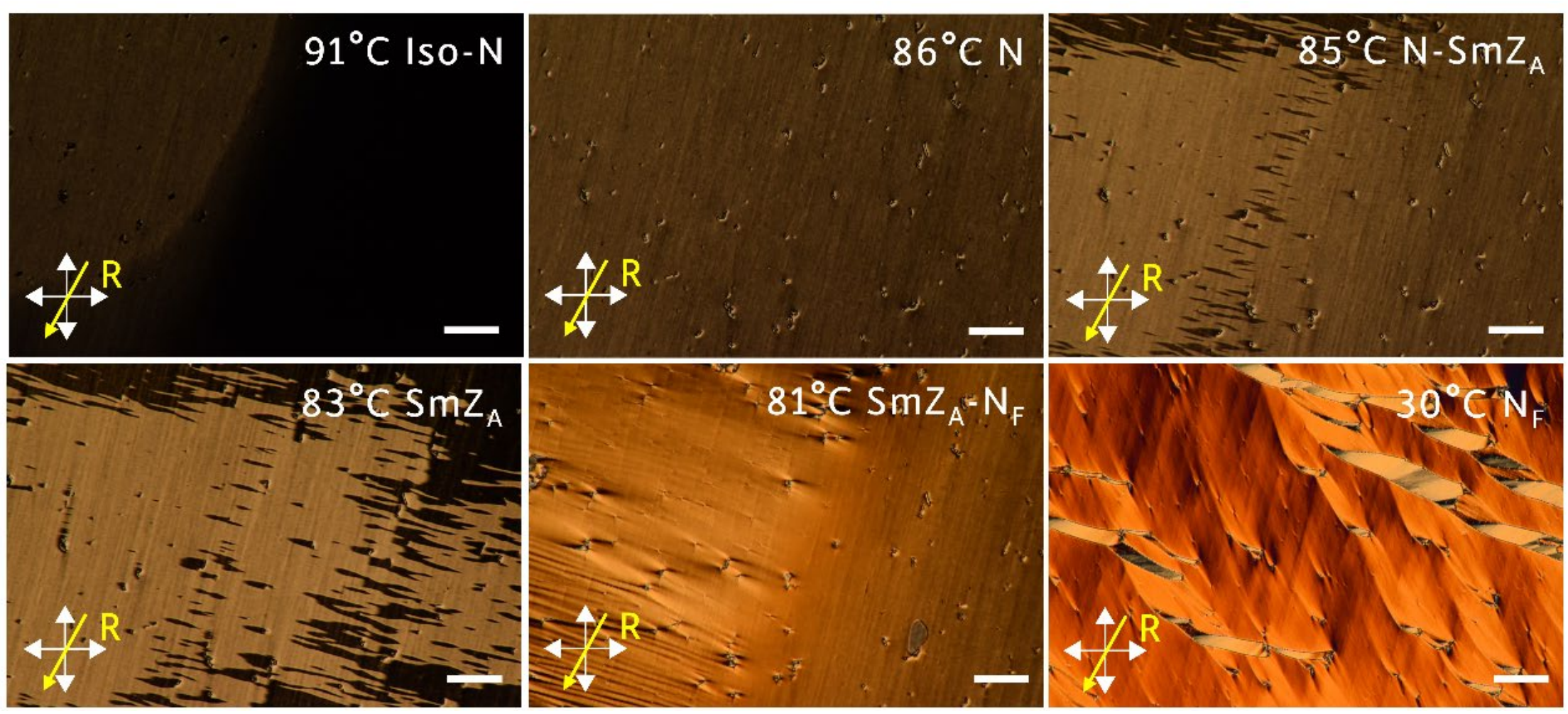

**Figure S3.** PLM textures of **RFM-3** observed under crossed polarizers in a 2-μm-thick planar LC cell. R denotes the rubbing direction. Scale bar, 100 μm.

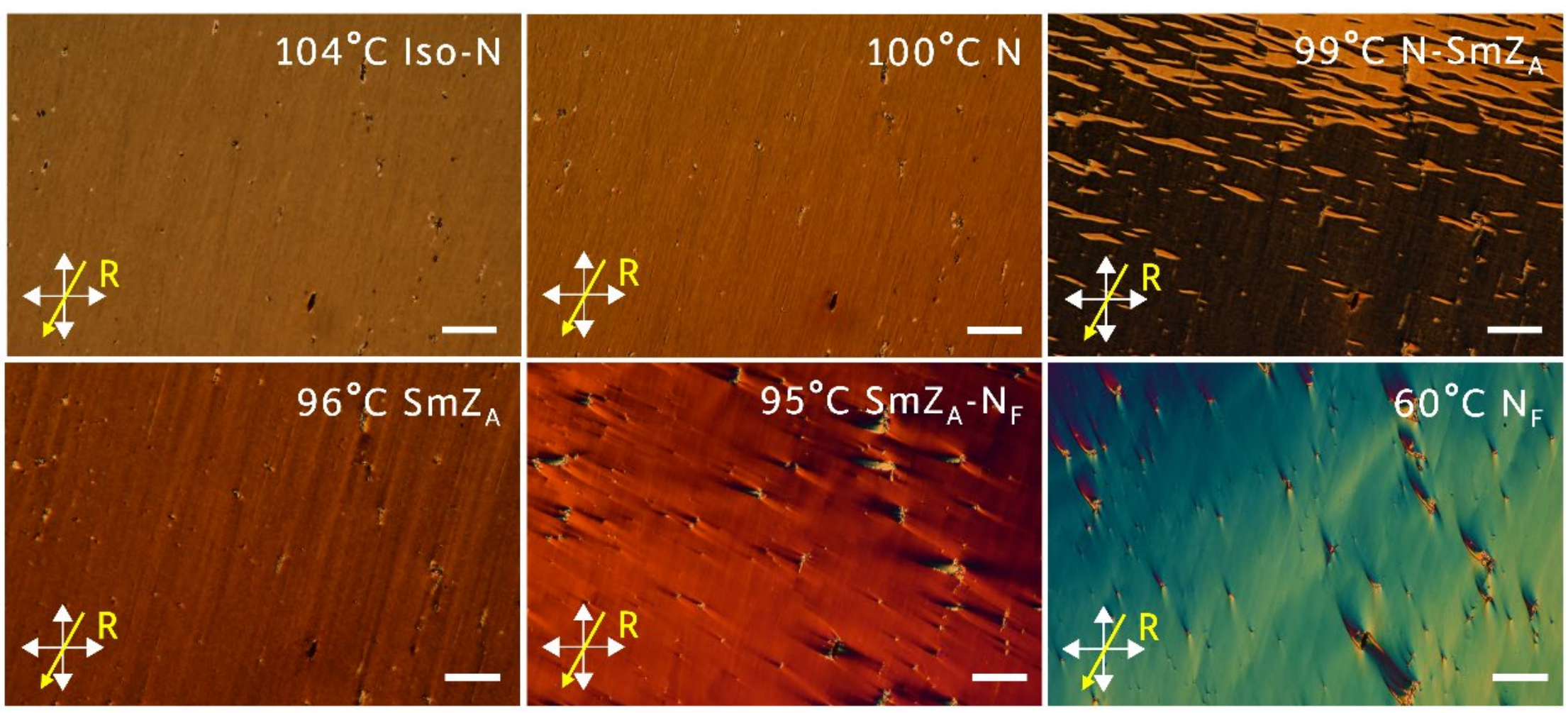

**Figure S4.** PLM textures of **RFM-4** observed under crossed polarizers in a 2-μm-thick planar LC cell. R denotes the rubbing direction. Scale bar, 100 μm.

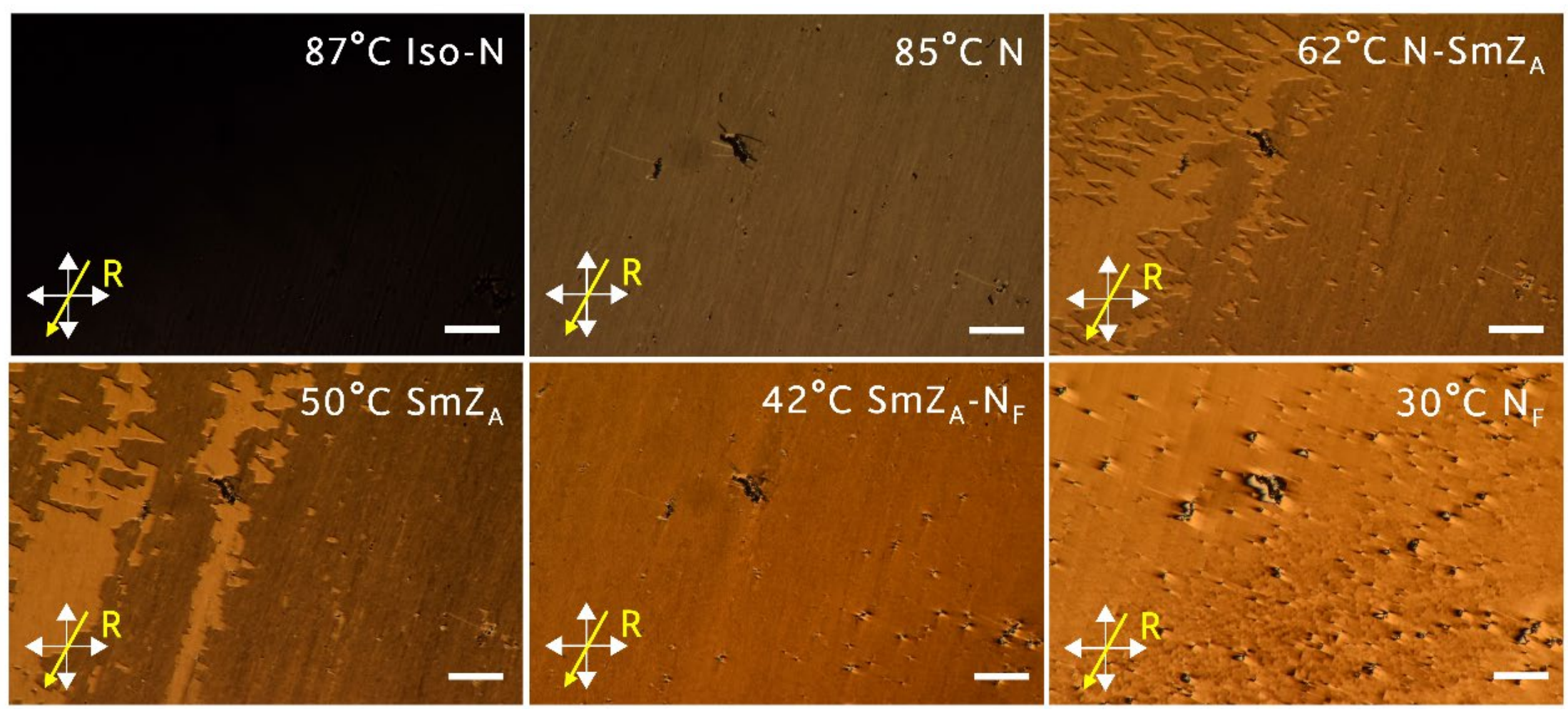

**Figure S5.** PLM textures of **RFM-5** observed under crossed polarizers in a 2-μm-thick planar LC cell. R denotes the rubbing direction. Scale bar, 100 μm.

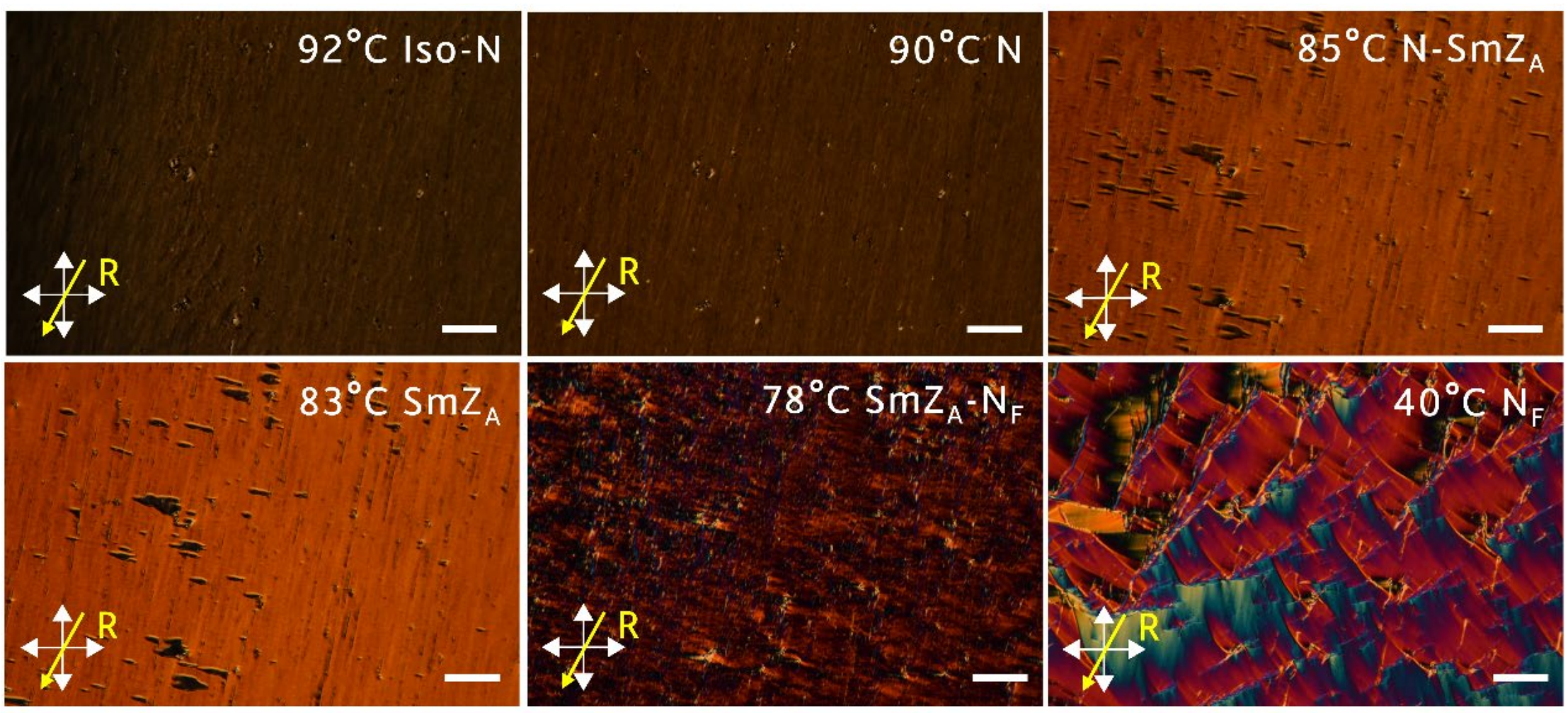

**Figure S6.** PLM textures of **RFM-6** observed under crossed polarizers in a 2-μm-thick planar LC cell. R denotes the rubbing direction. Scale bar, 100 μm.

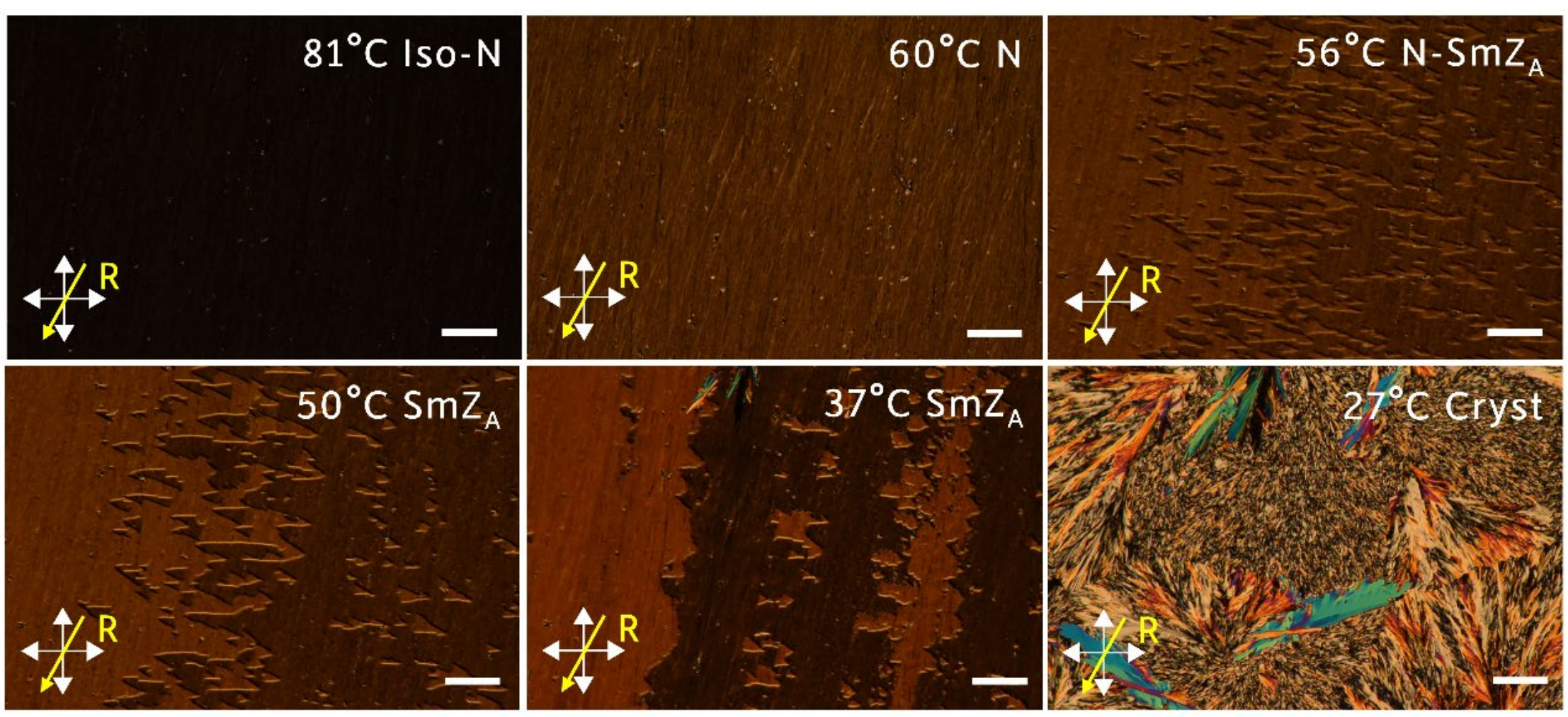

**Figure S7.** PLM textures of **RFM-7** observed under crossed polarizers in a 2-μm-thick planar LC cell. R denotes the rubbing direction. Scale bar, 100 μm.

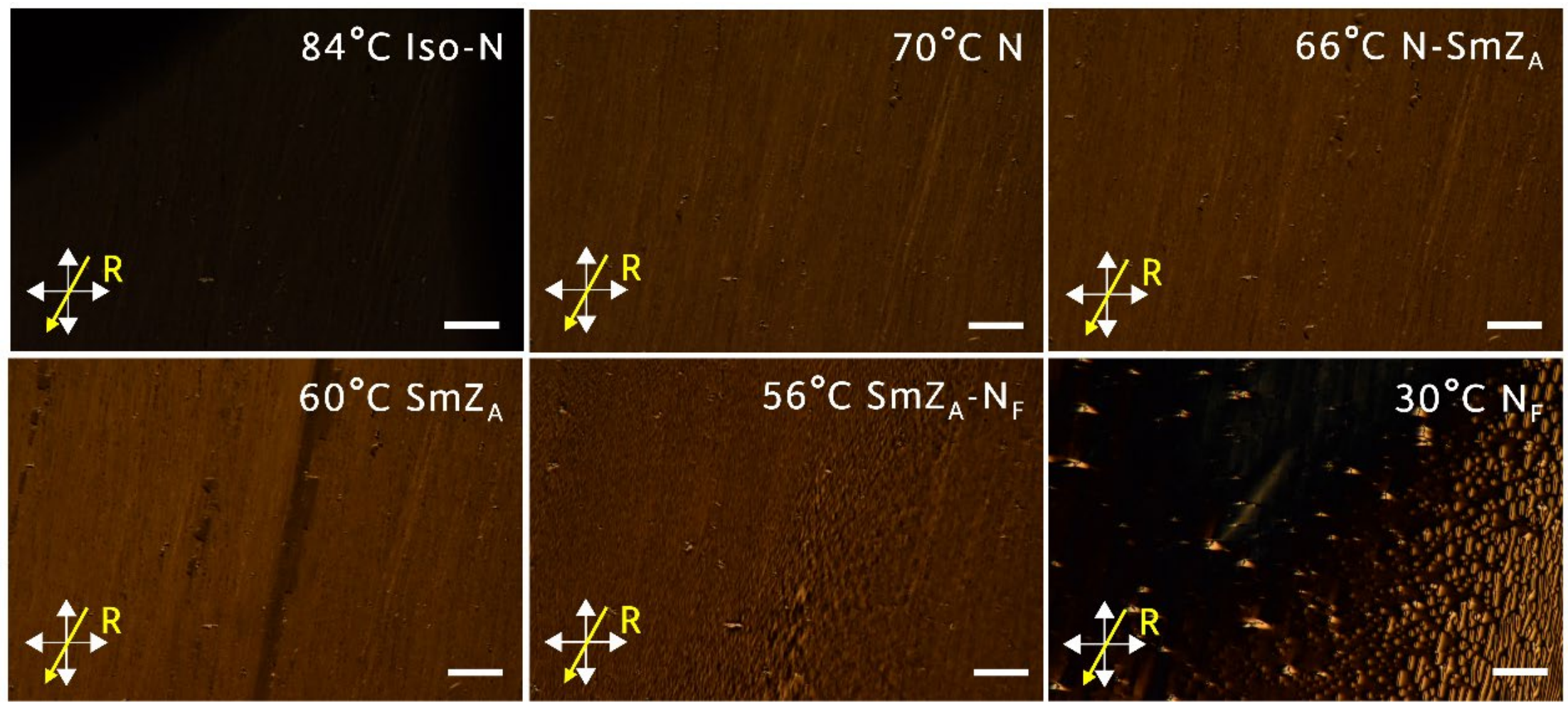

**Figure S8.** PLM textures of **RFM-8** observed under crossed polarizers in a 2-μm-thick planar LC cell. R denotes the rubbing direction. Scale bar, 100 μm.

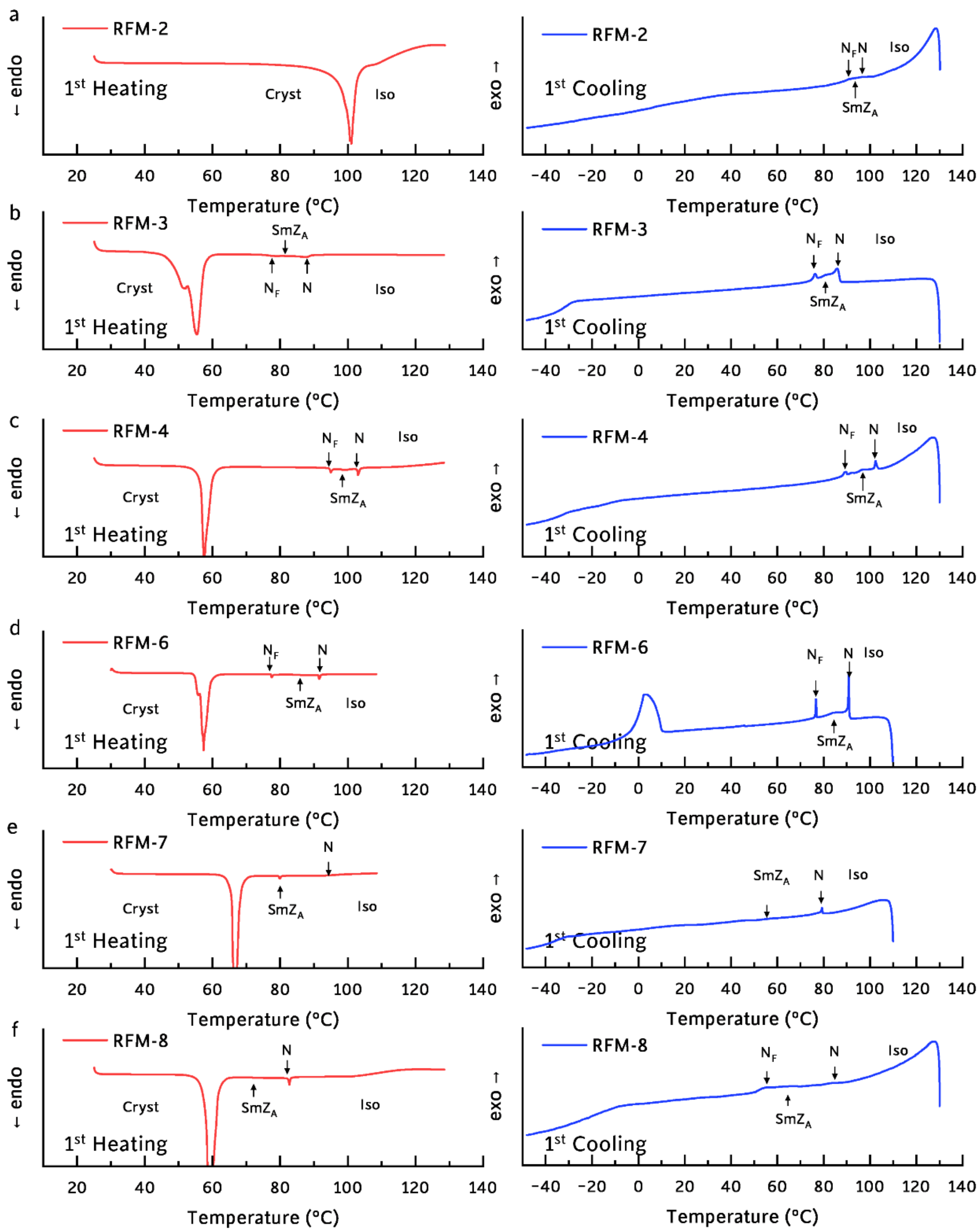

**Figure S9. a-f,** Differential scanning calorimetry (DSC) thermograms of RFMs recorded during the 1st heating and subsequent cooling scans. Scan rate, 10 K $min^{-1}$.

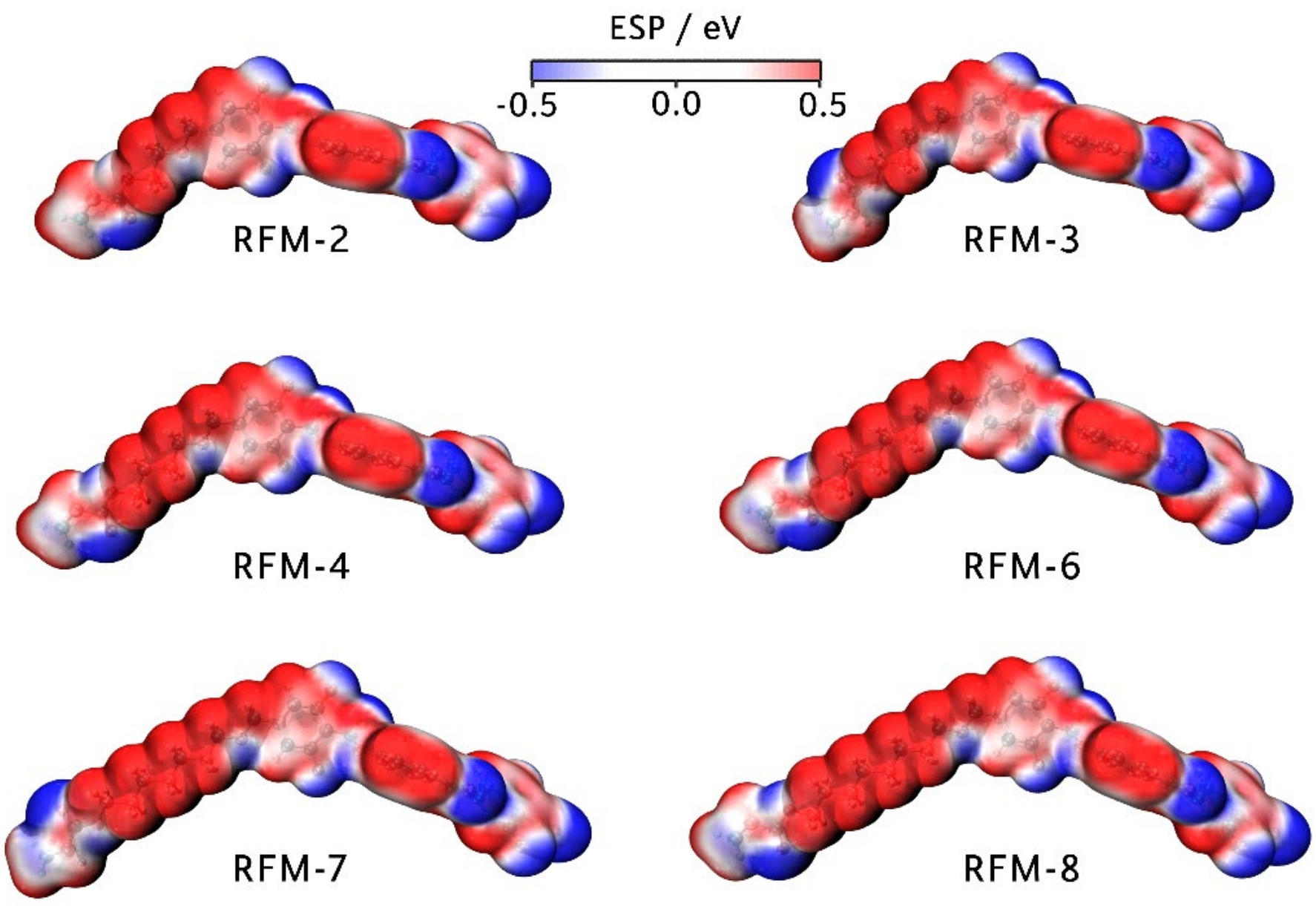

**Figure S10.** Optimized geometries of RFMs with three-dimensional electrostatic potential (ESP) maps calculated at the B3LYP-D3BJ/cc-pVTZ level. The ESP maps are displayed on the electron density iso-surface of 0.0004 a.u.

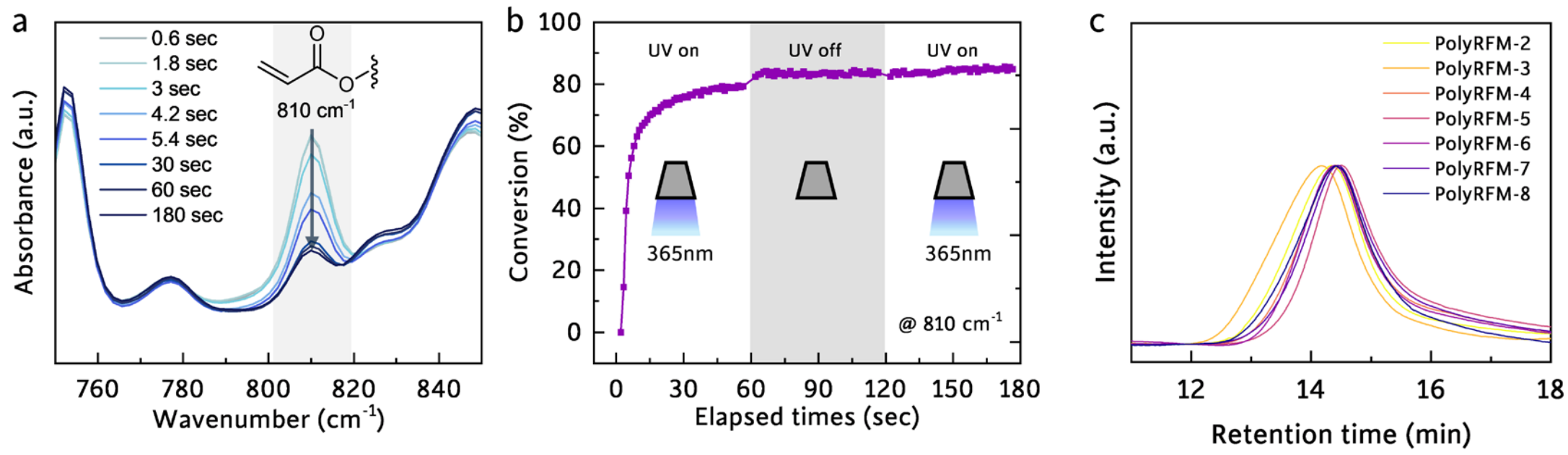

**Figure S11.** FTIR monitoring of *in situ* photopolymerization and SEC characterization of the resulting LCPs. **a,** Reaction completion during *in situ* photopolymerization was monitored by FTIR by tracking the decrease in the acrylate band at 810 $cm^{-1}$. **b**, Conversion of acrylate functional groups during the photopolymerization of **RFM-5** into **PolyRFM-5**, calculated from FTIR spectra. **c,** Size-exclusion chromatography (SEC) traces of the resulting LCPs synthesized by *in situ* photopolymerization.

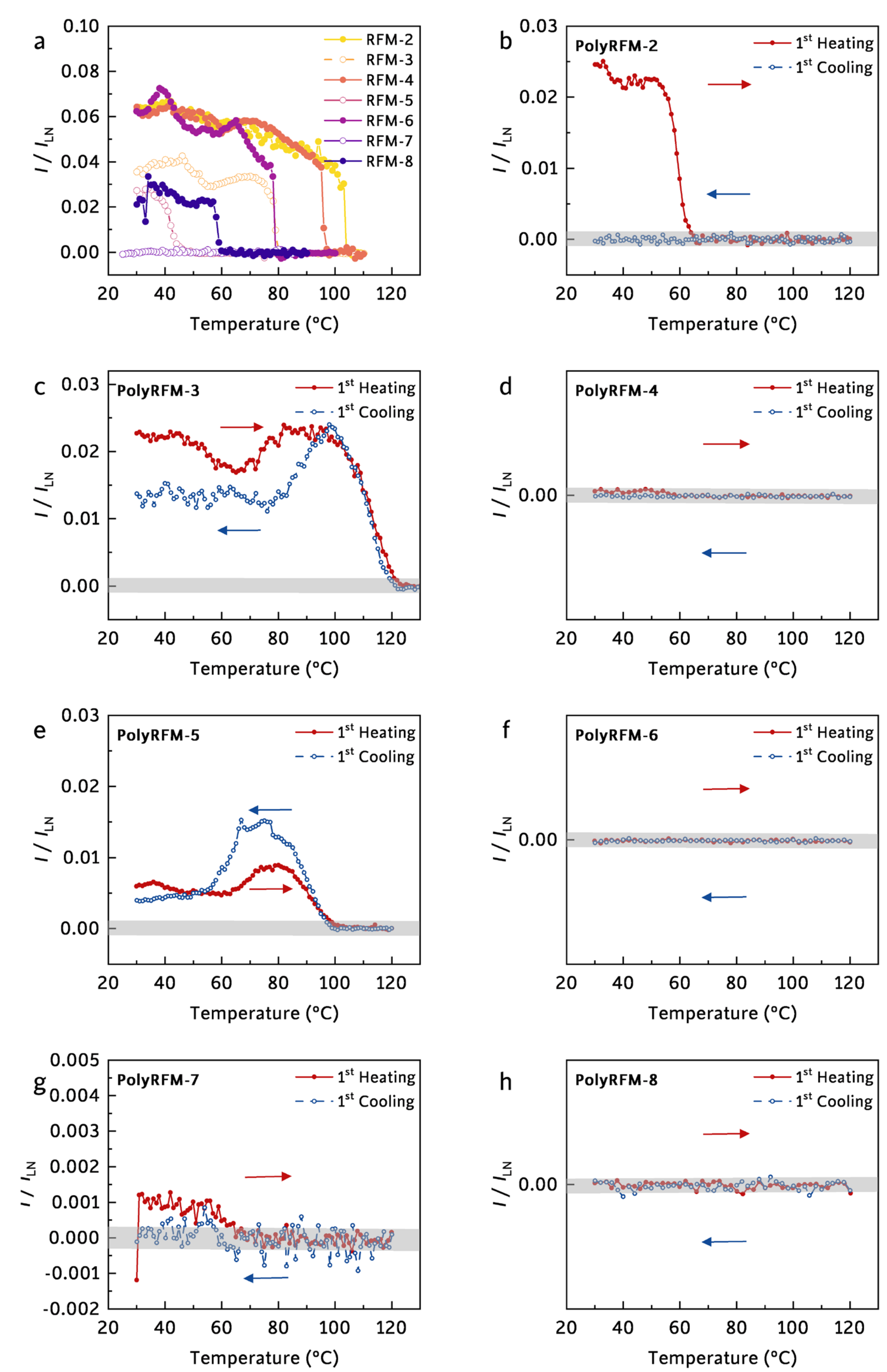

**Figure S12.** Temperature-dependent second-harmonic generation (SHG) intensity. **a**, Temperature dependence of the SHG signals of RFMs in 2-μm-thick planar LC cells during the 1$^{st}$ cooling from isotropic phase. **b-h**, Temperature dependence of SHG signals of PolyRFMs in the same 2-μm-thick planar LC cells after *in situ* photopolymerization, during the 1$^{st}$ heating and cooling scans. An x-cut lithium niobate (LN) wafer was used as the reference sample.

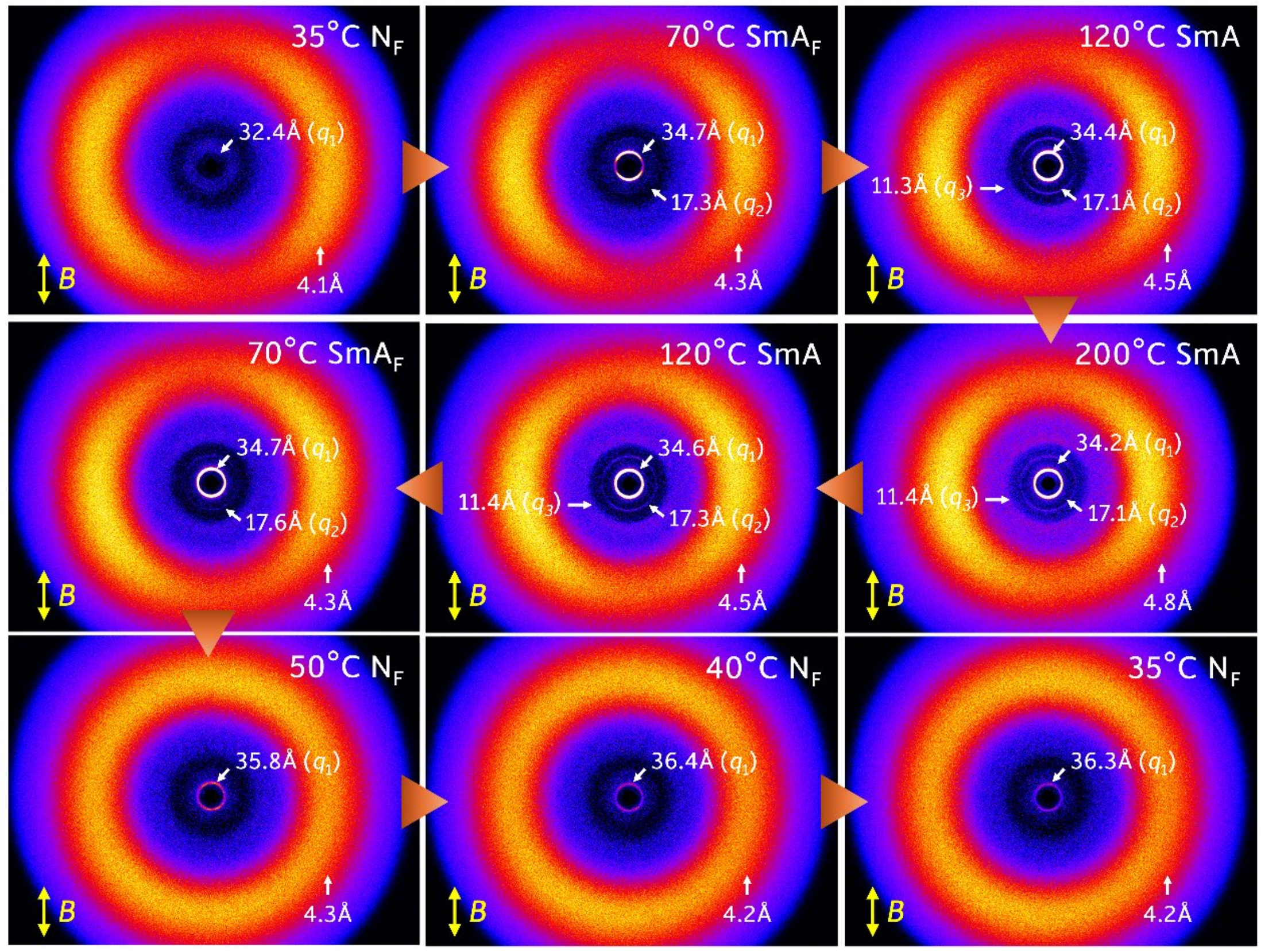

**Figure S13.** Two-dimensional wide-angle X-ray diffraction (2D WAXD) patterns of the PolyRFM-5 film in the SmA, $SmA_F$, and $N_F$ phases recorded during the first heating and cooling scans. The uniaxial alignment of ferro-LCP film was retained upon *in situ* photopolymerization of magnetically aligned **RFM-5** in its $N_F$ phase (at 35 °C). **B** denotes the direction of the applied magnetic field (~ 2.5 kG). The director of the ferroelectric LCP mesogens deviates from **B** after thermal annealing upon cooling from the high-temperature SmA phase.

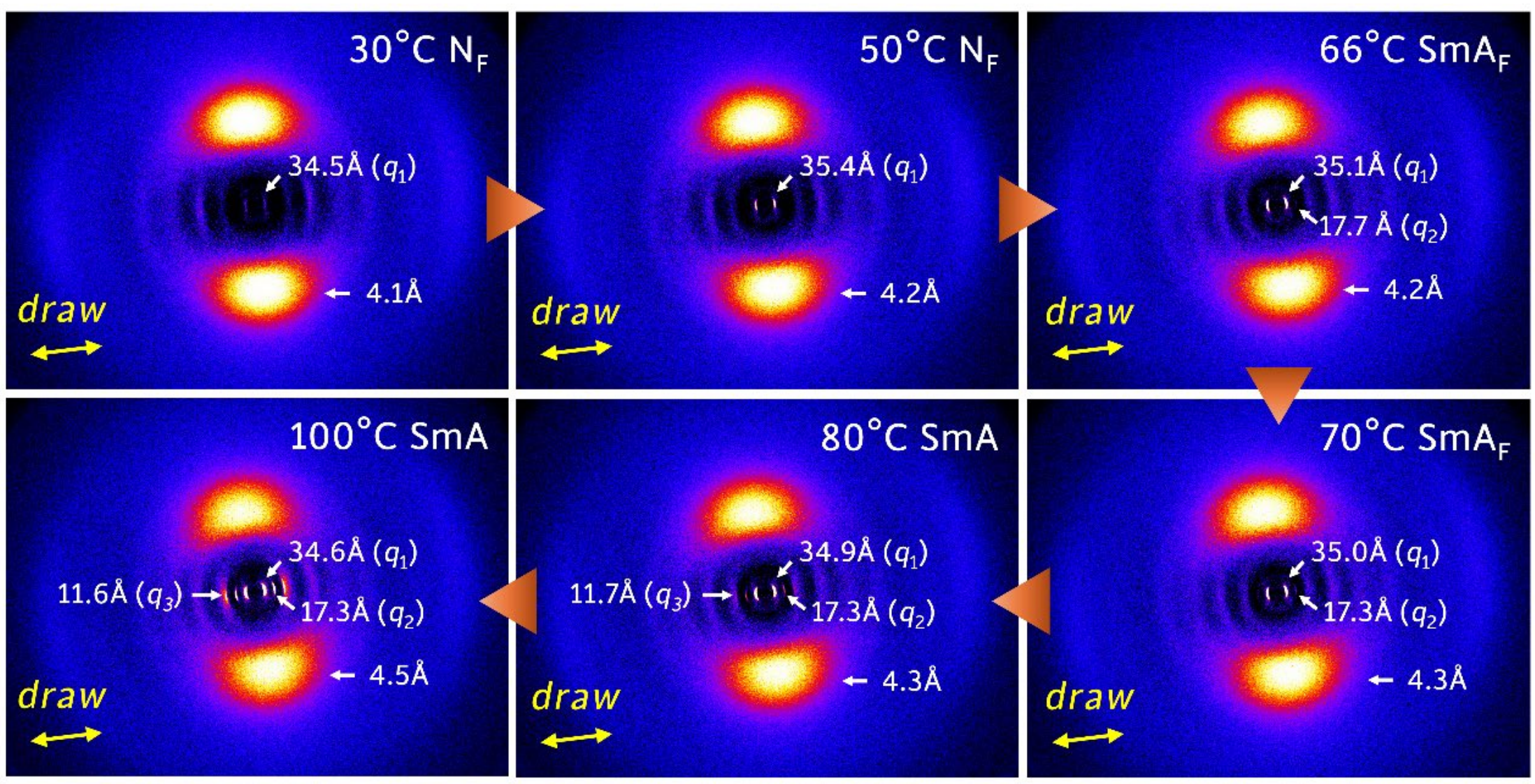

**Figure S14.** 2D WAXD patterns of a PolyRFM-5 fiber in the SmA, SmA$_F$, and N$_F$ phase during the first heating scan. The fiber was drawn from a ferroelectric LCP film at 120 °C in the SmA phase and subsequently quenched. The arrow denotes the drawing direction. The director of ferro-LCP mesogens is uniaxially aligned along the drawing direction throughout the first heating scan.

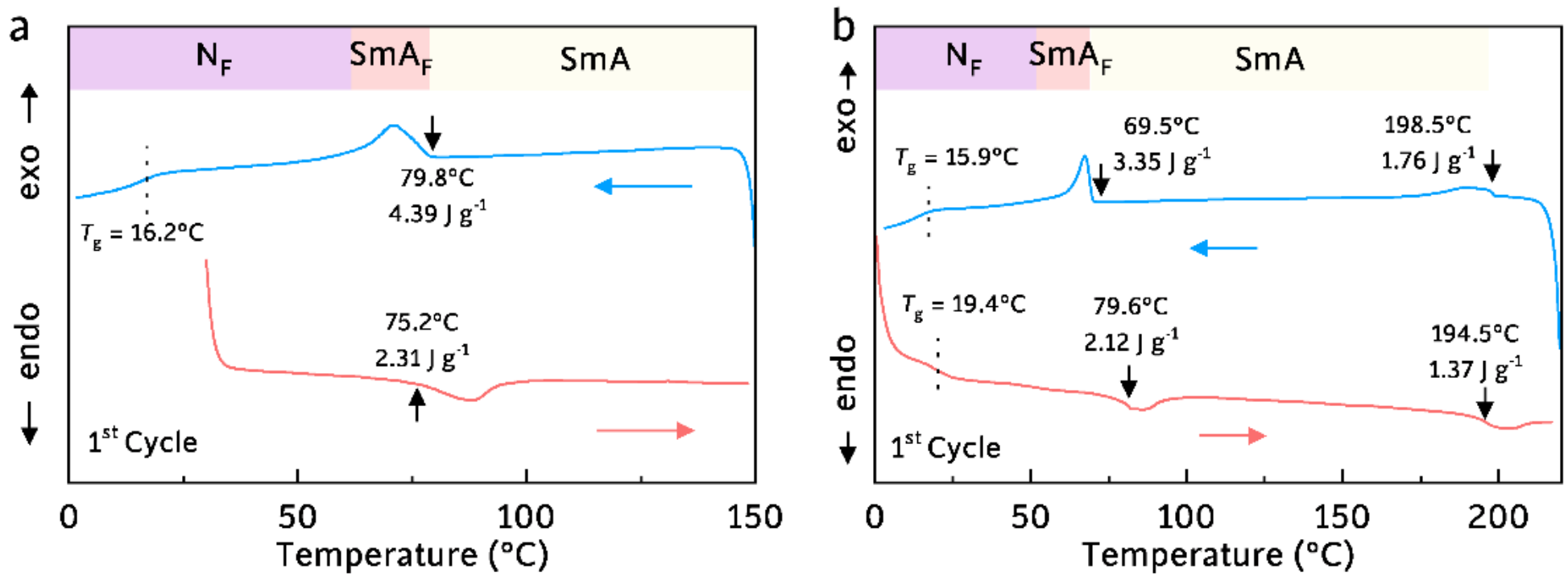

**Figure S15.** DSC thermograms of PolyRFM-5 recorded during the 1$^{st}$ heating and cooling cycle at different scan rates. **a,** Scan rate, 10 K min$^{-1}$. **b,** Scan rate, 20 K min$^{-1}$.

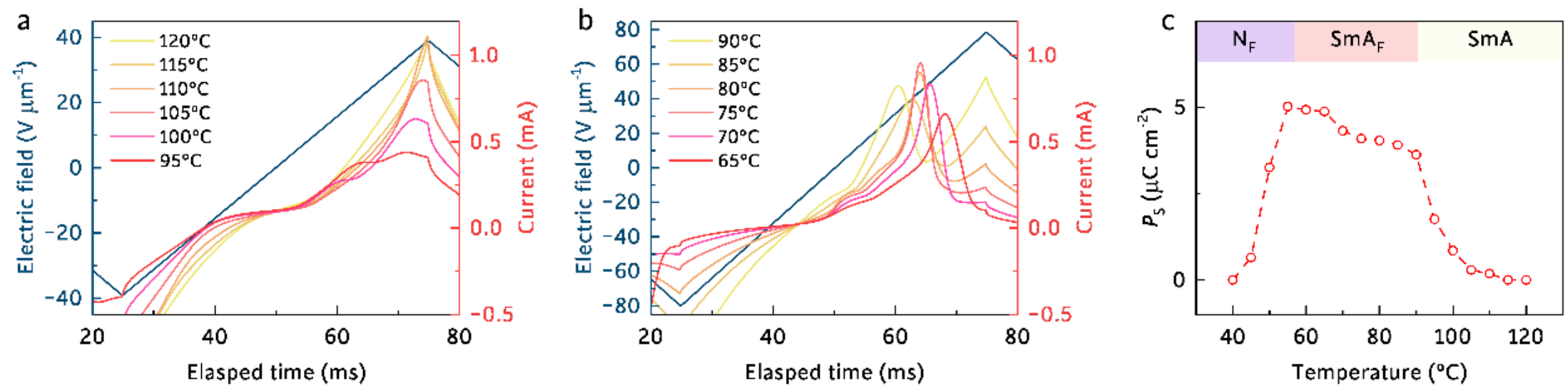

**Figure S16.** Polarization reversal current response and spontaneous polarization of PolyRFM-5. **a,b,** Polarization reversal current response traces measured for **PolyRFM-5** at different temperatures. **c,** Temperature dependence of the spontaneous polarization ($P_S$) of **PolyRFM-5**.

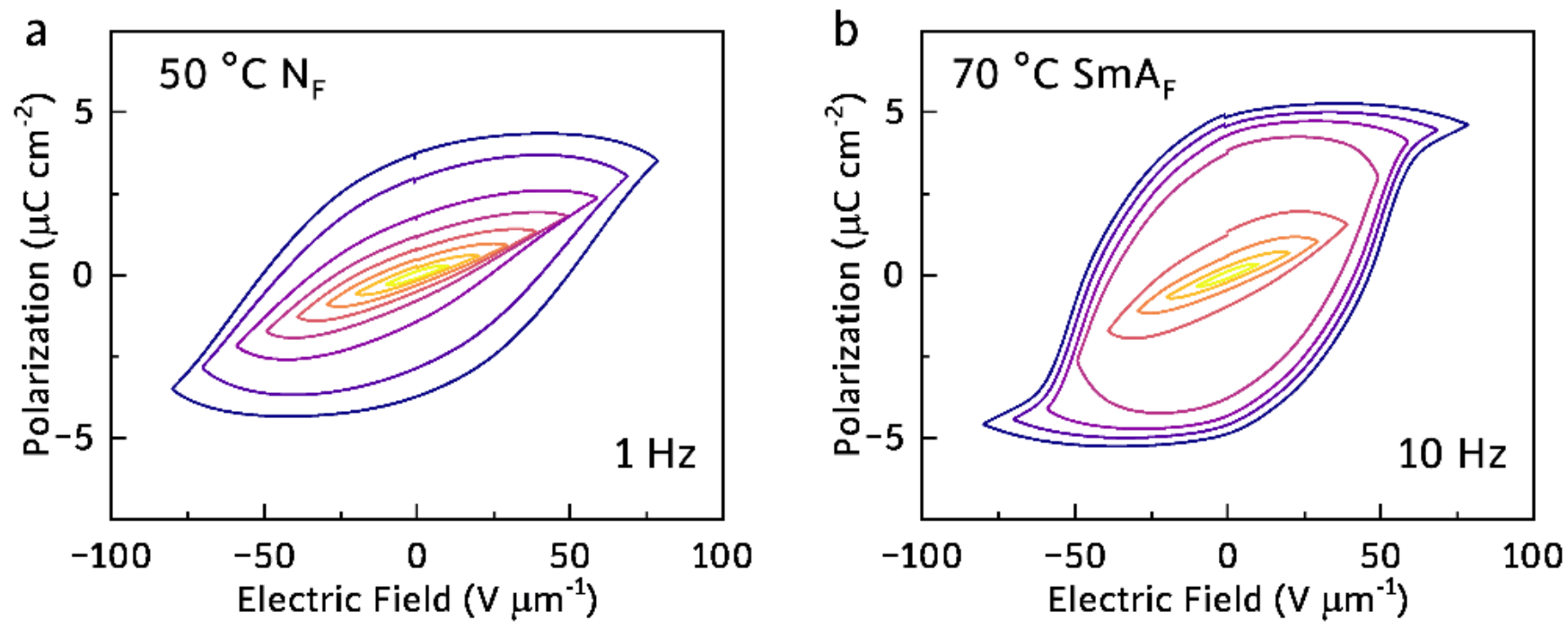

**Figure S17.** *P–E* hysteresis loops of PolyRFM-5. **a,b,** *P–E* hysteresis loops measured for **PolyRFM-5** at 50 °C and 1 Hz in the $N_F$ phase (**a**) and at 70 °C 10 and 10 Hz in the $SmA_F$ phase (**b**).

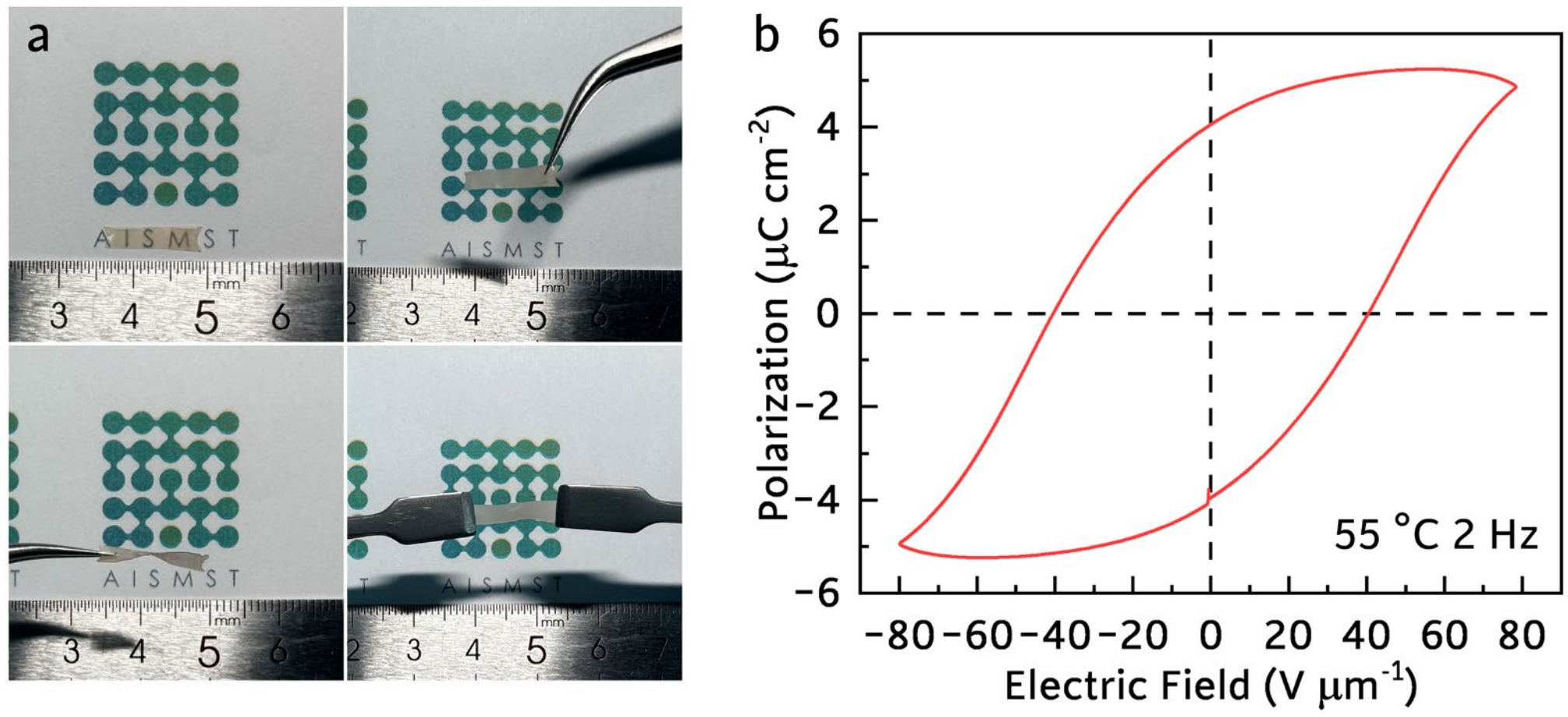

**Figure S18.** Slightly crosslinked ferro-LCP film. **a,** Schematic illustration of the fabrication of a uniaxially aligned **RFM-5**/HDDA mixture containing 3 wt% HDDA in a homemade 100-μm-thick LC cell with parallel rubbing. After *in situ* photopolymerization in the $N_F$ phase, the uniaxially aligned **PolyRFM-5** film containing 3 wt% HDDA was peeled off from the glass substrates. The resulting ferroelectric LCP film exhibited high flexibility. **b**, *P–E* hysteresis loop of **PolyRFM-5** containing 3 wt% HDDA.

**Table S1. Molecular characteristics of the PolyRFMs**

| LCP | $M_n$ [a] (kDa) | $M_w$ [a] (kDa) | $D_p$ [b] | $Đ$ [c] | *Ferroelectricity* [d] |
|---|---|---|---|---|---|
| PolyRFM-2 | 155.2 | 319.7 | 285 | 2.06 | × |
| PolyRFM-3 | 172.6 | 474.7 | 310 | 2.75 | ✓ |
| PolyRFM-4 | 136.6 | 250.1 | 239 | 1.83 | × |
| PolyRFM-5 | 109.6 | 180.9 | 187 | 1.65 | ✓ |
| PolyRFM-6 | 129.1 | 228.6 | 215 | 1.77 | × |
| PolyRFM-7 | 114.5 | 216.4 | 187 | 1.89 | ✓ |
| PolyRFM-8 | 111.3 | 262.8 | 177 | 2.36 | × |

[a] The number-average and weight-average molecular weights of LCPs were determined using SEC.

[b] The degrees of polymerization ($D_p$) was calculated by $D_p = M_n / M_{monomer}$.

[c] The polydispersity index ($Đ = M_w/M_n$) was determined from SEC.

[d] The ferroelectricity was determined from SHG active.

## 3. Synthesis and characterization

### 3.1 Synthesis of reactive ferroelectric mesogens (RFMs)

NaH (60 wt%)
DMF, 0°C-RT

n=2 1a
n=3 1b
n=4 1c
n=5 1d
n=6 1e
n=7 1f
n=8 1g

**2-((3,5-difluorobenzyl)oxy)ethan-1-ol (1a):** To a dried 100 mL round-bottom flask was added ethane-1,2-diol (14.99 g, 241.52 mmol) followed by anhydrous N,N-dimethylformamide (DMF, 20 mL). The mixture was cooled to 0 °C using an ice-water bath. Sodium hydride (NaH, 60 wt% dispersion in mineral oil, 2.03 g, 50.72 mmol) was added portion wise over 5 min while maintaining the internal temperature below 5 °C. After stirring at 0 °C for 1 h, a solution of 1-(bromomethyl)-3,5-difluorobenzene (10.00 g, 48.30 mmol) in 5 mL anhydrous DMF was added dropwise via syringe over 10 min. The reaction mixture was then allowed to warm gradually to ambient temperature and stirred for 16 h. Upon completion (monitored by TLC), the reaction was diluted with water (30 mL), and extracted with ethyl acetate (EA, 3 × 30 mL). The combined organic layers were washed sequentially with water (2 × 20 mL) and saturated aqueous sodium chloride solution (20 mL), dried over anhydrous sodium sulfate, filtered, and concentrated under reduced pressure. The crude residue was purified by column chromatography to afford the target product (pale yellow liquid). Yield: 7.73 g (85.0%). $^{1}$H NMR (500 MHz, DMSO-$d_6$) δ 7.10 (td, 1H), 7.07 – 7.02 (m, 2H), 4.69 (td, 1H), 4.52 (s, 2H), 3.64 – 3.53 (m, 2H), 3.48 (dd, 2H). $^{19}$F NMR (471 MHz, DMSO-$d_6$) δ -110.08 (t).

**3-((3,5-difluorobenzyl)oxy)propan-1-ol (1b):** Yield: 7.90 g (88.0%). $^{1}$H NMR (500 MHz, DMSO-$d_6$) δ 7.12 (tt, 1H), 7.02 (h, 2H), 4.47 (s, 2H), 4.43 (td, 1H), 3.51 (d, 2H), 3.33 (d, 2H), 1.73 – 1.64 (m, 2H). $^{19}$F NMR (471 MHz, DMSO-$d_6$) δ -109.99 (t).

**4-((3,5-difluorobenzyl)oxy)butan-1-ol (1c):** Yield: 8.10 g (84.5%). $^{1}$H NMR (500 MHz, DMSO-$d_6$) δ 7.12 (tt, 1H), 7.02 (h, 2H), 4.47 (s, 2H), 4.43 (td, 1H), 3.51 (d, 2H), 3.33 (d, 2H), 1.73 – 1.64 (m, 2H). $^{19}$F NMR (471 MHz, DMSO-$d_6$) δ -109.99 (t).

**5-((3,5-difluorobenzyl)oxy)pentan-1-ol (1d):** Yield: 8.22 g (86.0%); $^{1}$H NMR (500 MHz, DMSO-$d_6$) δ 7.12 (t, 1H), 7.02 (d, 2H), 4.47 (s, 2H), 4.34 (tz, 1H), 3.43 (t, 2H), 3.40 – 3.36 (m, 2H), 1.55 (p, 2H), 1.42 (p, 2H), 1.34 (tt, 2H). $^{19}$F NMR (471 MHz, DMSO-$d_6$) δ -109.94 (t).

**6-((3,5-difluorobenzyl)oxy)hexan-1-ol (1e):** Yield: 8.40 g (85.0%); $^{1}$H NMR (500 MHz, Chloroform-$d$) δ 6.81 – 6.76 (m, 2H), 6.62 (tt, $J$ = 9.0, 2.4 Hz, 1H), 4.39 (s, 2H),

3.53 (t, $J$ = 6.6 Hz, 2H), 3.41 (t, $J$ = 6.6 Hz, 2H), 1.57 (dq, $J$ = 7.9, 6.5 Hz, 2H), 1.49 (p, $J$ = 6.8 Hz, 2H), 1.37 – 1.27 (m, 4H). $^{19}$F NMR (471 MHz, Chloroform-*d*) δ -109.99 (t).

**7-((3,5-difluorobenzyl)oxy)heptan-1-ol (1f):** Yield: 8.66 g (86.2%); $^{1}$H NMR (500 MHz, Chloroform-*d*) δ 6.82 (t, $J$ = 6.2 Hz, 2H), 6.65 (tt, $J$ = 9.0, 2.3 Hz, 1H), 4.42 (s, 2H), 3.56 (t, $J$ = 6.5 Hz, 2H), 3.43 (t, $J$ = 6.6 Hz, 2H), 1.58 (p, $J$ = 6.6 Hz, 2H), 1.51 (p, $J$ = 6.7 Hz, 2H), 1.39 – 1.29 (m, 6H). $^{19}$F NMR (471 MHz, Chloroform-*d*) δ -109.99 (t).

**8-((3,5-difluorobenzyl)oxy)octan-1-ol (1g):** Yield: 9.10 g (89.1%); $^{1}$H NMR (500 MHz, Chloroform-*d*) δ 6.85 (d, $J$ = 12.5 Hz, 2H), 6.69 (tt, $J$ = 9.0, 2.2 Hz, 1H), 4.45 (s, 2H), 3.61 (d, $J$ = 6.7 Hz, 2H), 3.46 (t, $J$ = 6.6 Hz, 2H), 1.57 (ddt, $J$ = 20.9, 13.8, 6.8 Hz, 6H), 1.40 – 1.32 (m, 6H). $^{19}$F NMR (471 MHz, Chloroform-*d*) δ -110.01 (t).

Si–Cl
Imidazole
DMF, 0°C-RT, overnight
HO ... TBSO

n=2 1a | n=2 2a
n=3 1b | n=3 2b
n=4 1c | n=4 2c
n=5 1d | n=5 2d
n=6 1e | n=6 2e
n=7 1f | n=7 2f
n=8 1g | n=8 2g

**tert-butyl(2-((3,5-difluorobenzyl)oxy)ethoxy)dimethylsilane (2a):** To a dried 100 mL was added compound 1a (7.70 g, 40.92 mmol) and imidazole (5.57 g, 81.84 mmol), followed by anhydrous DMF (50 mL). The mixture was cooled to 0 °C using an ice-water bath. The *tert*-butyldimethylsilyl chloride (TBSCl) was added portion wise over 5 min while maintaining the internal temperature below 5 °C. The reaction was gradually warmed to ambient temperature and stirred for overnight. Upon completion (monitored by TLC), the mixture was diluted with water, and extracted with EAs (3 × 30 mL). The combined organic layers were washed with water (2 × 20 mL) and saturated aqueous sodium chloride solution (20 mL), dried over anhydrous sodium sulfate, filtered, and concentrated under reduced pressure. The crude residue was purified by flash column chromatography on silica gel to afford target compound 2a. Yield: 11.07 g (89.5%). $^{1}$H NMR (500 MHz, DMSO-$d_6$) δ 7.06 (td, 1H), 6.99 (h, 2H), 4.50 (s, 2H), 3.70 (td, 2H), 3.50 – 3.44 (m, 2H), 0.84 – 0.78 (m, 9H), -0.01 (d, 6H). $^{19}$F NMR (471 MHz, DMSO-$d_6$) δ -110.09 (t). 

**tert-butyl(3-((3,5-difluorobenzyl)oxy)propoxy)dimethylsilane (2b):** Yield: 11.06 g (89.0%). $^{1}$H NMR (500 MHz, DMSO-$d_6$) δ 7.08 (tt, 1H), 6.98 (h, 2H), 4.44 (s, 2H), 3.63 (t, 2H), 3.46 (t, 2H), 1.69 (p, 2H), 0.80 (s, 9H), -0.03 (s, 6H). $^{19}$F NMR (471 MHz, DMSO-$d_6$) δ -110.03 (t).

**tert-butyl(4-((3,5-difluorobenzyl)oxy)butoxy)dimethylsilane (2c):** Yield: 10.22 g

(86.0%). $^{1}$H NMR (500 MHz, Chloroform-*d*) δ 6.89 – 6.83 (m, 2H), 6.70 (ddd, 1H), 4.47 (s, 2H), 3.64 (t, 2H), 3.50 (t, 2H), 1.73 – 1.55 (m, 4H), 0.89 (s, 9H), 0.04 (d, 6H). $^{19}$F NMR (471 MHz, Chloroform-*d*) δ -110.01 (d).

**tert-butyl((5-((3,5-difluorobenzyl)oxy)pentyl)oxy)dimethylsilane (2d):** Yield: 10.58 g (88.5 %). $^{1}$H NMR (500 MHz, Chloroform-*d*) δ 6.84 (h, 2H), 6.68 (tt, 1H), 4.45 (s, 2H), 3.60 (t, 2H), 3.47 (t, 2H), 1.63 (p, 2H), 1.57 – 1.49 (m, 2H), 1.41 (tdd, 2H), 0.88 (s, 9H), 0.03 (s, 6H). $^{19}$F NMR (471 MHz, Chloroform-*d*) δ -109.99 (d).

**tert-butyl((6-((3,5-difluorobenzyl)oxy)hexyl)oxy)dimethylsilane (2e):** Yield: 10.06 g (89.2%). $^{1}$H NMR (500 MHz, DMSO-$d_6$) δ 7.09 (tt, 1H), 7.03 – 6.97 (m, 2H), 4.46 (s, 2H), 3.61 – 3.52 (m, 2H), 3.42 (td, 2H), 1.53 (p, 2H), 1.44 (p, 2H), 1.37 – 1.21 (m, 4H), 0.84 (d, 9H), -0.00 (d, 6H). $^{19}$F NMR (471 MHz, DMSO-$d_6$) δ -109.99 (d).

**tert-butyl((7-((3,5-difluorobenzyl)oxy)heptyl)oxy)dimethylsilane (2f):** Yield: 10.56 g (87.4%). $^{1}$H NMR (500 MHz, DMSO-$d_6$) δ 7.07 (td, 1H), 6.97 (h, 2H), 4.42 (s, 2H), 3.51 (t, 2H), 3.38 (t, 2H), 1.50 (p, 2H), 1.42 – 1.34 (m, 2H), 1.32 – 1.18 (m, 6H), 0.81 (s, 9H), -0.03 (s, 6H). $^{19}$F NMR (471 MHz, DMSO-$d_6$) δ -110.04 (d).

**tert-butyl((8-((3,5-difluorobenzyl)oxy)octyl)oxy)dimethylsilane (2g):** Yield: 10.96 g (88.3%). $^{1}$H NMR (500 MHz, DMSO-$d_6$) δ 7.12 – 7.05 (m, 1H), 6.99 (h, 2H), 4.44 (s, 2H), 3.40 (t, 2H), 3.34 (t, 2H), 1.52 (p, 2H), 1.40 (m, 4H), 1.24 (q, 6H), 0.83 (d, 9H), -0.01 (d, 6H). $^{19}$F NMR (471 MHz, DMSO-$d_6$) δ -110.03 (d).

(1) LDA, THF, $N_2$, -78, 1h
(2) DMF, -78°C - RT, overnight

| | |
|---|---|
| n=2 2a | n=2 3a |
| n=3 2b | n=3 3b |
| n=4 2c | n=4 3c |
| n=5 2d | n=5 3d |
| n=6 2e | n=6 3e |
| n=7 2f | n=7 3f |
| n=8 2g | n=8 3g |

**4-((2-((tert-butyldimethylsilyl)oxy)ethoxy)methyl)-2,6-difluorobenzaldehyde (3a):** To a dried 100 mL round-bottom flask under nitrogen atmosphere was added compound 2a (11.00 g, 36.37mmol), followed by anhydrous tetrahydrofuran (THF, 40 mL). The solution was cooled to -78 °C, and after stirring at this temperature for 30 min, a 2.0 M solution of lithium diisopropylamide (LDA in THF, 1.05 equiv.) was added dropwise via syringe over 10 min, maintaining the internal temperature below -75 °C. The mixture was stirred at -78 °C for an additional 30 min, then anhydrous DMF (4 equiv.) was added dropwise at the same temperature. The reaction was gradually warmed to ambient temperature and stirred for 2 h. Upon completion (monitored by TLC), the mixture was carefully quenched with saturated aqueous ammonium chloride solution (100 mL) at 0 °C, diluted with water (30 mL), and extracted with EA (3 × 30 mL). The combined organic layers were washed with water (2 × 20 mL) and saturated aqueous sodium chloride solution (20 mL), dried over anhydrous sodium sulfate, filtered, and concentrated under reduced pressure. The crude residue was purified by flash column chromatography on silica gel (petroleum ether/ethyl acetate = 9:1, v/v) to afford target compound 3a as a pale yellow oil. Yield: 10.16 g (84.5%). $^{1}$H NMR (500 MHz, DMSO-$d_6$) δ 10.14 (s, 1H), 7.15 (d, 2H), 4.57 (s, 2H), 3.72 (t, 2H), 3.51 (t, 2H), 0.83 (s, 9H), 0.01 (s, 6H). $^{19}$F NMR (471 MHz, DMSO-$d_6$) δ -110.09 (t).

**4-((3-((tert-butyldimethylsilyl)oxy)propoxy)methyl)-2,6-difluorobenzaldehyde (3b):** Yield: 9.55 g (85.0%). $^{1}$H NMR (500 MHz, Chloroform-*d*) δ 10.21 (s, 1H), 6.56 (d, 2H), 4.44 (s, 2H), 3.72 (t, 2H), 3.58 (t, 2H), 1.82 (q, 2H), 0.87 (d, 9H), 0.03 (d, 6H). $^{19}$F NMR (471 MHz, Chloroform-*d*) δ -117.18 (d).

**4-((4-((tert-butyldimethylsilyl)oxy)butoxy)methyl)-2,6-difluorobenzaldehyde (3c):** Yield: 10.62 g (82.0%). $^{1}$H NMR (500 MHz, DMSO-$d_6$) δ 10.19 (s, 1H), 7.17 (d, 2H), 4.54 (s, 2H), 3.60 (t, 2H), 3.48 (s, 2H), 3.48 (d, 2H), 1.65 – 1.48 (m, 4H), 0.85 (d, 9H), 0.02 (d, 6H). $^{19}$F NMR (471 MHz, DMSO-$d_6$) δ -115.45 (d).

**4-(((5-((tert-butyldimethylsilyl)oxy)pentyl)oxy)methyl)-2,6-difluorobenzaldehyde (3d):** Yield: 9.88 g (83 %). $^{1}$H NMR (500 MHz, Chloroform-*d*) δ 10.32 (s, 1H), 6.97 (d, 2H), 4.51 (s, 2H), 3.62 (t, 2H), 3.52 (t, 2H), 1.71 – 1.63 (m, 2H), 1.55 (h, 2H), 1.49 – 1.40 (m, 2H), 0.89 (d, 9H), 0.05 (s, 6H). $^{19}$F NMR (471 MHz, Chloroform-*d*) δ -114.51 (d).

**4-(((6-((tert-butyldimethylsilyl)oxy)hexyl)oxy)methyl)-2,6-difluorobenzaldehyde (3e):** Yield: 10.22 g (81.0%). $^{1}$H NMR (500 MHz, DMSO-$d_6$) δ 10.17 (s, 1H), 7.14 (d, 2H), 4.51 (s, 2H), 3.57 – 3.50 (m, 2H), 3.44 (dd, 2H), 1.53 (dt, 2H), 1.42 (q, 2H), 1.37 – 1.25 (m, 6H), 0.82 (d, 9H), -0.01 (d, 6H). $^{19}$F NMR (471 MHz, DMSO-$d_6$) δ -115.46 (d).

**4-(((7-((tert-butyldimethylsilyl)oxy)heptyl)oxy)methyl)-2,6-difluorobenzaldehyde (3f):** Yield: 10.62 g (80.0%). $^{1}$H NMR (500 MHz, DMSO-$d_6$) δ 10.18 (s, 1H), 7.16 (d, 2H), 4.53 (s, 2H), 3.55 (t, 2H), 3.45 (t, 2H), 1.55 (p, 2H), 1.43 (t, 2H), 1.30 (q, 6H), 0.84 (d, 9H), 0.00 (d, 6H). $^{19}$F NMR (471 MHz, DMSO-$d_6$) δ -115.46 (d).

**4-(((8-((tert-butyldimethylsilyl)oxy)octyl)oxy)methyl)-2,6-difluorobenzaldehyde (3g):** Yield: 10.02 g (78.0%). $^{1}$H NMR (500 MHz, DMSO-$d_6$) δ 10.20 (s, 1H), 7.15 (d, 2H), 4.54 (s, 2H), 3.56 (t, 2H), 3.45 (t, 2H), 1.55 (p, 2H), 1.43 (t, 2H), 1.37 – 1.25 (m, 8H), 0.84 (d, 9H), 0.00 (d, 6H). $^{19}$F NMR (471 MHz, DMSO-$d_6$) δ -115.43 (d).

AcOH : THF : $H_2O$ = 3: 1 : 1

RT, overnight

n=2 3a
n=3 3b
n=4 3c
n=5 3d
n=6 3e
n=7 3f
n=8 3g

n=2 4a
n=3 4b
n=4 4c
n=5 4d
n=6 4e
n=7 4f
n=8 4g

**2,6-difluoro-4-((2-hydroxyethoxy)methyl)benzaldehyde (4a):** To a 100 mL round-bottom flask was added compound 3a (10.00 g, 30.26 mmol), followed by a mixed solvent of acetic acid/THF/H2O (3:1:1, v/v, 50mL). The reaction was stirred for 12 h. Upon completion (monitored by TLC), the mixture was quenched with saturated aqueous ammonium chloride solution (15 mL), diluted with water (30 mL), and extracted with EA (3 × 30 mL). The combined organic layers were washed with water (2 × 20 mL) and saturated aqueous sodium chloride solution (20 mL), dried over anhydrous sodium sulfate, filtered, and concentrated under reduced pressure. The crude residue was purified by column chromatography on silica gel (petroleum ether/ethyl acetate = 3:1, v/v) to afford target compound 4a. Yield: 5.53 g (84.5%). $^{1}$H NMR (500 MHz, Chloroform-*d*) δ 10.29 (s, 1H), 6.97 (d, *J* = 9.5 Hz, 2H), 4.58 (s, 2H), 3.81 (s, 2H), 3.66 – 3.63 (m, 2H). $^{19}$F NMR (471 MHz, Chloroform-*d*) δ -114.26 (d).

**2,6-difluoro-4-((3-hydroxypropoxy)methyl)benzaldehyde (4b):** Yield: 3.40 g (35.8%). $^{1}$H NMR (500 MHz, Chloroform-*d*) δ 10.32 (s, 1H), 6.97 (d, *J* = 9.4 Hz, 2H), 4.55 (s, 2H), 3.82 (t, *J* = 5.8 Hz, 2H), 3.70 (t, *J* = 5.9 Hz, 2H), 1.92 (p, *J* = 5.9 Hz, 2H). $^{19}$F NMR (471 MHz, Chloroform-*d*) δ -114.25 (d).

**2,6-difluoro-4-((4-hydroxybutoxy)methyl)benzaldehyde (4c):** Yield: 3.12 g (33.4%). $^{1}$H NMR (500 MHz, Chloroform-*d*) δ 10.32 (d, *J* = 1.1 Hz, 1H), 6.97 (d, *J* = 9.5 Hz, 2H), 4.53 (s, 2H), 3.69 (t, *J* = 6.1 Hz, 2H), 3.57 (t, *J* = 6.0 Hz, 2H), 1.80 – 1.64 (m, 2H), 1.25 (t, *J* = 7.1 Hz, 2H). $^{19}$F NMR (471 MHz, Chloroform-*d*) δ -114.34 (d).

**2,6-difluoro-4-(((5-hydroxypentyl)oxy)methyl)benzaldehyde (4d):** Yield: 3.51 g (38.6%). $^{1}$H NMR (500 MHz, Chloroform-*d*) δ 10.32 (s, 1H), 6.97 (d, *J* = 9.6 Hz, 2H), 4.51 (s, 2H), 3.67 (t, *J* = 6.5 Hz, 2H), 3.52 (t, *J* = 6.4 Hz, 2H), 1.82 – 1.55 (m, 2H), 1.55 – 1.40 (m, 4H). $^{19}$F NMR (471 MHz, Chloroform-*d*) δ -114.48 (d).

**2,6-difluoro-4-(((6-hydroxyhexyl)oxy)methyl)benzaldehyde (4e):** Yield: 3.77 g (40.0%). $^{1}$H NMR (500 MHz, Chloroform-*d*) δ 10.31 (s, 1H), 6.97 (d, *J* = 9.4 Hz, 2H), 4.51 (s, 2H), 3.65 (t, *J* = 6.6 Hz, 2H), 3.51 (t, *J* = 6.5 Hz, 2H), 1.68 – 1.64 (m, 2H), 1.61 – 1.57 (m, 2H), 1.44 – 1.39 (m, 4H). $^{19}$F NMR (471 MHz, Chloroform-*d*) δ -114.51 (d).

**2,6-difluoro-4-(((7-hydroxyheptyl)oxy)methyl)benzaldehyde (4f):** Yield: 3.16 g (33.9%). $^{1}$H NMR (500 MHz, Chloroform-*d*) δ 10.31 (s, 1H), 6.97 (d, *J* = 9.5 Hz, 2H), 4.51 (s, 2H), 3.64 (t, *J* = 6.6 Hz, 2H), 3.50 (t, *J* = 6.5 Hz, 2H), 1.68 – 1.37 (m, 10H). $^{19}$F NMR (471 MHz, Chloroform-*d*) δ -114.52 (d).

**2,6-difluoro-4-(((8-hydroxyoctyl)oxy)methyl)benzaldehyde (4g):** Yield: 2.99 g (31.5%). $^{1}$H NMR (500 MHz, Chloroform-*d*) δ 10.32 (s, 1H), 6.97 (d, *J* = 9.7 Hz, 2H), 4.51 (s, 2H), 3.64 (t, *J* = 6.6 Hz, 2H), 3.50 (t, *J* = 6.5 Hz, 2H), 1.59 (dq, *J* = 35.7, 6.9 Hz, 6H), 1.38 (dd, *J* = 14.2, 6.5 Hz, 6H). $^{19}$F NMR (471 MHz, Chloroform-*d*) δ -114.53 (d).

2,6-Lutidine
DCM, $N_2$, -30°C - RT, overnight

n=2 4a
n=3 4b
n=4 4c
n=5 4d
n=6 4e
n=7 4f
n=8 4g

n=2 5a
n=3 5b
n=4 5c
n=5 5d
n=6 5e
n=7 5f
n=8 5g

**2-((3,5-difluoro-4-formylbenzyl)oxy)ethyl acrylate (5a):** To a dried 100 mL round-bottom flask under nitrogen atmosphere was added compound 4a (5.50 g, 25.43 mmol), followed by anhydrous dichloromethane (DCM, 20 mL). The solution was cooled to -30 °C, and after stirring at this temperature for 30 min, 2,6-lutidune (2.0 equiv.) was added quickly. Then appropriate amount of acryloyl chloride (1.2 equiv.) was added dropwise via syringe, maintaining the internal temperature below -25 °C. After stirring at this temperature for 1 h, the reaction was gradually warmed to ambient temperature. Upon completion (monitored by TLC), the mixture was quenched with saturated sodium chloride solution, and extracted with DCM (3 × 20 mL). The combined organic layers were dried over anhydrous sodium sulfate, filtered, and concentrated under reduced pressure. The crude residue was purified by column chromatography on silica gel (petroleum ether/ethyl acetate = 95:5, v/v) to afford target compound 5a as a pale yellow oil. Yield: 6.12 g (89.0%). $^{1}$H NMR (500 MHz, Chloroform-*d*) δ 10.31 (s, 1H), 6.98 (d, *J* = 9.4 Hz, 2H), 6.45 (dd, *J* = 17.3, 1.3 Hz, 1H), 6.13 (dd, *J* = 17.2, 10.5 Hz, 1H), 5.87 (dd, *J* = 10.4, 1.3 Hz, 1H), 4.59 (s, 2H), 4.39 – 4.36 (m, 2H), 3.78 – 3.75 (m, 2H). $^{19}$F NMR (471 MHz, Chloroform-*d*) δ -114.23 (d).

**3-((3,5-difluoro-4-formylbenzyl)oxy)propyl acrylate (5b):** Yield: 2.30 g (58.5%); $^{1}$H NMR (500 MHz, Chloroform-*d*) δ 10.32 (s, 1H), 6.97 (d, *J* = 9.5 Hz, 2H), 6.41 (dd, *J* = 17.4, 1.4 Hz, 1H), 6.12 (dd, *J* = 17.3, 10.4 Hz, 1H), 5.84 (dd, *J* = 10.4, 1.4 Hz, 1H), 4.53 (s, 2H), 4.31 (t, *J* = 6.3 Hz, 2H), 3.62 (t, *J* = 6.1 Hz, 2H), 2.06 – 1.99 (m, 2H). $^{19}$F NMR (471 MHz, Chloroform-*d*) δ -114.35 (d).

**4-((3,5-difluoro-4-formylbenzyl)oxy)butyl acrylate (5c):** Yield: 2.20 g (55.0%); $^{1}$H NMR (500 MHz, Chloroform-*d*) δ 10.31 (d, *J* = 1.0 Hz, 1H), 6.96 (d, *J* = 9.6 Hz, 2H), 6.39 (dd, *J* = 17.3, 1.5 Hz, 1H), 6.11 (dd, *J* = 17.3, 10.4 Hz, 1H), 5.82 (dd, *J* = 10.5, 1.4 Hz, 1H), 4.51 (s, 2H), 4.19 (t, *J* = 6.4 Hz, 2H), 3.54 (t, *J* = 6.2 Hz, 2H), 1.83 – 1.69 (m, 4H). $^{19}$F NMR (471 MHz, Chloroform-*d*) δ -114.41 (d).

**5-((3,5-difluoro-4-formylbenzyl)oxy)pentyl acrylate (5d):** Yield: 2.77 g (61.2%); $^{1}$H NMR (500 MHz, Chloroform-*d*) δ 10.31 (s, 1H), 6.95 (d, *J* = 9.7 Hz, 2H), 6.39 (dd, *J* = 17.3, 1.4 Hz, 1H), 6.11 (dd, *J* = 17.3, 10.4 Hz, 1H), 5.81 (dd, *J* = 10.4, 1.5 Hz, 1H), 4.50 (s, 2H), 4.17 (t, *J* = 6.6 Hz, 2H), 3.51 (t, *J* = 6.4 Hz, 2H), 1.69 (tt, *J* = 14.5, 6.7 Hz, 4H), 1.48 (tt, *J* = 9.9, 6.2 Hz, 2H). $^{19}$F NMR (471 MHz, Chloroform-*d*) δ -114.45 (d).

**6-((3,5-difluoro-4-formylbenzyl)oxy)hexyl acrylate (5e):** Yield: 2.91 g (66.5%); $^{1}$H NMR (500 MHz, Chloroform-*d*) δ 10.32 (s, 1H), 6.97 (d, *J* = 9.8 Hz, 2H), 6.39 (d, *J* = 17.3 Hz, 1H), 6.11 (dd, *J* = 17.3, 10.7 Hz, 1H), 5.81 (d, *J* = 10.4 Hz, 1H), 4.51 (s, 2H), 4.16 (t, *J* = 6.7 Hz, 2H), 3.51 (t, *J* = 6.4 Hz, 2H), 1.67 (dq, *J* = 20.6, 6.8 Hz, 4H), 1.42 (dq, *J* = 8.4, 4.5 Hz, 4H). $^{19}$F NMR (471 MHz, Chloroform-*d*) δ -114.49 (d).

**7-((3,5-difluoro-4-formylbenzyl)oxy)heptyl acrylate (5f):** Yield: 2.35 g (55.0%); $^{1}$H NMR (500 MHz, Chloroform-*d*) δ 10.32 (s, 1H), 6.97 (d, *J* = 9.6 Hz, 2H), 6.40 (d, *J* = 17.3 Hz, 1H), 6.12 (dd, *J* = 17.3, 10.4 Hz, 1H), 5.82 (d, *J* = 10.4 Hz, 1H), 4.51 (s, 2H), 4.15 (t, *J* = 6.7 Hz, 2H), 3.51 (t, *J* = 6.5 Hz, 2H), 1.66 (dt, *J* = 15.9, 7.1 Hz, 4H), 1.40 (dd, *J* = 12.2, 7.4 Hz, 6H). $^{19}$F NMR (471 MHz, Chloroform-*d*) δ -114.50 (d).

**8-((3,5-difluoro-4-formylbenzyl)oxy)octyl acrylate (5g):** Yield: 2.70 g (62.0%); $^{1}$H NMR (500 MHz, Chloroform-*d*) δ 10.32 (s, 1H), 6.97 (d, *J* = 9.6 Hz, 2H), 6.39 (d, *J* = 17.3 Hz, 1H), 6.12 (dd, *J* = 17.3, 10.5 Hz, 1H), 5.81 (d, *J* = 10.4 Hz, 1H), 4.51 (s, 2H), 4.15 (t, *J* = 6.7 Hz, 2H), 3.50 (t, *J* = 6.5 Hz, 2H), 1.65 (dt, *J* = 14.9, 7.7 Hz, 4H), 1.38 (dd, *J* = 17.0, 8.3 Hz, 8H). $^{19}$F NMR (471 MHz, Chloroform-*d*) δ -114.51 (d).

$NaClO_2$ / $NaH_2PO_4$
2-Methyl-2-butene
THF : *t*-BuOH : $H_2O$ = 4 : 1 : 5
0°C - RT, overnight

n=2 5a → n=2 6a
n=3 5b → n=3 6b
n=4 5c → n=4 6c
n=5 5d → n=5 6d
n=6 5e → n=6 6e
n=7 5f → n=7 6f
n=8 5g → n=8 6g

**4-((2-(acryloyloxy)ethoxy)methyl)-2,6-difluorobenzoic acid (6a):** To a 100 mL round-bottom flask was added compound 5a (6.00 g, 22.21 mmol), followed by a mixed solvent of THF/*tert*-butanol (4:1, v/v, 20mL). The mixture was cooled to 0 °C using an ice-water bath, and then 2-methyl-2-butune (10.0 equiv.) was added. After stirring at this temperature for 20 min, a pre-mixed aqueous solution of sodium dihydrogen phosphate (8.0 equiv.) and sodium chlorite (2.0 equiv.) were added dropwise via a separatory funnel maintaining the ice bath at 0 °C. After stirring at this temperature for 1 h, the reaction was gradually warmed to ambient temperature and stirred for 5 h. Upon completion (monitored by TLC), the mixture was extracted with EA. The combined organic layers were dried over anhydrous sodium sulfate, filtered, and concentrated under reduced pressure. The crude residue was purified by column chromatography on silica gel (petroleum ether/ethyl acetate = 2:1, v/v) to afford target compound 6a. Yield: 6.07 g (95.5%). $^{1}$H NMR (500 MHz, Chloroform-*d*) δ 6.97 (d, *J* = 9.2 Hz, 2H), 6.46 (d, *J* = 17.3 Hz, 1H), 6.18 (dd, *J* = 17.3, 10.5 Hz, 1H), 5.88 (d, *J* = 6.2 Hz, 1H), 4.58 (s,

2H), 4.40 – 4.34 (m, 2H), 3.79 – 3.74 (m, 2H). $^{19}$F NMR (471 MHz, Chloroform-*d*) δ -108.15 (d).

**4-((3-(acryloyloxy)propoxy)methyl)-2,6-difluorobenzoic acid (6b):** Yield: 2.01 g (96.6%). $^{1}$H NMR (500 MHz, DMSO-*d*6) δ 13.80 (s, 1H), 7.09 (d, *J* = 9.0 Hz, 2H), 6.27 (dd, *J* = 17.3, 1.5 Hz, 1H), 6.12 (dd, *J* = 17.3, 10.3 Hz, 1H), 5.89 (dd, *J* = 10.4, 1.5 Hz, 1H), 4.48 (s, 2H), 4.17 (t, *J* = 6.4 Hz, 2H), 3.49 (t, *J* = 6.2 Hz, 2H), 1.88 (p, *J* = 6.3 Hz, 2H).$^{19}$F NMR (471 MHz, DMSO-$d_6$) δ -111.88 (d, *J* = 10.6 Hz).

**4-((4-(acryloyloxy)butoxy)methyl)-2,6-difluorobenzoic acid (6c):** Yield: 2.22 g (91.5%). $^{1}$H NMR (500 MHz, Chloroform-*d*) δ 6.99 – 6.94 (m, 2H), 6.41 (dd, *J* = 17.3, 1.4 Hz, 1H), 6.12 (dd, *J* = 17.3, 10.4 Hz, 1H), 5.83 (dd, *J* = 10.4, 1.5 Hz, 1H), 4.51 (s, 2H), 4.21 (t, *J* = 6.4 Hz, 2H), 3.54 (t, *J* = 6.2 Hz, 2H), 1.83 – 1.69 (m, 4H). $^{19}$F NMR (471 MHz, Chloroform-*d*) δ -107.97 (d, *J* = 9.4 Hz).

**4-(((5-(acryloyloxy)pentyl)oxy)methyl)-2,6-difluorobenzoic acid (6d):** Yield: 2.11 g (95.6%). $^{1}$H NMR (500 MHz, DMSO-$d_6$) δ 13.83 (s, 1H), 7.12 (d, 2H), 6.31 (dd, 1H), 6.16 (ddd, 1H), 5.92 (dd, 1H), 4.50 (s, 2H), 4.11 (t, 2H), 3.59 (td, 2H), 3.45 (t, 2H), 1.76 (td, 2H), 1.61 (dp, 4H), 1.40 (qd, 2H). $^{19}$F NMR (471 MHz, DMSO-$d_6$) δ -111.88 (d).

**4-(((6-(acryloyloxy)hexyl)oxy)methyl)-2,6-difluorobenzoic acid (6e):** Yield: 2.48 g (93.0%). $^{1}$H NMR (500 MHz, Chloroform-*d*) δ 6.94 (d, *J* = 9.2 Hz, 2H), 6.39 (dd, *J* = 17.3, 1.3 Hz, 1H), 6.12 (dd, *J* = 17.3, 10.4 Hz, 1H), 5.81 (dd, *J* = 10.4, 1.3 Hz, 1H), 4.49 (s, 2H), 4.20 – 4.13 (m, 2H), 3.49 (t, *J* = 6.5 Hz, 2H), 1.67 (dp, *J* = 20.1, 6.4, 6.0 Hz, 4H), 1.46 – 1.38 (m, 4H).$^{19}$F NMR (471 MHz, Chloroform-*d*) δ -108.57 (d, *J* = 10.8 Hz).

**4-(((7-(acryloyloxy)heptyl)oxy)methyl)-2,6-difluorobenzoic acid (6f):** Yield: 2.44 g (94.2%). $^{1}$H NMR (500 MHz, Chloroform-*d*) δ 6.97 (d, *J* = 9.1 Hz, 2H), 6.40 (dd, *J* = 17.3, 1.4 Hz, 1H), 6.12 (dd, *J* = 17.3, 10.4 Hz, 1H), 5.82 (dd, *J* = 10.4, 1.4 Hz, 1H), 4.50 (s, 2H), 4.16 (t, *J* = 6.7 Hz, 2H), 3.49 (t, *J* = 6.5 Hz, 2H), 1.66 (dt, *J* = 19.2, 6.9 Hz, 4H), 1.39 (dd, *J* = 9.7, 4.0 Hz, 6H). $^{19}$F NMR (471 MHz, Chloroform-*d*) δ -108.01 (d, *J* = 10.9 Hz).

**4-(((8-(acryloyloxy)octyl)oxy)methyl)-2,6-difluorobenzoic acid (6g):** Yield: 2.55 g (95.4%). $^{1}$H NMR (500 MHz, Chloroform-*d*) δ 6.94 (d, *J* = 9.1 Hz, 2H), 6.38 (dd, *J* = 17.4, 1.7 Hz, 1H), 6.11 (s, 1H), 5.86 – 5.73 (m, 1H), 4.48 (s, 2H), 4.18 – 4.10 (m, 2H), 3.47 (t, *J* = 6.8 Hz, 2H), 1.74 – 1.13 (m, 12H). $^{19}$F NMR (471 MHz, Chloroform-*d*) δ -108.48 (d, *J* = 10.7 Hz).

HO–C6H4–COOH → (*p*-TsOH / DHP; Ether, $N_2$, 0°C - RT, 4h) → THPO–C6H4–COOH (7)

**4-((tetrahydro-2H-pyran-2-yl)oxy)benzoic acid (7):** The synthetic procedure was reported in previous our works.[3]

EDCI / DMAP
DCM, 0°C - RT, overnight

7 → 8

**4-cyano-3,5-difluorophenyl 4-((tetrahydro-2H-pyran-2-yl)oxy)benzoate (8):** To a dried 500 mL round-bottom flask under nitrogen atmosphere was added compound 7 (50.00 g, 224.98 mmol), 2,6-difluoro-4-hydroxybenzonitrile (31.72 g, 204.53 mmol), and DMAP (2.50 g, 20.45 mmol) followed by anhydrous dichloromethane (DCM, 20 mL). The solution was cooled to 0 °C, and EDCl (58.81 g, 306.79 mmol) was added portion wise. The reaction was gradually warmed to ambient temperature and monitored by TLC. Upon completion, the mixture was quenched with saturated sodium chloride solution, and extracted with DCM (3 × 20 mL). The combined organic layers were dried over anhydrous sodium sulfate, filtered, and concentrated under reduced pressure. The crude residue was purified by column chromatography on silica gel (petroleum ether/DCM = 2:1, v/v) to afford target compound 8. Yield: 55.12 g (75.0%). $^{1}$H NMR (500 MHz, Chloroform-*d*) δ 8.09 (d, *J* = 9.0 Hz, 2H), 7.15 (d, *J* = 9.0 Hz, 2H), 7.04 (d, *J* = 8.0 Hz, 2H), 5.56 (s, 16H), 3.89 – 3.80 (m, 1H), 3.64 (dtd, *J* = 11.3, 4.0, 1.4 Hz, 1H), 2.08 – 1.96 (m, 1H), 1.91 (dq, *J* = 7.6, 3.3 Hz, 2H), 1.75 – 1.61 (m, 3H). $^{19}$F NMR (471 MHz, Chloroform-*d*) δ -102.13.

PPTS
MeOH:THF=1:1 $N_2$, Reflux, 6h

8 → 9

**4-cyano-3,5-difluorophenyl 4-hydroxybenzoate (9):** A round bottom flask was charged with compound 8 (55.00 g, 153.08 mmol) and PPTS (19.23 g, 76.54 mmol), followed by a mixed solvent of THF/methol = 1:1 (v/v, 100 mL). The solution was heated at 60 °C and stirred for 6 h. Upon completion, the mixture was quenched with saturated sodium chloride solution, and extracted with EA (3 × 100 mL). The combined organic layers were dried over anhydrous sodium sulfate, filtered, and concentrated under reduced pressure. The crude residue was recrystallization (petroleum ether/THF) and dried in the vacuum oven. Yield: 32.86 g (78.0%). $^{1}$H NMR (500 MHz, DMSO-$d_6$) δ 10.01 (s, 1H), 7.70 (dd, *J* = 9.5, 5.3 Hz, 4H), 6.84 (d, *J* = 8.7 Hz, 2H). $^{19}$F NMR (471 MHz, Chloroform-*d*) δ -102.11 (d).

9
EDCI / DMAP
DCM, 0°C - RT, TLC

n=2 6a → n=2 RFM-2
n=3 6b → n=3 RFM-3
n=4 6c → n=4 RFM-4
n=5 6d → n=5 RFM-5
n=6 6e → n=6 RFM-6
n=7 6f → n=7 RFM-7
n=8 6g → n=8 RFM-8

**4-((4-cyano-3,5-difluorophenoxy)carbonyl)phenyl 4-((2-(acryloyloxy)ethoxy)methyl)-2,6-difluorobenzoate (RFM-2):** To a dried 500 mL round-bottom flask under nitrogen atmosphere was added compound 6a (6.00 g, 20.96 mmol), compound 9 (5.49 g, 19.96 mmol), and DMAP (0.12 g, 1.00 mmol) followed by anhydrous dichloromethane (DCM, 20 mL). The solution was cooled to 0 °C, and EDCl (5.74 g, 29.94 mmol) was added portion wise. After the solution becomes clear (around 10 min), the reaction was monitored by TLC. The reaction shall be terminated immediately when transesterification byproducts were detected. Upon completion, the mixture was quickly quenched with saturated sodium chloride solution, and extracted with DCM (3 × 20 mL). The combined organic layers were dried over anhydrous sodium sulfate, filtered, and concentrated under reduced pressure. The crude residue was purified by column chromatography on silica gel (petroleum ether/DCM = 3:1, v/v) to afford target compound RFM-2. The obtained product was sequentially recrystallized from isopropanol and acetonitrile/water, followed by drying in the vacuum oven at 40 °C. Yield: 4.79 g (44.2%). $^{1}$H NMR (500 MHz, Chloroform-*d*) δ 8.23 (d, $J$ = 8.6 Hz, 2H), 7.44 (d, $J$ = 8.6 Hz, 2H), 7.08 (d, $J$ = 8.1 Hz, 2H), 7.05 (d, $J$ = 9.4 Hz, 2H), 6.46 (dd, $J$ = 17.3, 1.3 Hz, 1H), 6.18 (dd, $J$ = 17.3, 10.4 Hz, 1H), 5.88 (dd, $J$ = 10.4, 1.3 Hz, 1H), 4.62 (s, 2H), 4.42 – 4.37 (m, 2H), 3.82 – 3.77 (m, 2H). $^{19}$F NMR (471 MHz, Chloroform-*d*) δ -101.79 (d, $J$ = 8.8 Hz), -108.12 (d, $J$ = 11.5 Hz). $^{13}$C NMR (126 MHz, Chloroform-*d*) δ 166.16, 164.76 – 162.55 (m), 162.65, 161.38 (dd, $J$ = 259.1, 6.0 Hz), 159.15 (d, $J$ = 2.5 Hz), 155.97 (t, $J$ = 13.6 Hz), 155.27, 146.77 (t, $J$ = 9.9 Hz), 132.30, 131.52, 128.14, 125.81, 122.41, 110.70 – 110.40 (m), 108.84, 108.31 (t, $J$ = 16.8 Hz), 107.09 (dd, $J$ = 23.1, 4.2 Hz), 90.28 (t, $J$ = 19.6 Hz), 71.31 (t, $J$ = 2.2 Hz), 68.94, 63.43.

**4-((4-cyano-3,5-difluorophenoxy)carbonyl)phenyl 4-((3-(acryloyloxy)propoxy)methyl)-2,6-difluorobenzoate (RFM-3):** Yield: 2.10 g (40.2%). $^{1}$H NMR (500 MHz, Chloroform-*d*) δ 8.24 (d, $J$ = 8.8 Hz, 2H), 7.44 (d, $J$ = 8.8 Hz, 2H), 7.08 (d, $J$ = 8.1 Hz, 2H), 7.03 (d, $J$ = 9.6 Hz, 2H), 6.41 (dt, $J$ = 17.2, 1.1 Hz, 1H), 6.13 (dd, $J$ = 17.3, 10.4 Hz, 1H), 5.84 (dt, $J$ = 10.4, 1.1 Hz, 1H), 4.55 (s, 2H), 4.32 (t, $J$ = 6.4 Hz, 2H), 3.63 (t, $J$ = 6.1 Hz, 2H), 2.04 (p, $J$ = 6.2 Hz, 2H). $^{19}$F NMR (471 MHz, Chloroform-*d*) δ -101.76 (d, $J$ = 11.2 Hz), -108.24 (d, $J$ = 11.2 Hz). $^{13}$C NMR (126 MHz, Chloroform-*d*) δ 166.27, 163.62 (dd, $J$ = 261.5, 6.7 Hz), 162.65, 161.34 (dd, $J$ = 259.0, 6.0 Hz), 159.17 (t, $J$ = 2.4 Hz), 155.95 (t, $J$ = 13.6 Hz), 155.28, 147.10 (t, $J$ = 9.8 Hz), 132.28, 130.98, 128.41, 125.78, 122.41, 110.62 – 110.29 (m), 108.82, 108.17 (t, $J$ = 16.7 Hz), 107.07 (dd, $J$ = 23.0, 4.3 Hz), 90.28 (t, $J$ = 19.6 Hz), 71.24 (d, $J$ = 2.1 Hz), 67.59, 61.55, 29.07.

**4-((4-cyano-3,5-difluorophenoxy)carbonyl)phenyl 4-((4-(acryloyloxy)butoxy)methyl)-2,6-difluorobenzoate (RFM-4):** Yield: 2.01 g (40.8%); $^{1}$H NMR (500 MHz, Chloroform-*d*) δ 8.22 (d, $J$ = 8.8 Hz, 2H), 7.43 (d, $J$ = 8.8 Hz, 2H), 7.06 (d, $J$ = 8.0 Hz, 2H), 7.02 (d, $J$ = 9.6 Hz, 2H), 6.39 (dd, $J$ = 17.3, 1.5 Hz, 1H), 6.11 (dd, $J$ = 17.3, 10.4 Hz, 1H), 5.82 (dd, $J$ = 10.4, 1.5 Hz, 1H), 4.53 (s, 2H), 4.20 (t, $J$ = 6.3 Hz, 2H), 3.56 (t, $J$ = 6.2 Hz, 2H), 1.85 – 1.70 (m, 4H). $^{19}$F NMR (471 MHz, Chloroform-*d*) δ -101.77 (d, $J$ = 11.0 Hz), -108.27 (d, $J$ = 11.9 Hz). $^{13}$C NMR (126 MHz, Chloroform-*d*) δ 166.35, 164.66 (d, $J$ = 6.7 Hz), 162.65, 162.58 (d, $J$ = 6.7 Hz), 162.38

(d, $J$ = 6.0 Hz), 160.32 (d, $J$ = 6.0 Hz), 159.18 (t, $J$ = 2.3 Hz), 155.95 (t, $J$ = 13.7 Hz), 155.29, 147.31 (t, $J$ = 10.0 Hz), 132.27, 130.82, 128.54, 125.77, 122.41, 110.74 – 110.32 (m), 108.82, 108.12 (t, $J$ = 16.7 Hz), 107.07 (dd, $J$ = 23.3, 4.4 Hz), 90.28 (t, $J$ = 19.5 Hz), 71.16 (d, $J$ = 2.3 Hz), 70.67, 64.27, 26.28, 25.58.

**4-((4-cyano-3,5-difluorophenoxy)carbonyl)phenyl 4-(((5-(acryloyloxy)pentyl)oxy)methyl)-2,6-difluorobenzoate (RFM-5):** Yield: 1.96 g (42.8%); $^{1}$H NMR (500 MHz, Chloroform-*d*) δ 8.23 (d, $J$ = 8.7 Hz, 2H), 7.44 (d, $J$ = 8.7 Hz, 2H), 7.08 (d, $J$ = 8.1 Hz, 2H), 7.03 (d, $J$ = 9.6 Hz, 2H), 6.40 (dd, $J$ = 17.3, 1.3 Hz, 1H), 6.12 (dd, $J$ = 17.4, 10.4 Hz, 1H), 5.82 (dd, $J$ = 10.5, 1.4 Hz, 1H), 4.54 (s, 2H), 4.18 (t, $J$ = 6.6 Hz, 2H), 3.54 (t, $J$ = 6.4 Hz, 2H), 1.78 – 1.65 (m, 4H), 1.57 – 1.45 (m, 2H). $^{19}$F NMR (471 MHz, Chloroform-*d*) δ -101.79 (d, $J$ = 9.2 Hz), -108.32 (d, $J$ = 11.1 Hz). $^{13}$C NMR (126 MHz, Chloroform-*d*) δ 166.41, 163.64 (dd, $J$ = 261.6, 6.6 Hz), 162.67, 161.36 (dd, $J$ = 259.0, 6.0 Hz), 159.35 – 158.98 (m), 155.98 (t, $J$ = 13.6 Hz), 155.31, 147.49 (t, $J$ = 9.9 Hz), 132.29, 130.74, 128.62, 125.79, 122.43, 110.90 – 110.28 (m), 108.84, 108.10 (t, $J$ = 16.7 Hz), 107.09 (dd, $J$ = 23.1, 4.4 Hz), 90.29 (t, $J$ = 19.6 Hz), 71.17 (d, $J$ = 2.9 Hz), 71.06, 64.49, 29.37, 28.54, 22.78.

**4-((4-cyano-3,5-difluorophenoxy)carbonyl)phenyl 4-(((6-(acryloyloxy)hexyl)oxy)methyl)-2,6-difluorobenzoate (RFM-6):** Yield: 1.89 g (38.2%); $^{1}$H NMR (500 MHz, Chloroform-*d*) δ 8.23 (d, $J$ = 8.8 Hz, 2H), 7.44 (d, $J$ = 8.7 Hz, 1H), 7.07 (d, $J$ = 8.1 Hz, 1H), 7.03 (d, $J$ = 9.6 Hz, 2H), 6.39 (dd, $J$ = 17.4, 1.4 Hz, 1H), 6.11 (dd, $J$ = 17.4, 10.5 Hz, 1H), 5.81 (dd, $J$ = 10.4, 1.4 Hz, 1H), 4.54 (s, 2H), 4.16 (t, $J$ = 6.7 Hz, 2H), 3.53 (t, $J$ = 6.5 Hz, 2H), 1.78 – 1.62 (m, 4H), 1.55 – 1.36 (m, 4H). $^{19}$F NMR (471 MHz, Chloroform-*d*) δ -101.79 (d, $J$ = 11.0 Hz), -108.34 (d, $J$ = 11.3 Hz). $^{13}$C NMR (126 MHz, Chloroform-*d*) δ 166.43, 163.64 (dd, $J$ = 261.8, 6.7 Hz), 162.67, 161.37 (dd, $J$ = 259.0, 6.0 Hz), 159.22 (t, $J$ = 2.2 Hz), 155.98 (t, $J$ = 13.7 Hz), 155.32, 147.56 (t, $J$ = 10.0 Hz), 132.29, 130.67, 128.67, 125.79, 122.43, 110.78 – 110.18 (m), 108.84, 108.08 (t, $J$ = 16.7 Hz), 107.09 (dd, $J$ = 23.1, 4.3 Hz), 90.29 (t, $J$ = 19.6 Hz), 71.22, 71.14 (d, $J$ = 2.3 Hz), 64.61, 29.65, 28.67, 25.94, 25.90.

**4-((4-cyano-3,5-difluorophenoxy)carbonyl)phenyl 4-(((7-(acryloyloxy)heptyl)oxy)methyl)-2,6-difluorobenzoate (RFM-7):** Yield: 1.69 g (41.9%); $^{1}$H NMR (500 MHz, Chloroform-*d*) δ 8.26 – 8.18 (m, 2H), 7.48 – 7.41 (m, 2H), 7.08 (d, $J$ = 8.1 Hz, 2H), 7.03 (d, $J$ = 9.7 Hz, 2H), 6.39 (dd, $J$ = 17.3, 1.5 Hz, 1H), 6.11 (dd, $J$ = 17.3, 10.4 Hz, 1H), 5.81 (dd, $J$ = 10.3, 1.5 Hz, 1H), 4.54 (s, 2H), 4.15 (t, $J$ = 6.7 Hz, 2H), 3.52 (t, $J$ = 6.5 Hz, 2H), 1.67 (tq, $J$ = 12.7, 6.6 Hz, 4H), 1.46 – 1.36 (m, 6H). $^{19}$F NMR (471 MHz, Chloroform-*d*) δ -101.79 (d, $J$ = 10.9 Hz), -108.36 (d, $J$ = 11.2 Hz). $^{13}$C NMR (126 MHz, Chloroform-*d*) δ 166.44, 163.65 (dd, $J$ = 261.6, 6.7 Hz), 162.67, 161.37 (dd, $J$ = 258.9, 5.9 Hz), 159.22 (d, $J$ = 2.3 Hz), 155.98 (t, $J$ = 13.6 Hz), 155.32, 147.62 (t, $J$ = 9.9 Hz), 132.30, 130.63, 128.70, 125.79, 122.44, 110.76 – 110.15 (m), 108.84, 108.07 (t, $J$ = 16.7 Hz), 107.09 (dd, $J$ = 23.0, 4.3 Hz), 90.30 (t, $J$ = 19.6 Hz), 71.33, 71.15 (d, $J$ = 3.0 Hz), 64.69, 29.67, 29.15, 28.65, 26.15, 25.97.

**4-((4-cyano-3,5-difluorophenoxy)carbonyl)phenyl 4-(((8-(acryloyloxy)octyl)oxy)methyl)-2,6-difluorobenzoate (RFM-8):** Yield: 1.58 g

(40.9%); $^1$H NMR (500 MHz, Chloroform-*d*) δ 8.25 – 8.20 (m, 2H), 7.45 – 7.41 (m, 2H), 7.06 (d, *J* = 8.1 Hz, 2H), 7.02 (d, *J* = 9.6 Hz, 2H), 6.37 (dd, *J* = 17.3, 1.5 Hz, 1H), 6.10 (dd, *J* = 17.3, 10.4 Hz, 1H), 5.79 (dd, *J* = 10.4, 1.4 Hz, 1H), 4.52 (s, 2H), 4.13 (t, *J* = 6.7 Hz, 2H), 3.51 (t, *J* = 6.5 Hz, 2H), 1.70 – 1.60 (m, 4H), 1.42 – 1.32 (m, 8H). $^{19}$F NMR (471 MHz, Chloroform-*d*) δ -101.79 (d, *J* = 11.0 Hz), -108.37 (d, *J* = 11.5 Hz). $^{13}$C NMR (126 MHz, Chloroform-*d*) δ 166.46, 163.64 (dd, *J* = 261.5, 6.7 Hz), 162.67, 161.37 (dd, *J* = 258.9, 6.0 Hz), 159.23 (t, *J* = 2.5 Hz), 155.98 (t, *J* = 13.7 Hz), 155.32, 147.65 (t, *J* = 9.9 Hz), 132.29, 130.61, 128.71, 125.78, 122.43, 110.77 – 110.15 (m), 108.84, 108.05 (t, *J* = 16.7 Hz), 107.09 (dd, *J* = 22.9, 4.3 Hz), 90.29 (t, *J* = 19.6 Hz), 71.38, 71.13 (t, *J* = 2.4 Hz), 64.75, 29.72, 29.41, 29.27, 28.69, 26.18, 25.96.

## 3.2 Spectra of RFMs

$^1$H NMR of RFM-2

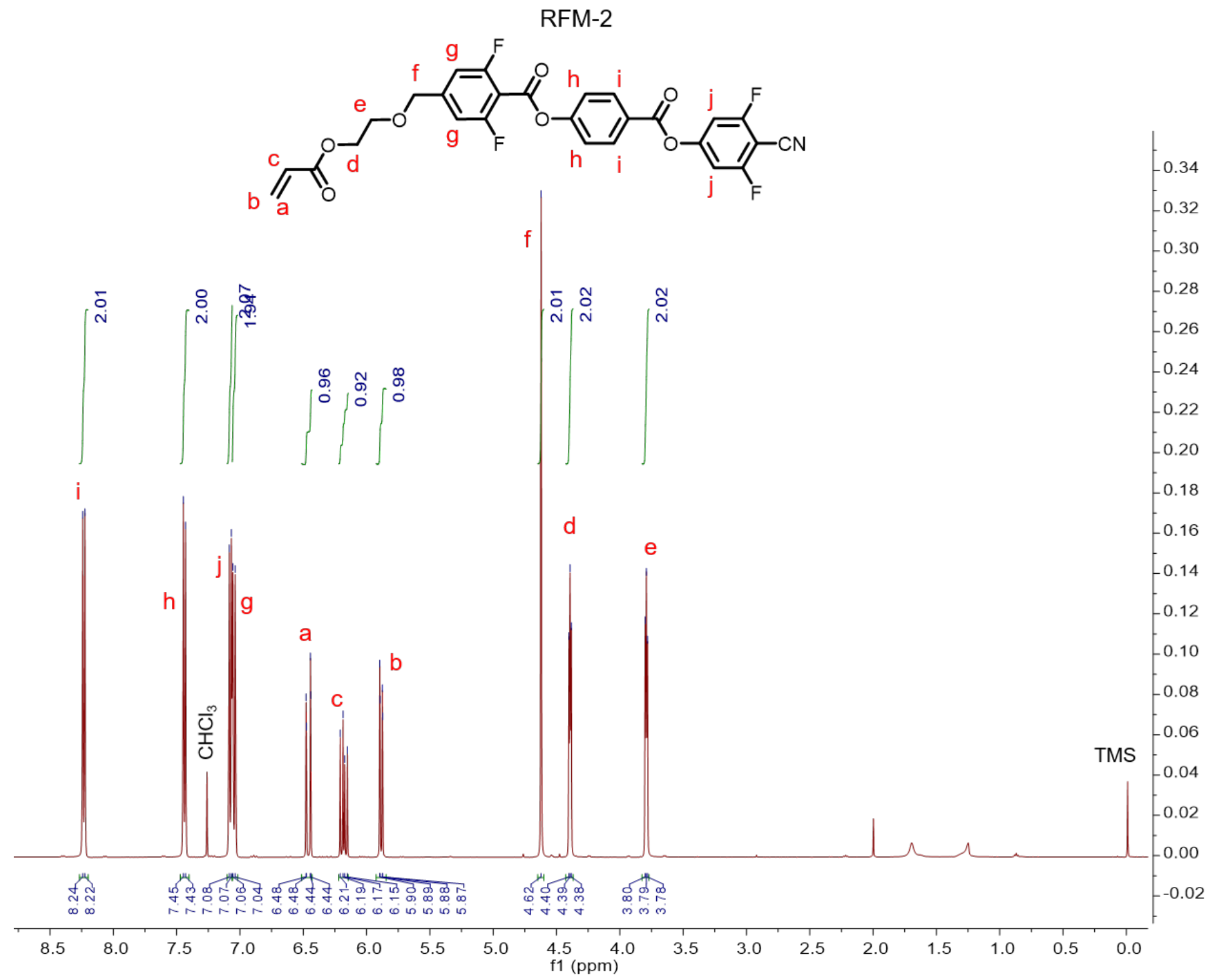

[19]F NMR of RFM-2

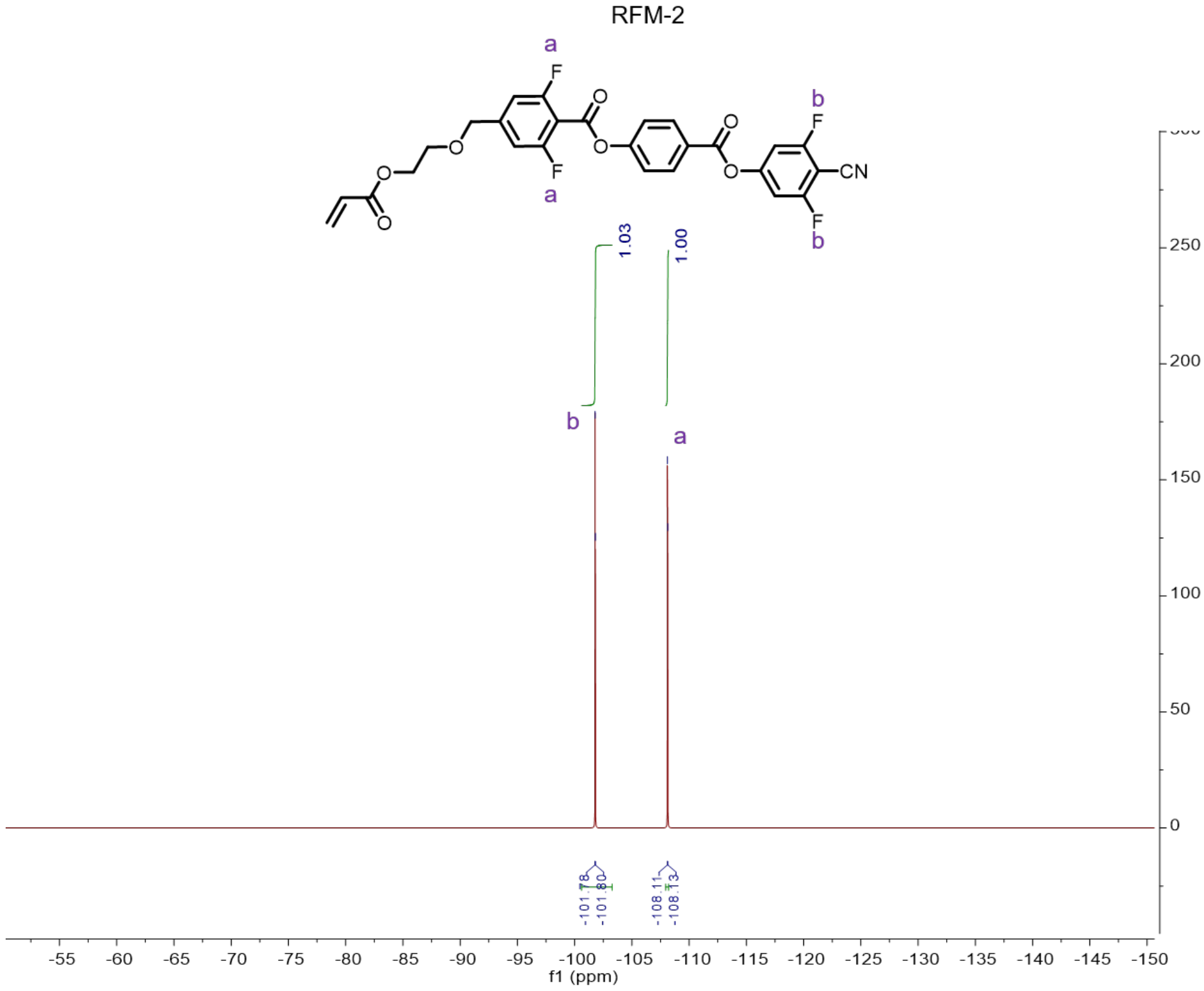

[13]C NMR of RFM-2

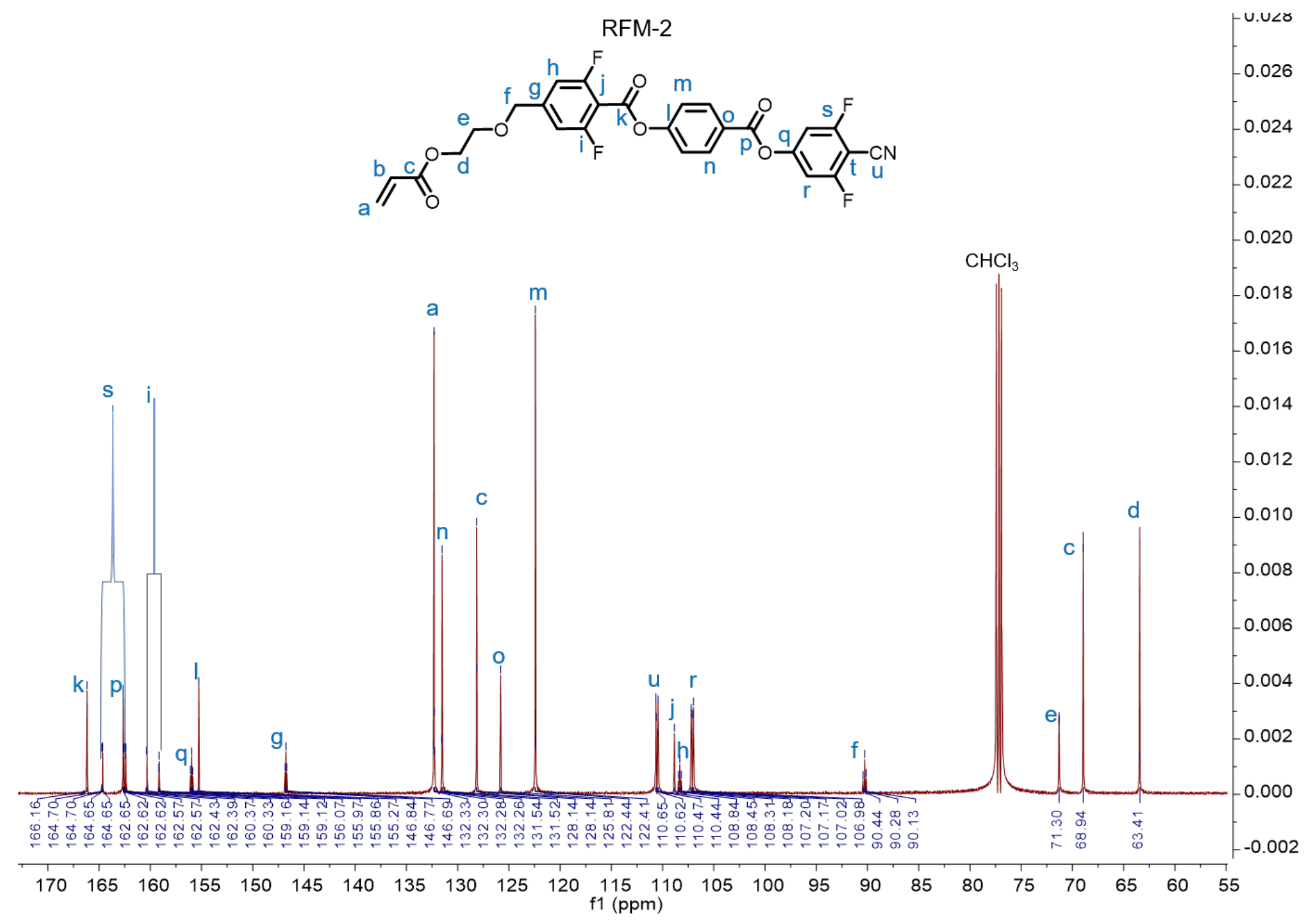

[1]H NMR of RFM-3

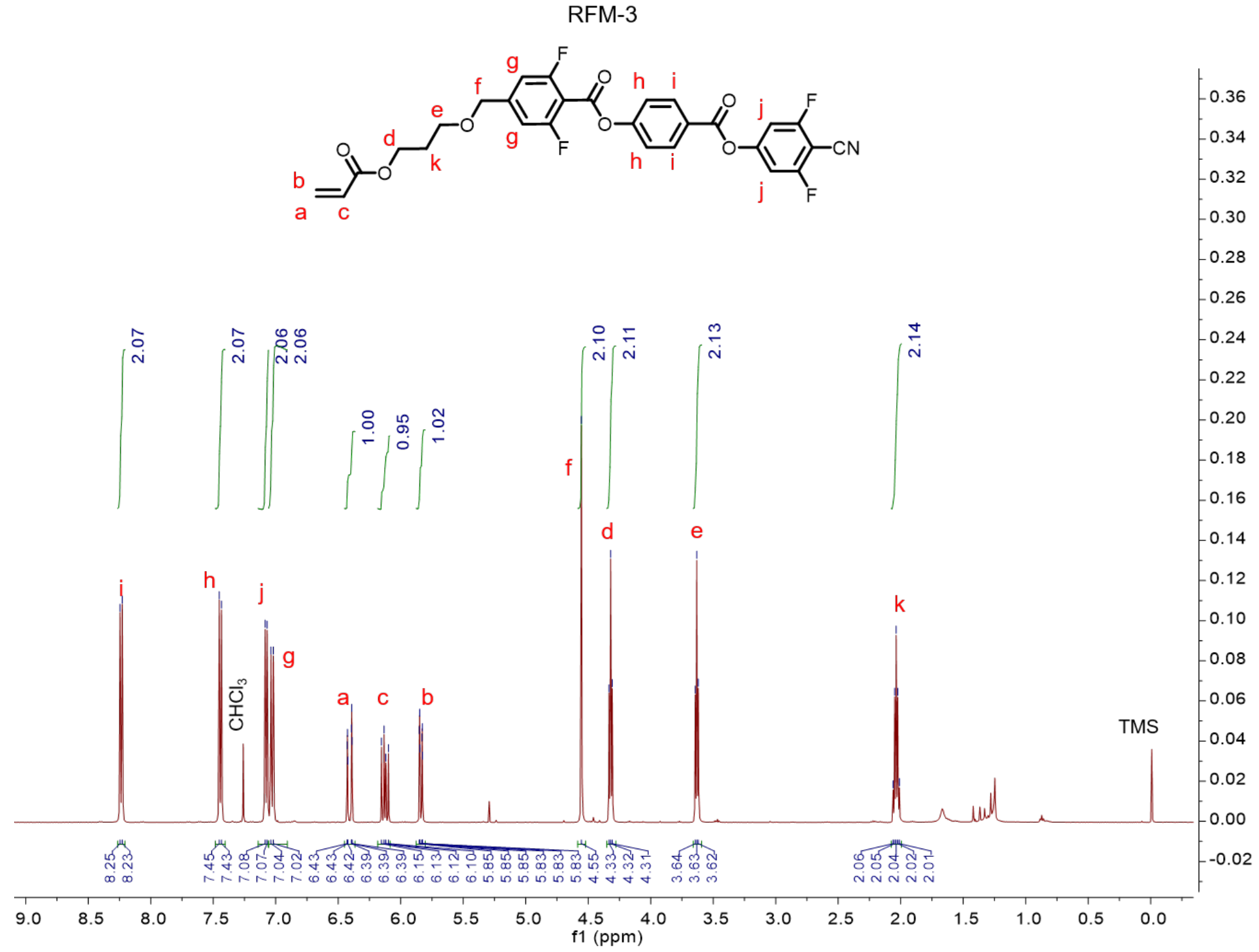

[19]F NMR of RFM-3

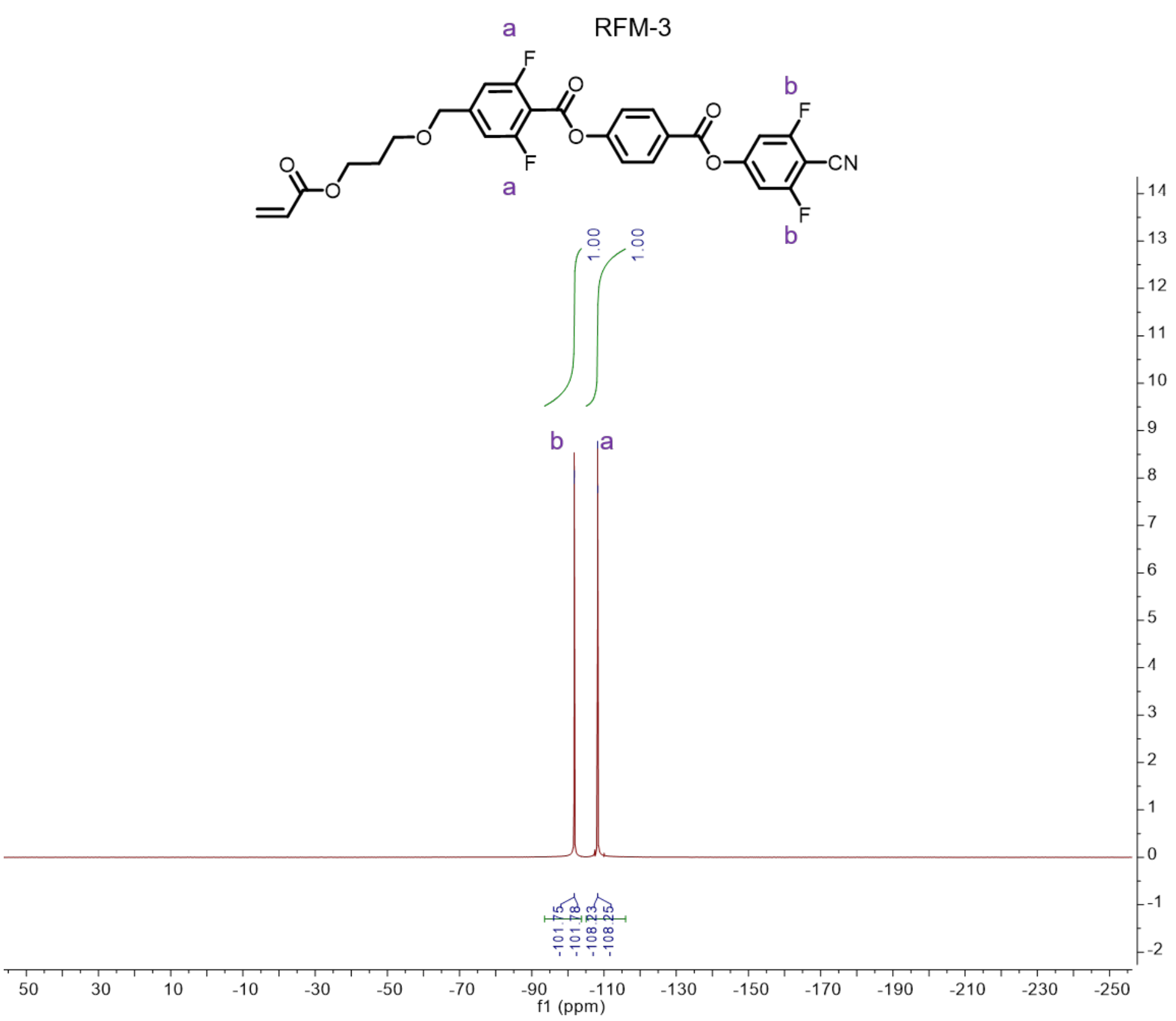

$^{13}C$ NMR of RFM-3

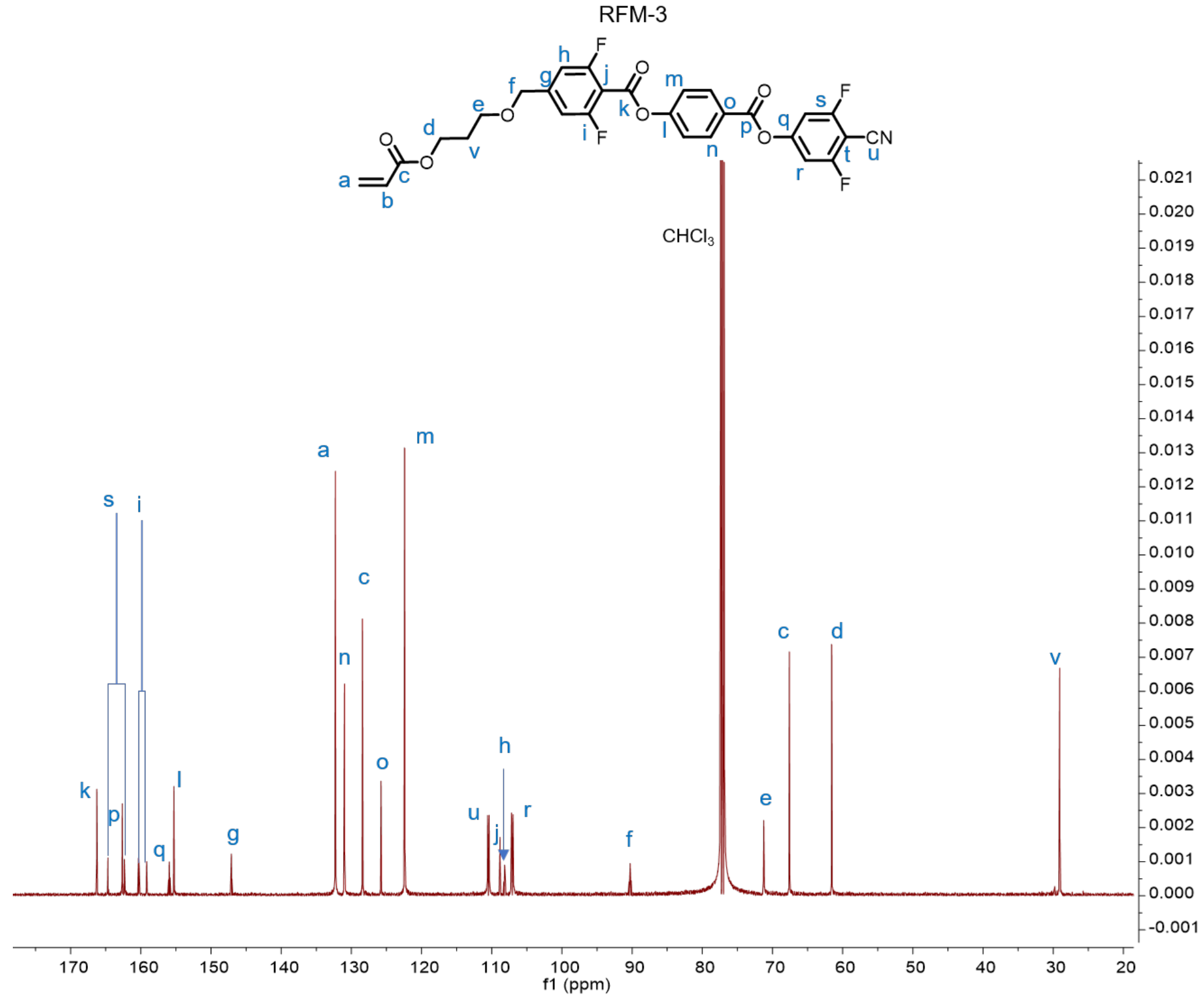

$^{1}H$ NMR of RFM-4

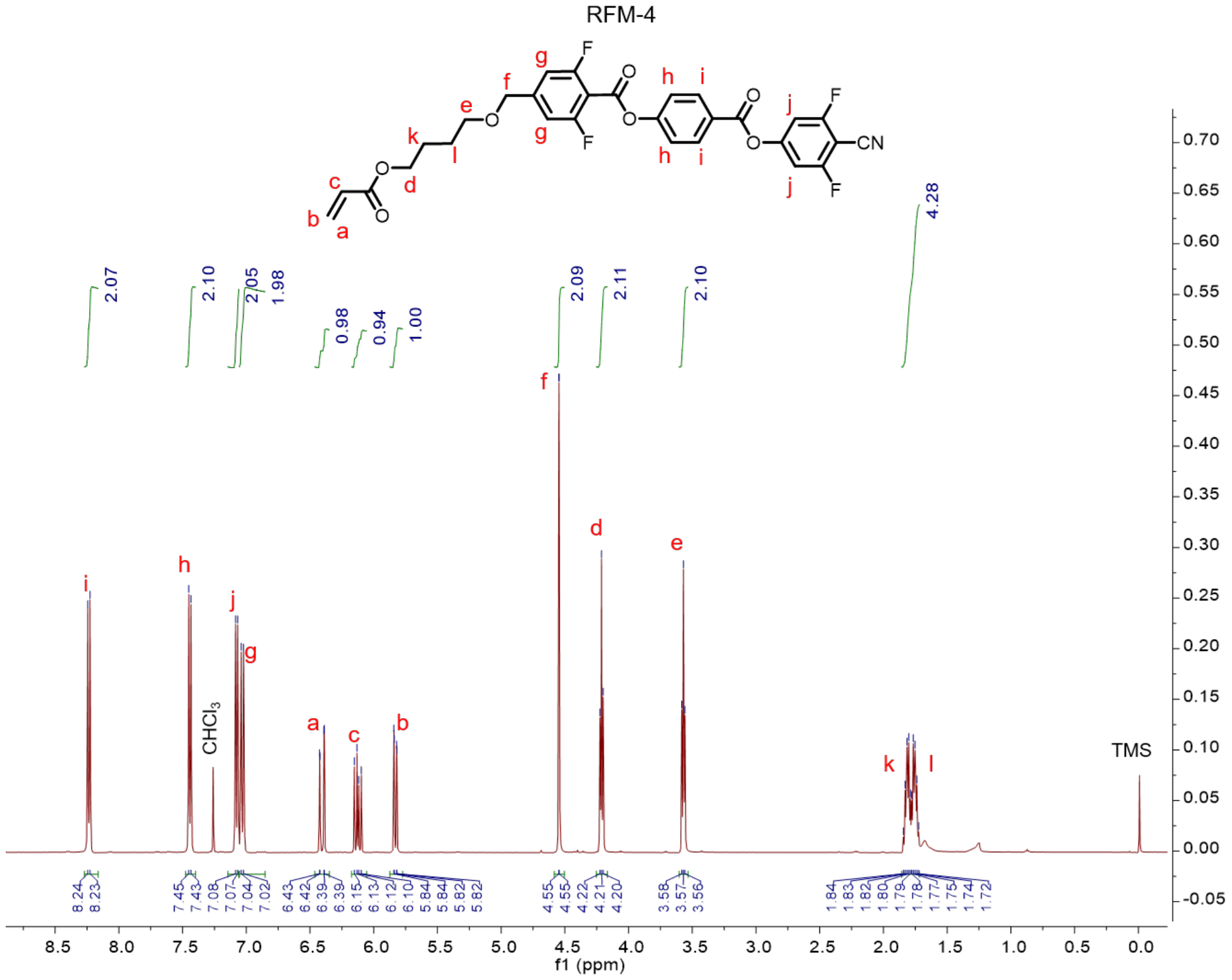

$^{19}F$ NMR of RFM-4

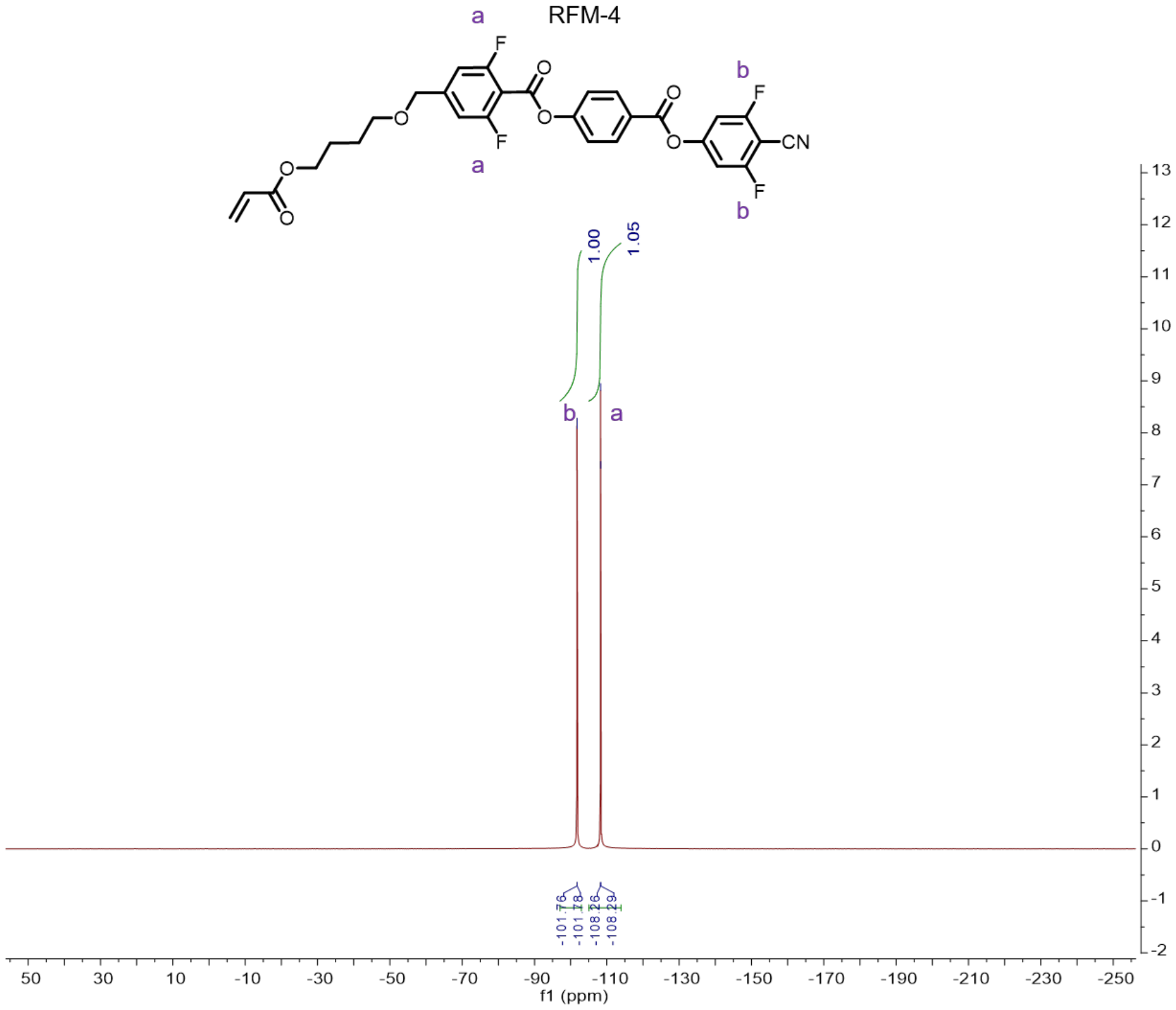

$^{13}C$ NMR of RFM-4

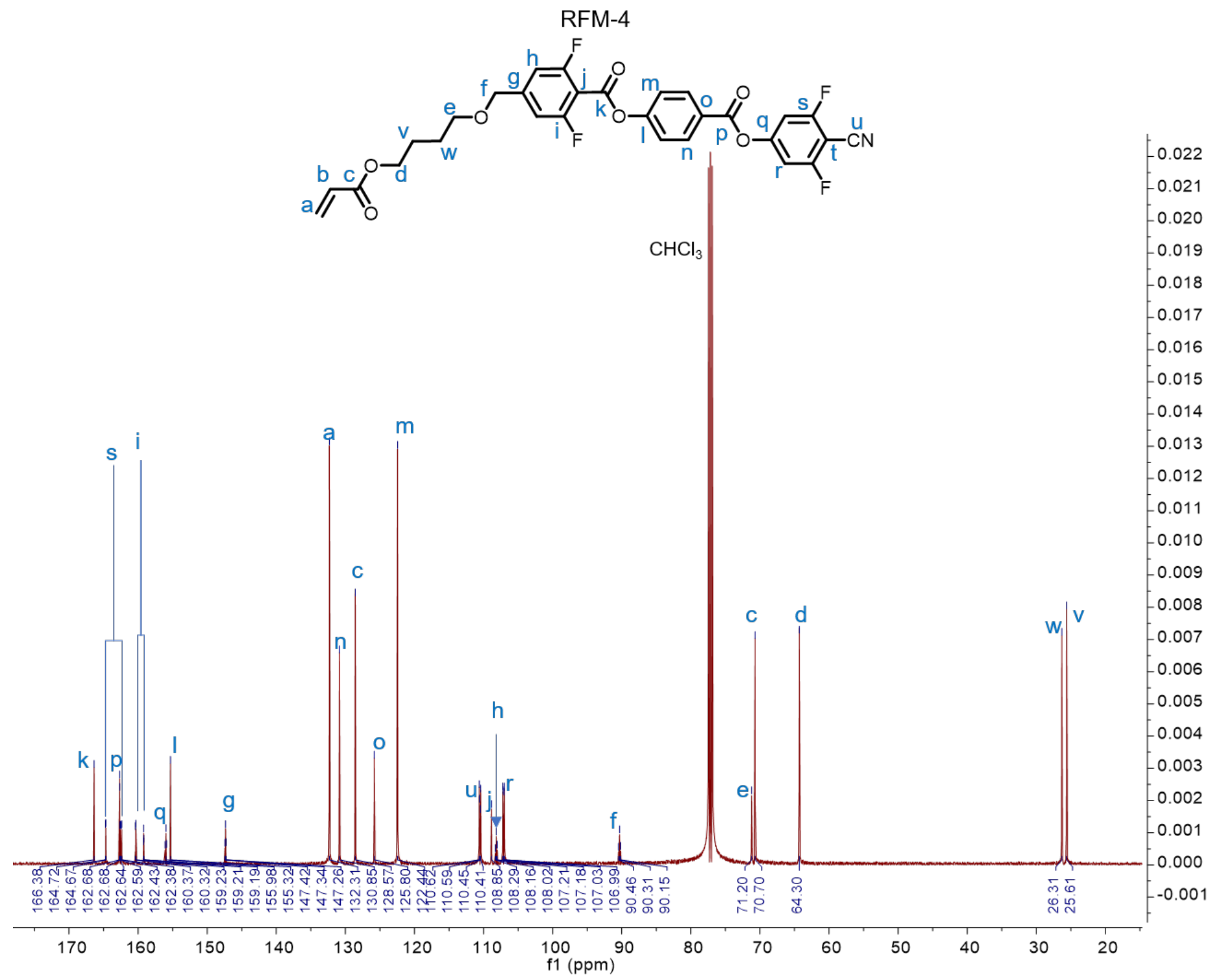

$^1$H NMR of RFM-5

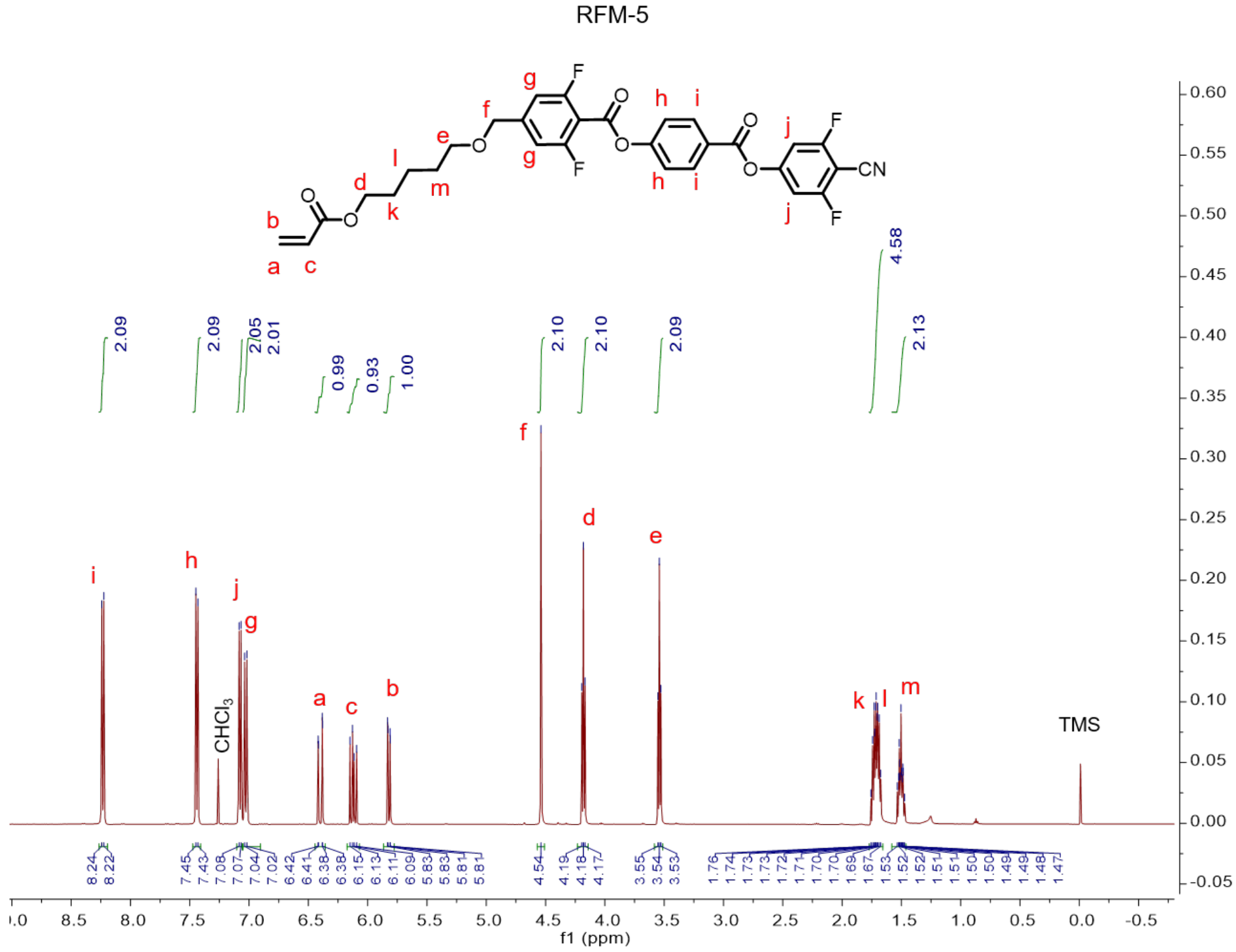

$^{19}$F NMR of RFM-5

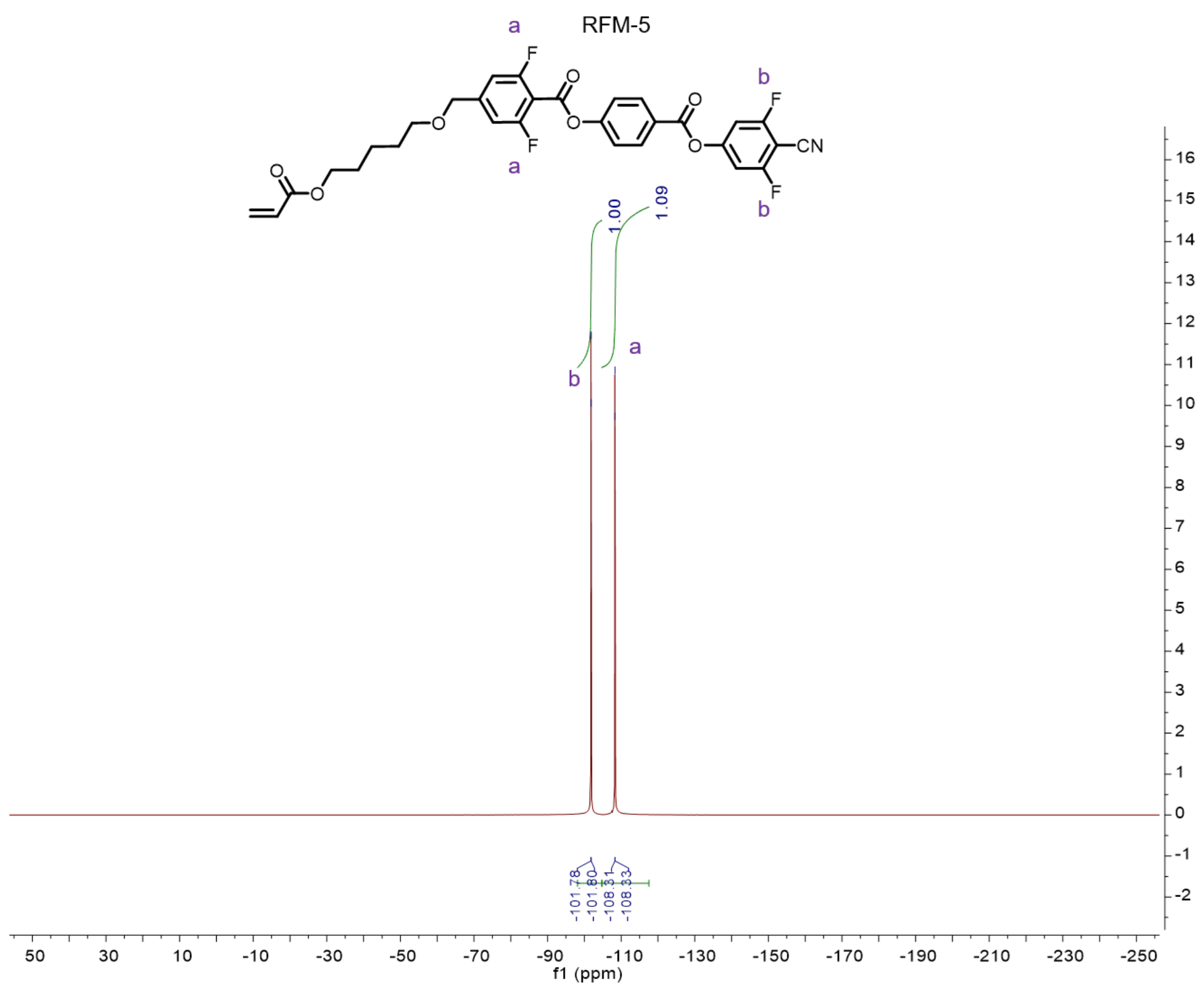

$^{13}C$ NMR of RFM-5

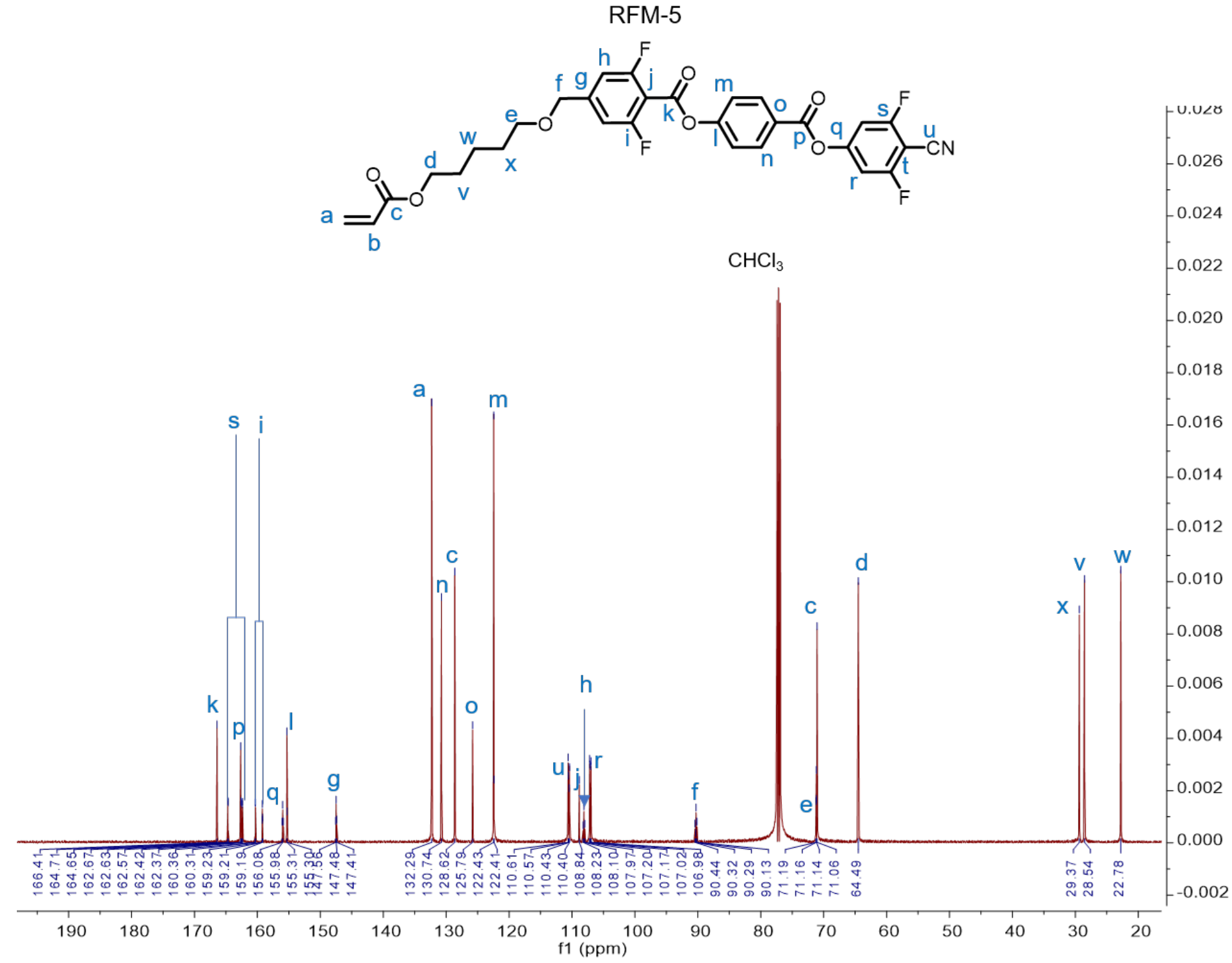

$^{1}H$ NMR of RFM-6

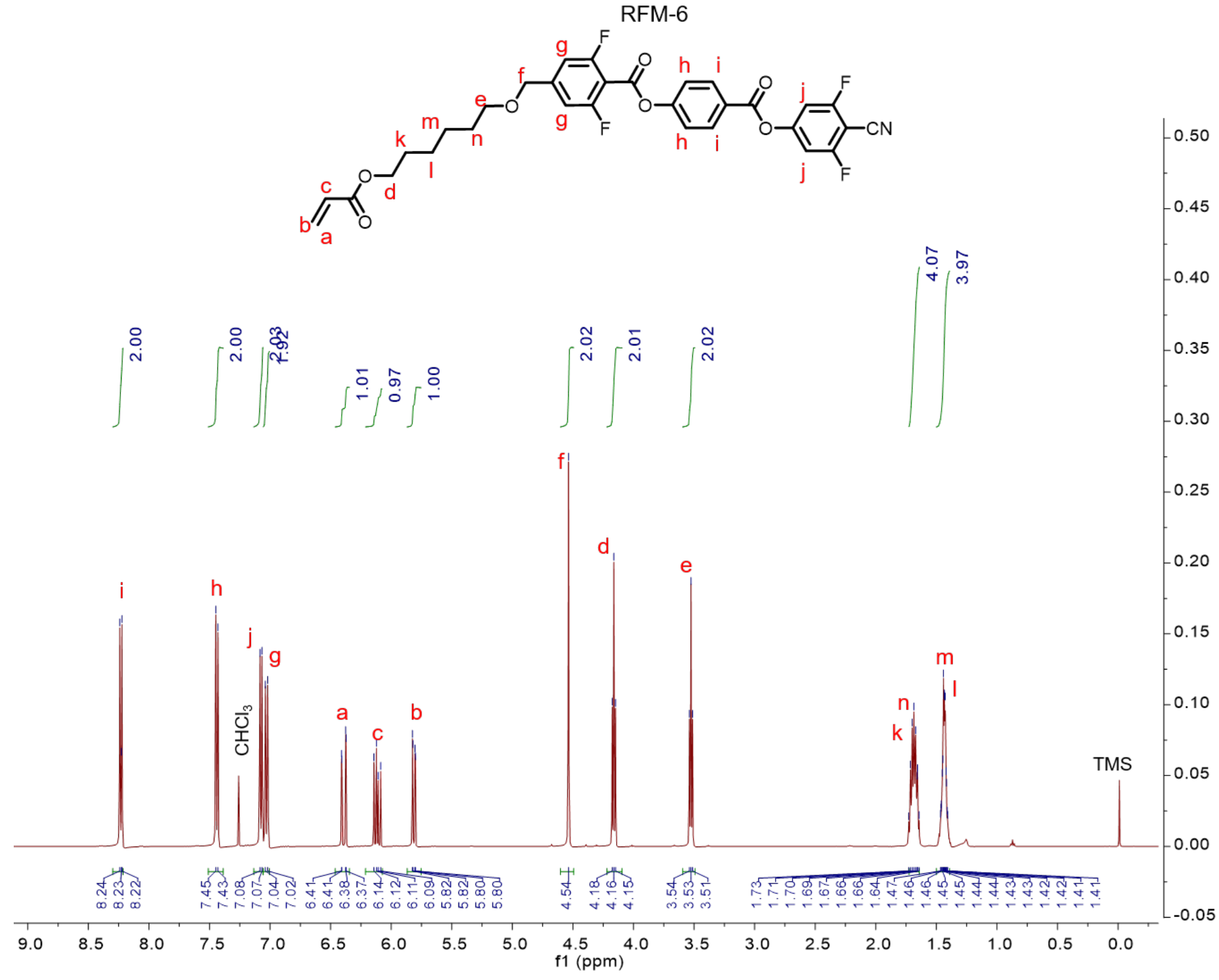

$^{19}F$ NMR of RFM-6

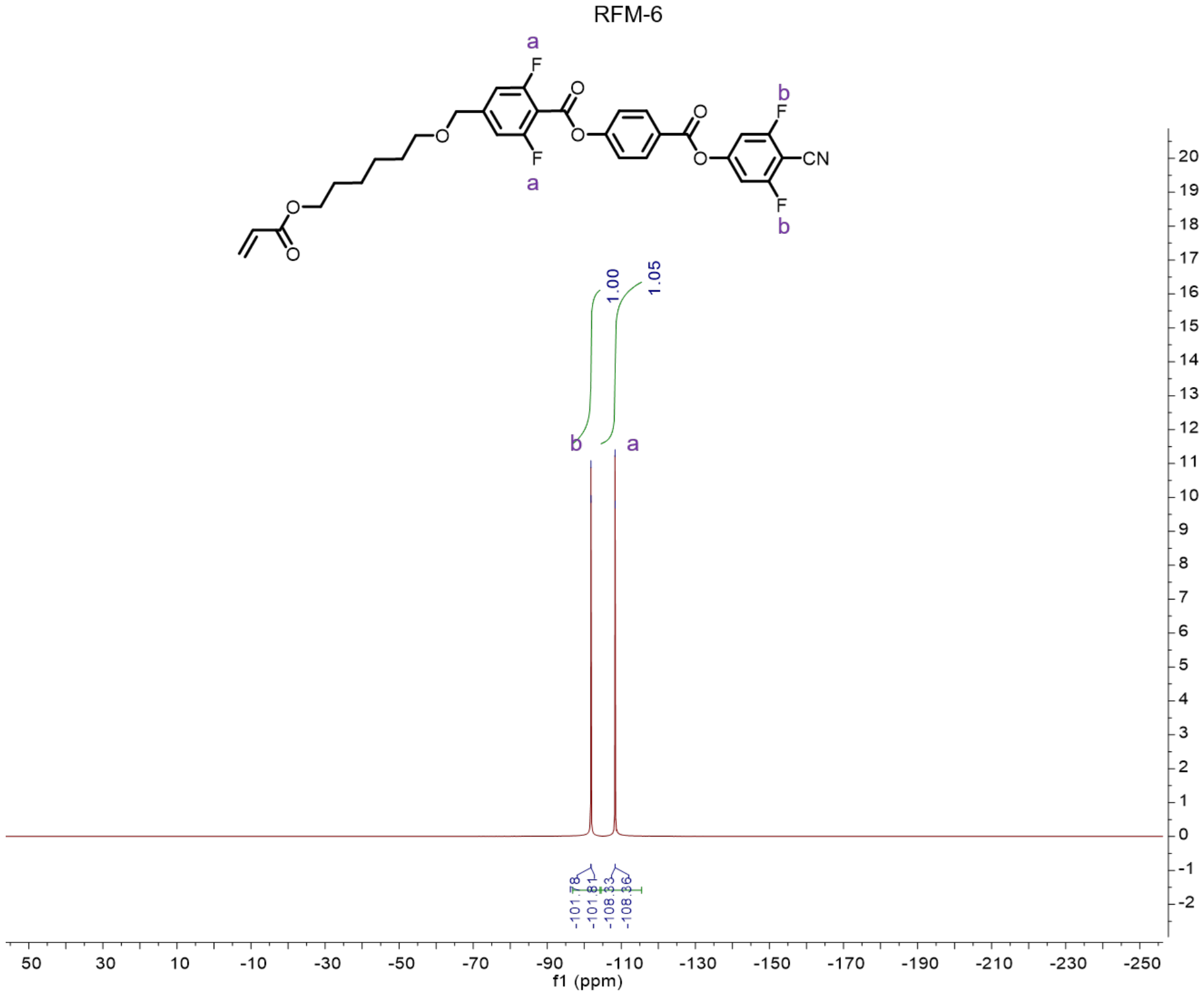

$^{13}C$ NMR of RFM-6

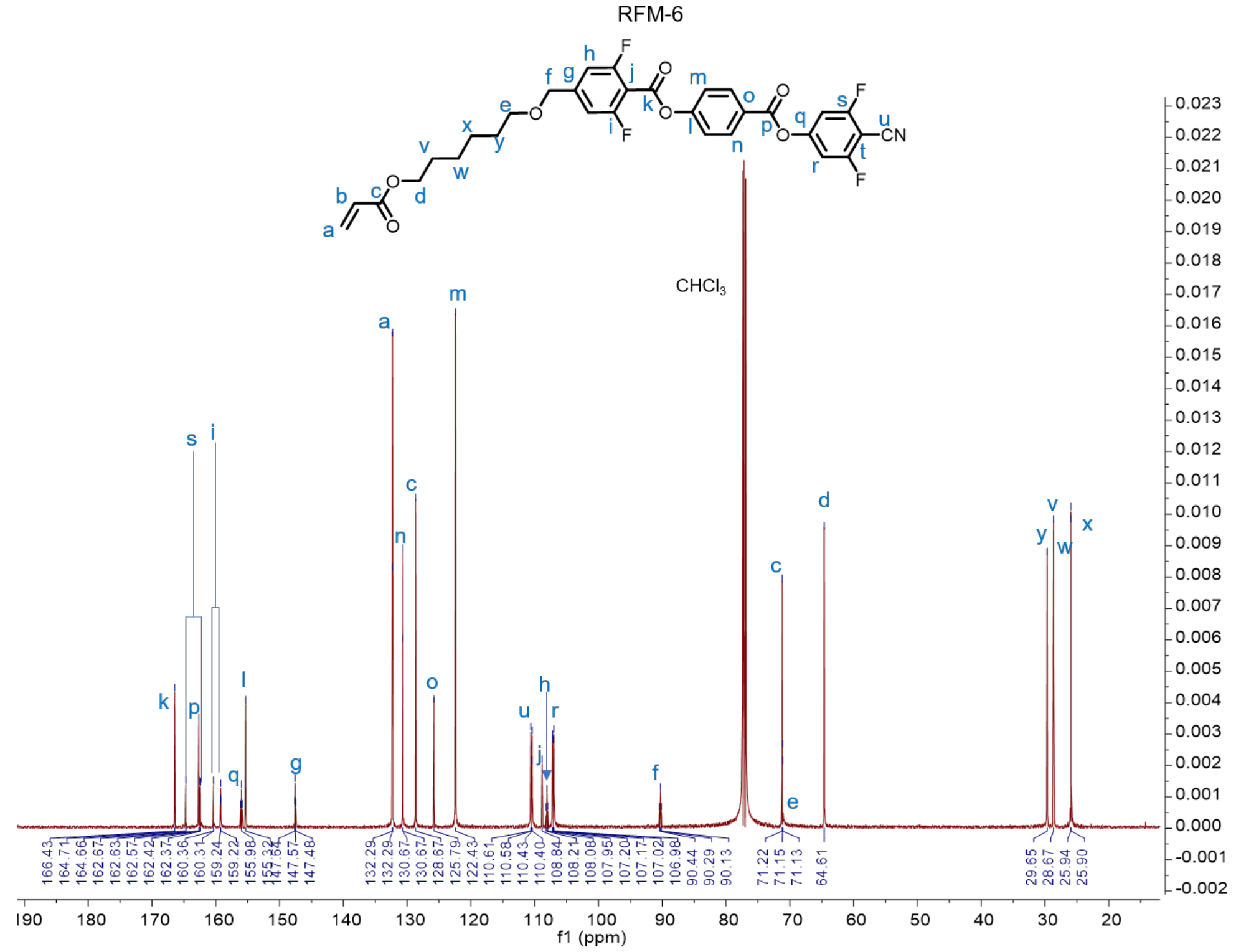

# $^1H$ NMR of RFM-7

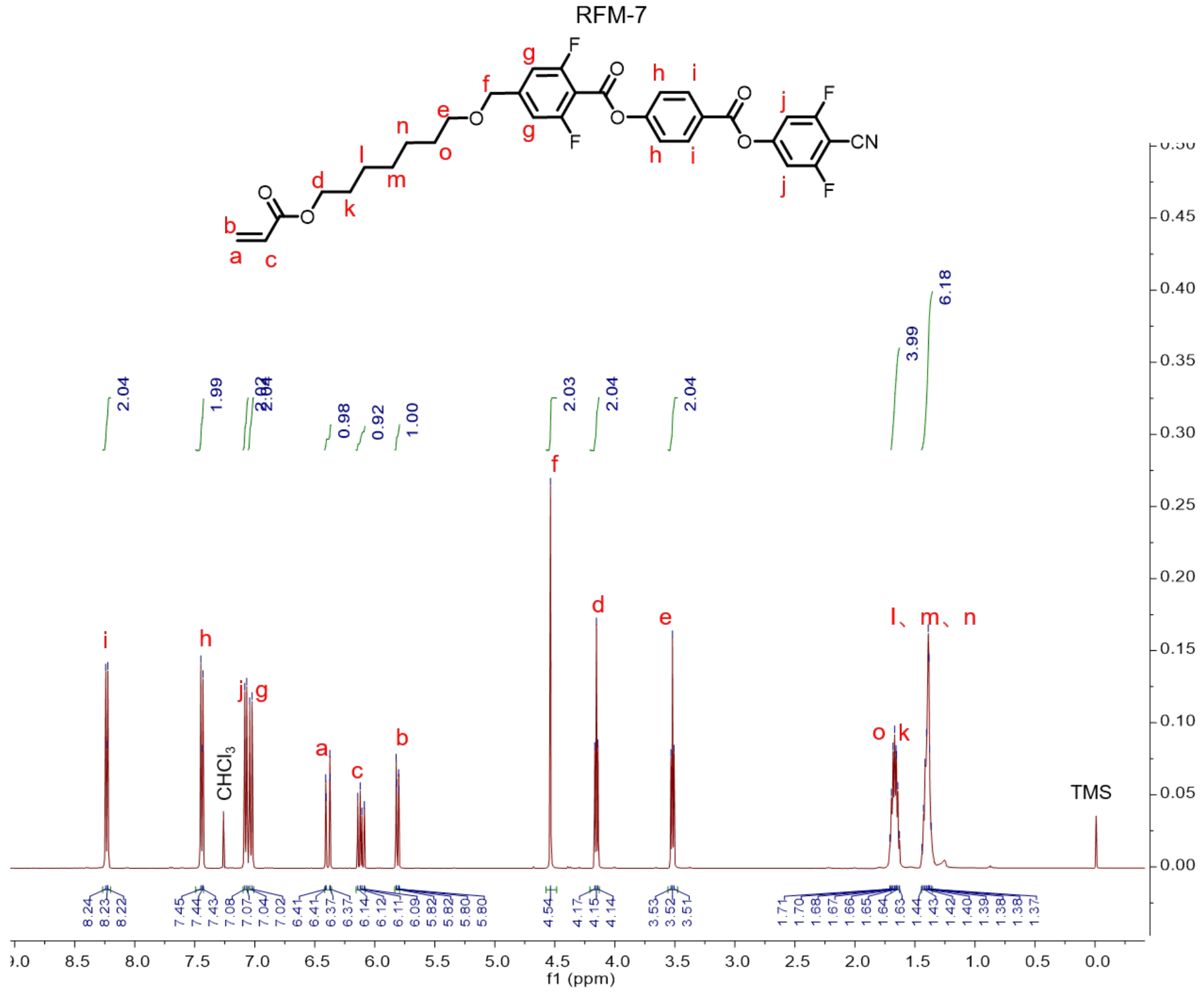

# $^{19}F$ NMR of RFM-7

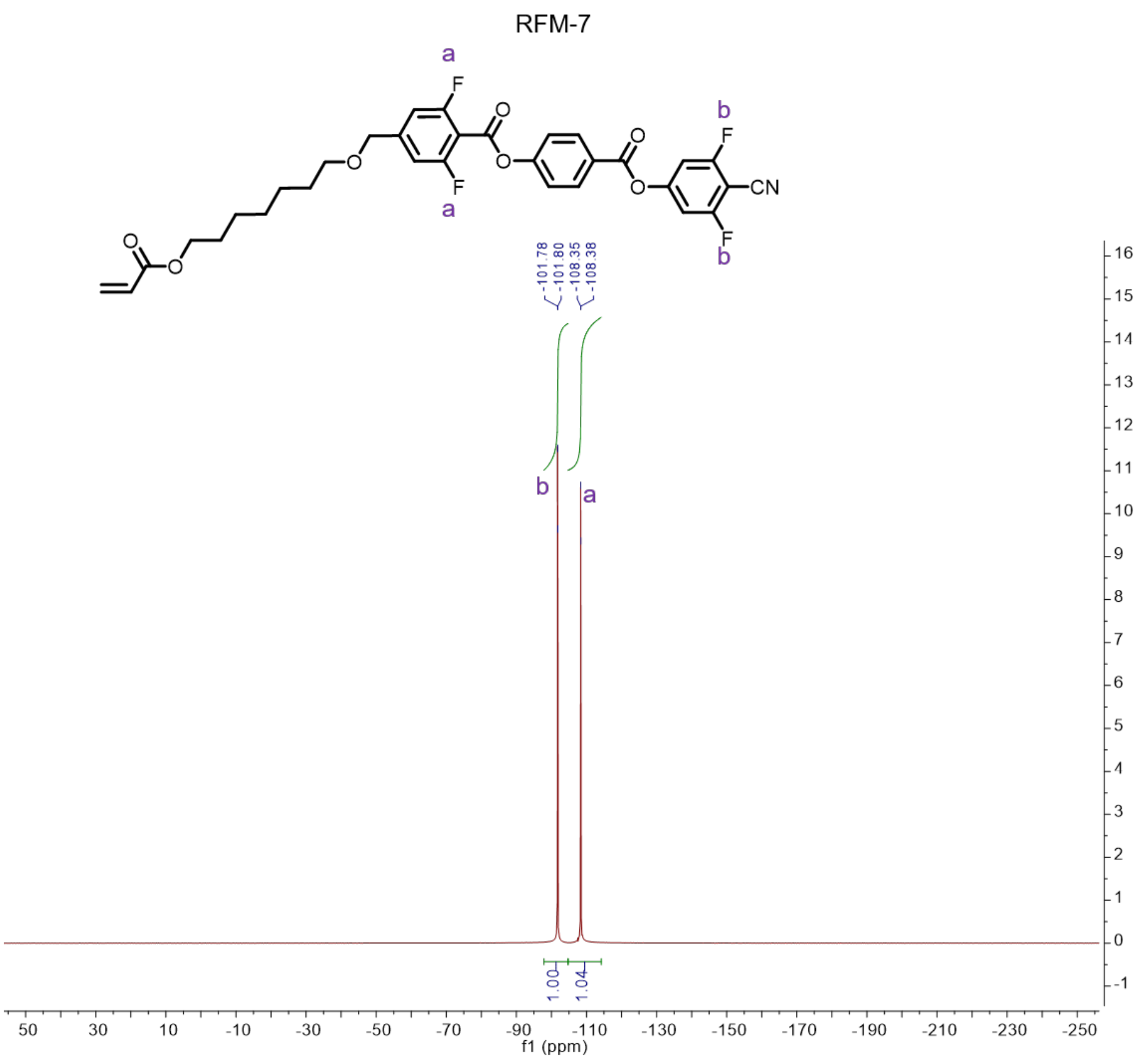

$^{13}C$ NMR of RFM-7

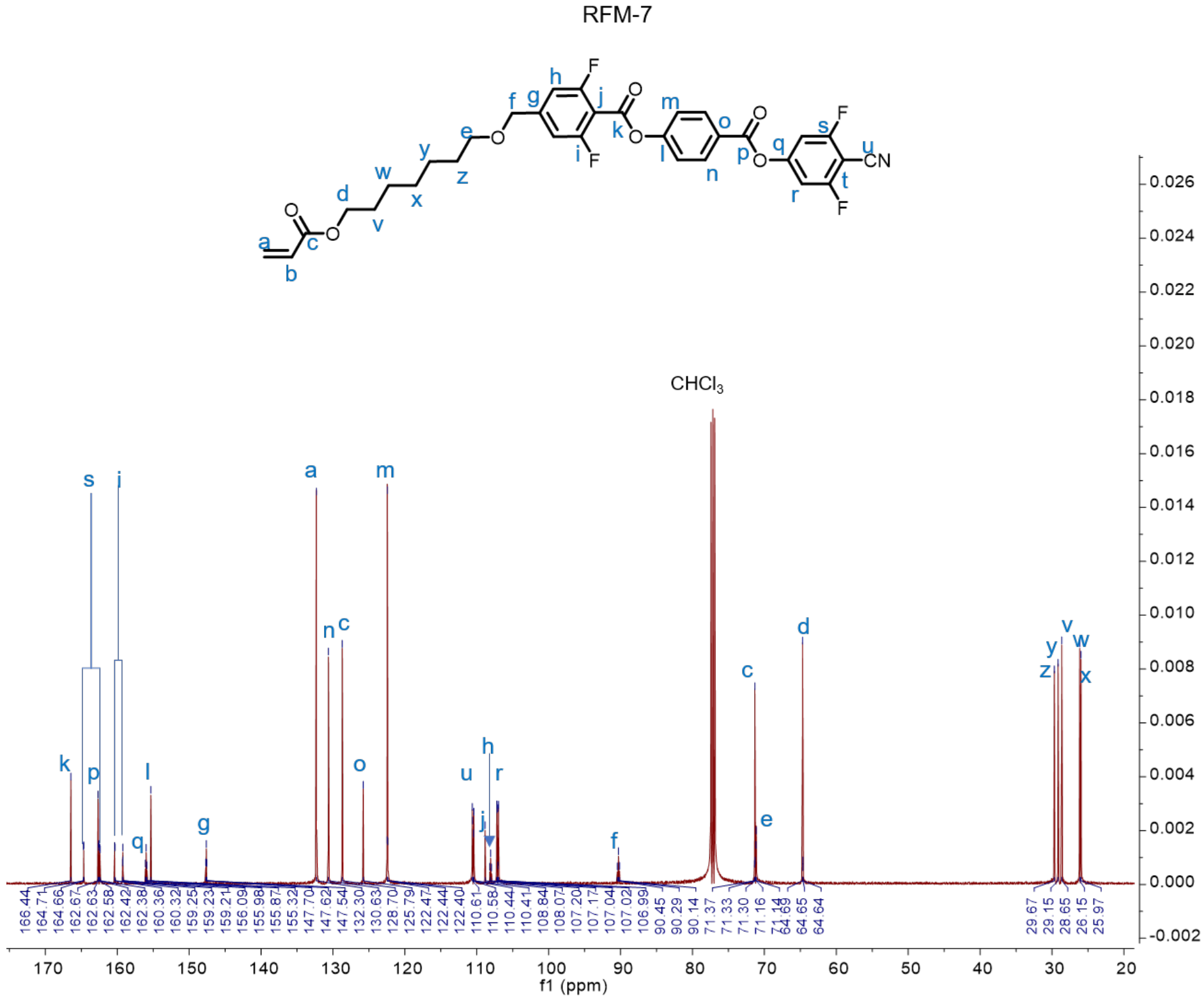

$^{1}H$ NMR of RFM-8

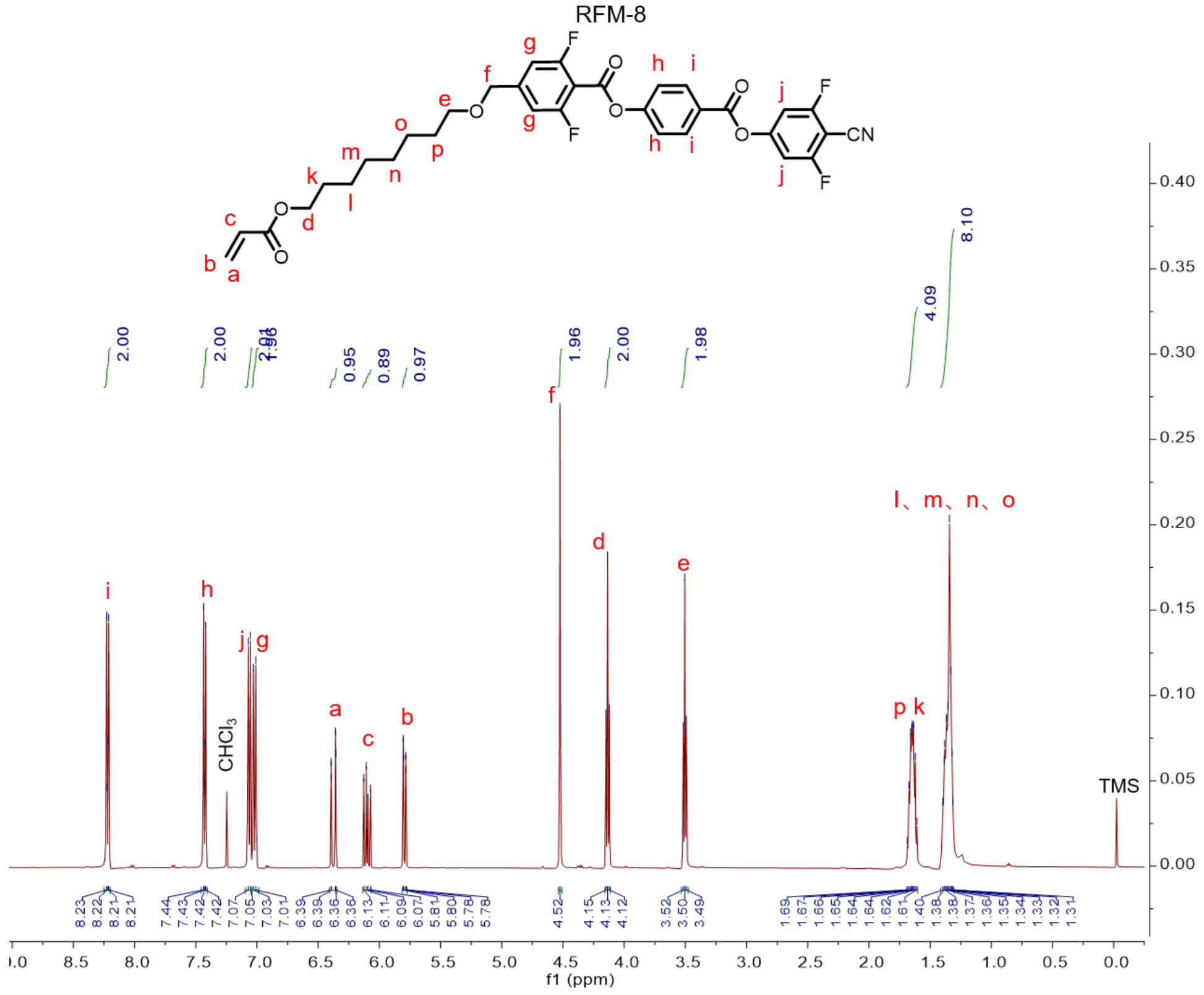

$^{19}F$ NMR of RFM-8

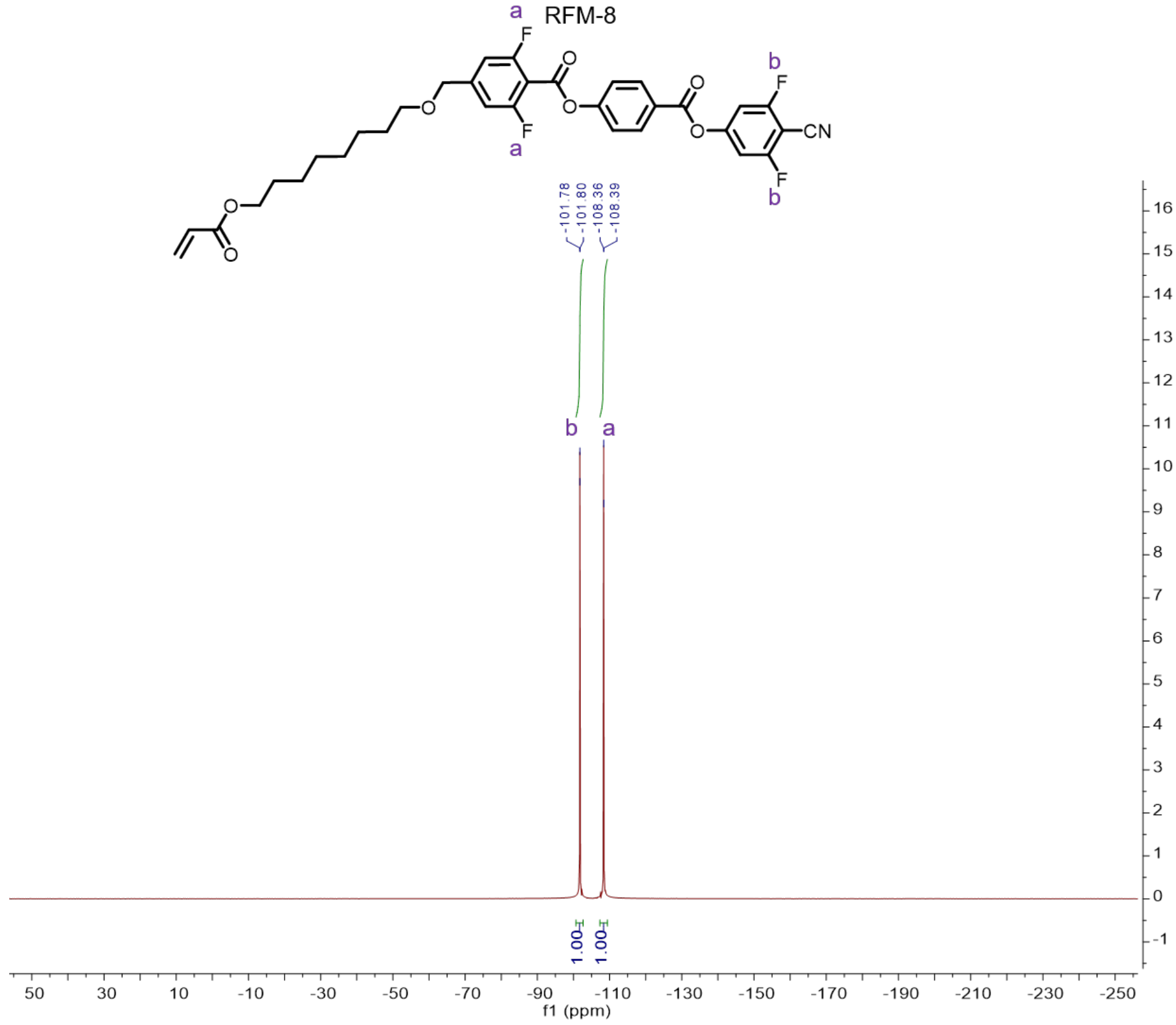

$^{13}C$ NMR of RFM-8

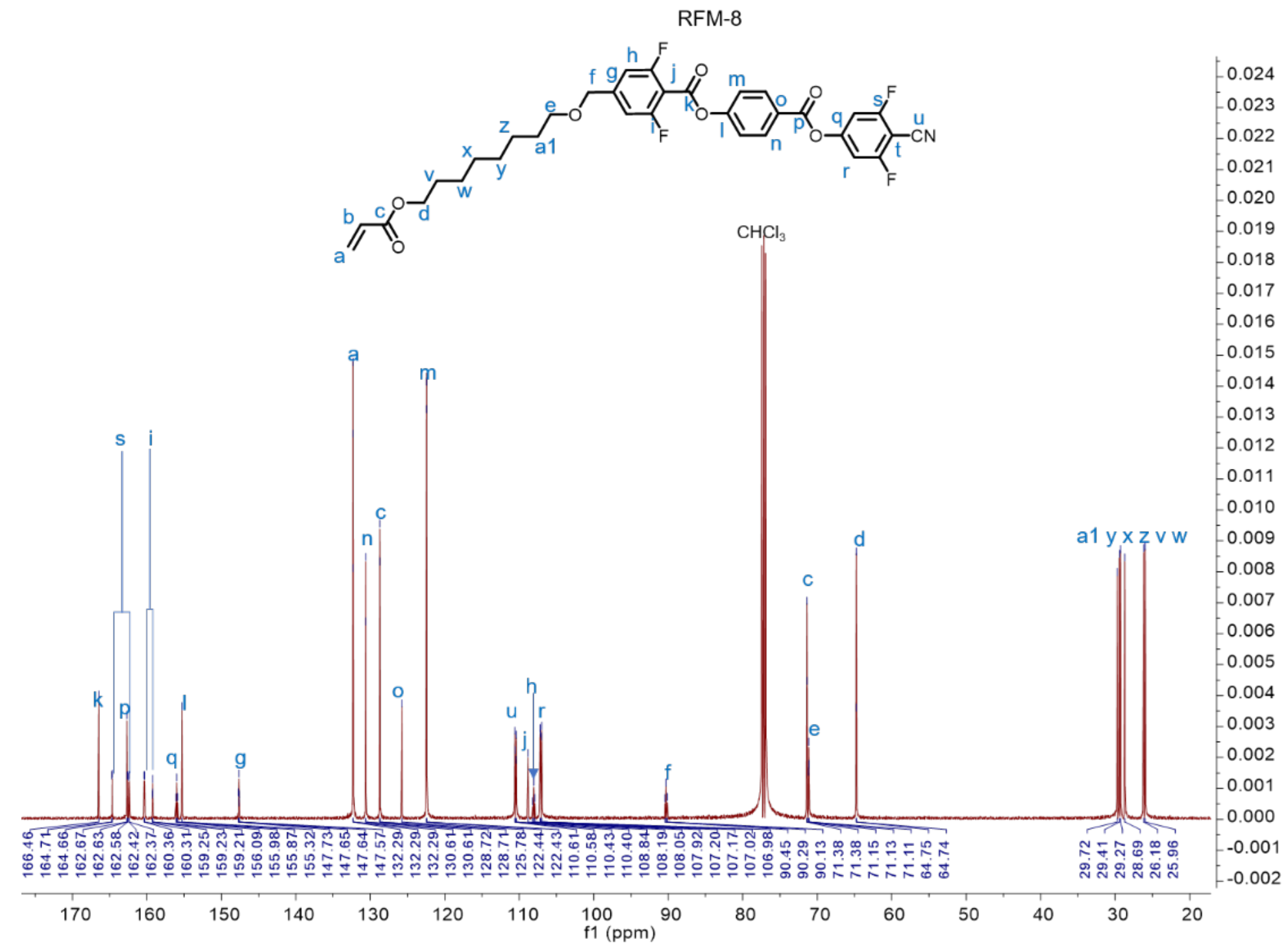